\documentclass[fleqn,usenatbib]{mnras}

\usepackage{newtxtext,newtxmath}

\usepackage[T1]{fontenc}

\DeclareRobustCommand{\VAN}[3]{#2}
\let\VANthebibliography\thebibliography
\def\thebibliography{\DeclareRobustCommand{\VAN}[3]{##3}\VANthebibliography}

\usepackage{graphicx}	
\usepackage{amsmath}	

\usepackage{amssymb}	
\usepackage{subfigure}
\usepackage{float,lscape}

\newcommand{\Msol}{\ensuremath{\mathrm{M}_{\odot}}}
\newcommand{\Lsol}{\ensuremath{\mathrm{L}_{\odot}}}

\title[A first look at MIGHTEE quasars]{MIGHTEE: A first look at MIGHTEE quasars}

\author[S. V. White et al.]{Sarah V. White$^{1,2}$\thanks{E-mail: sarahwhite.astro@gmail.com }, Ivan Delvecchio$^{3}$, Nathan Adams$^{4,5}$, Ian Heywood$^{1,5}$, Imogen H. Whittam$^{5,6}$,
\newauthor    Catherine L. Hale$^{5,7}$, Neo Namane$^{1}$, Rebecca A.\,A. Bowler$^{4}$, Jordan D. Collier$^{8}$  
\\
$^{1}$ Department of Physics and Electronics, Rhodes University, PO Box 94, Makhanda, 6140, South Africa\\
$^{2}$ South African Astronomical Observatory (SAAO), PO Box 9, Observatory, 7935, South Africa\\
$^{3}$ INAF – Osservatorio di Astrofisica e Scienza dello Spazio di Bologna, Via Gobetti 93/3, I-40129 Bologna, Italy\\
$^{4}$ Jodrell Bank Centre for Astrophysics, University of Manchester, Oxford Road, Manchester, M13 9PL, UK\\
$^{5}$ Astrophysics, University of Oxford, Denys Wilkinson Building, Keble Road, Oxford OX1 3RH, UK \\
$^{6}$ Department of Physics, University of the Western Cape, Bellville 7535, South Africa\\
$^{7}$ School of Physics and Astronomy, Institute for Astronomy, University of Edinburgh, Royal Observatory, Blackford Hill, EH9 3HJ Edinburgh, UK \\
$^{8}$ The Inter-University Institute for Data Intensive Astronomy (IDIA), University of Cape Town, Private Bag X3, Rondebosch, 7701, South Africa \\
}

\date{Accepted 2025 July 15. Received 2025 July 15; in original form 2025 May 30}

\pubyear{2015}

\begin{document}
\label{firstpage}
\pagerange{\pageref{firstpage}--\pageref{lastpage}}
\maketitle

\begin{abstract}
In this work we study a robust, $K_{s}$-band complete, spectroscopically-confirmed sample of 104 unobscured (Type-1) quasars within the COSMOS and XMM-LSS fields of the MeerKAT International GHz Tiered Extragalactic Exploration (MIGHTEE) Survey, at $0.60 < z_{\mathrm{spec}} < 3.41$. The quasars are selected via $gJK_{s}$ colour-space and, with 1.3-GHz flux-densities reaching rms $\approx$ 3.0\,\textmu Jy\,beam$^{-1}$, we find a radio-loudness fraction of 5 per cent. Thanks to the deep, multiwavelength datasets that are available over these fields, the properties of radio-loud and radio-quiet quasars can be studied in a statistically-robust way, with the emphasis of this work being on the active-galactic-nuclei (AGN)-related and star-formation-related contributions to the total radio emission. We employ multiple star-formation-rate estimates for the analysis so that our results can be compared more-easily with others in the literature, and find that the fraction of sources that have their radio emission dominated by the AGN crucially depends on the SFR estimate that is derived from the radio luminosity. When redshift dependence is not taken into account, a larger fraction of sources is classed as having their radio emission dominated by the AGN. When redshift dependence {\it is} considered, a larger fraction of our sample is tentatively classed as `starbursts'. We also find that the fraction of (possible) starbursts increases with redshift, and provide multiple suggestions for this trend. 
\end{abstract}





\section{Introduction}

Radio observations give us an unbiased view of the Universe because they enable star formation and black-hole accretion to be detected, without being affected by dust obscuration along the line-of-sight \citep[e.g.][]{Condon1992, McAlpine2015}. Both of these physical processes are important for governing how galaxies form and grow over cosmic time \citep[e.g.][]{Madau1996, Ueda2003, Wolf2003, HopkinsBeacom2006, Malefahlo2022}, but their contributions to the total radio emission need to be carefully disentangled through multiwavelength analyses \citep[e.g.][]{Simpson2006, Smolcic2008, Padovani2014, Whittam2022}. Such techniques are proving ever-more crucial as we have sensitive, next-generation radio telescopes -- such as MeerKAT \citep{Jonas2016} -- that are probing to $\sim$10\,\textmu Jy flux-density levels, and so enabling firm detections \citep[e.g.][]{Best2023, Whittam2024} of star-forming galaxies \citep[e.g.][]{Condon1984} and `radio-quiet' active galactic nuclei (AGN) \citep{Wilman2008}. The latter account for over 90 per cent of the AGN population \citep{Hooper1996} and lack the powerful radio-jets that are prominent in the population of `radio-loud' AGN\footnote{The terms `radio-quiet' and `radio-loud' arise from the dichotomy initially thought to exist in the AGN population \citep{Peacock1986, Xu1999}. However, there are different definitions for radio loudness, historically this being a straight-forward boundary of $\sim10^{24.5}$\,W\,Hz$^{-1}$ at 5\,GHz \citep{Miller1990}. More common, and perhaps more physically meaningful, is the ratio between radio flux-density and optical flux-density \citep{Kellermann1989}, assuming an AGN origin for both. This common definition is the one that we apply to our sample (Section~\ref{sec:radioloudness}).}. Previous analyses of radio-quiet AGN have relied on the stacking of data \citep{Kimball2011, Condon2013}, but individual detections are important for robustly mapping the cosmic histories of star formation and black-hole accretion out to high redshift \citep[e.g.][]{Hopkins2006}.

Regarding the radio-loud AGN population, several studies have investigated the impact of radio jets on their host galaxies, with the associated kinetic-energy generating turbulence in the surrounding medium \citep[e.g.][]{Mukherjee2018} or directly impacting upon molecular clouds \citep[e.g.][]{Croft2006}, triggering their collapse and so promoting star formation \citep[e.g.][]{Kalfountzou2014}. However, there is also evidence of gas being dispersed by relativistic jets and outflows \citep{Combes2013, Morganti2013}, cutting off the fuel that would otherwise accrete onto the central supermassive black-hole or collapse to form stars. Furthermore, the lower density within the cavities generated through the interaction of jets with the surrounding gas \citep[e.g.][]{Fabian2000} gives them buoyancy, and as they `rise', kinetic energy is likely dissipated as heat \citep{McNamara2007}. Through heating of the galaxy halo and interstellar medium (ISM), mechanical feedback may therefore suppress star formation \citep{Croton2006, Croston2007, Best2007}. It is believed that this `self-regulation' through AGN feedback can explain the observed co-evolution \citep{Ferrarese2000, Gebhardt2000, Gultekin2009} of the supermassive black-hole with its host galaxy \citep{King2003, Croton2006, Silk2012}. 

As for the radio-quiet AGN population, we are yet to establish which physical processes are taking place within these sources \citep[see the review by][and references therein]{Panessa2019}, and therefore the influence that these processes may have on their host galaxies. However, a common occurrence in both (high-excitation; \citealt{Best2012}) radio-loud and radio-quiet AGN is the presence of an accretion disc, which is producing sufficient thermal energy to photoionise the surrounding gas, giving rise to the broad-emission-line and narrow-emission-line regions \citep{Antonucci1993, Urry1995, Peterson1997}. This heating may also warm the gas reservoir to such an extent that star formation in the host galaxy is suppressed, and so we need to consider feedback mechanisms across the entire AGN population \citep{Silk1998, Simpson2006, White2023}.   

Perhaps the best-understood subset of AGN, in terms of theoretical modelling of efficient accretion \citep[e.g.][]{Shakura1973} onto the supermassive black-hole at the centre, are quasars \citep[e.g.][]{LyndenBell1969, Shen2009, Down2010}. In these sources, the light from the nuclear region strongly outshines the host galaxy, giving these sources a pointlike appearance in the optical--to--mid-infrared, with luminosities that allow them to be detected out to high redshift \citep[$z > 7$; e.g.,][]{Wang2021}. In terms of the radio emission from these sources, it was previously believed that radio-loud quasars and radio-quiet quasars have different origins to their radio emission, with the general consensus (up to approximately a decade ago) being that the radio emission from the latter is dominated by star formation \citep{, Kimball2011, Condon2013, Bonzini2015}. However, \citet{White2015, White2017} were the first to present evidence (via datasets spanning the optical {\it through} to the radio) that the {\it accretion-related} activity is dominating the total radio emission for large samples of quasars. This finding was supported by studies of `quasar feedback', with (for example) \citet{Zakamska2014} finding that the outflow velocities (measured via the [O\textsc{iii}] emission-line) strongly correlate with the radio luminosity of (obscured, Type-2) radio-quiet quasars, selected in the optical from the Sloan Digital Sky Survey (SDSS; \citealt{York2000}). 

This has been explored further by \citet{Hwang2018}, who find that the trend observed for the $z < 0.8$ quasars studied by \citet{Zakamska2014} extends out towards $2 < z < 4$, as seen in their sample of extremely-red quasars (ERQs). The latter are selected via their high mid-infrared--to--optical ratios  \citep{Ross2015, Hamann2017}, and whilst this may offer insights into the impacts of quasar evolution \citep[e.g.][]{Sanders1988, Hopkins2008} and orientation \citep[e.g.][]{Urry1995} on the observed properties, ERQs are less representative of the wider quasar population. Furthermore, the analysis of the gas kinematics is not of sufficient spatial resolution to rule out the presence of small-scale radio-jets (as opposed to the quasar-driven winds suggested by \citealt{Zakamska2014}). \citet{Jarvis2021} also conducted radio follow-up of a [O{\sc iii}]-luminous quasar sample\footnote{They apply an [O{\sc iii}] luminosity cut of $L_{\mathrm{[O\textsc{iii}]}} > 10^{42.11}$\,ergs\,s$^{-1}$, which
represents the 1.3 per cent most-luminous targets once a redshift cut of $z < 0.2$ is applied to their parent sample of SDSS quasars \citep{Mullaney2013}.}, at high spatial resolution ($\sim$0.3--1.0\,arcsec) with the Karl G. Jansky Very Large Array (JVLA). They reach a median radio-luminosity of  $L_{\mathrm{1.4\,GHz}} = 5.9 \times 10^{23}$\,W\,Hz$^{-1}$ and conclude that at least 57 per cent of the quasars have AGN-dominated radio-emission. If small-scale radio-jets {\it are} present -- as has been observed in radio-quiet quasars via very-long baseline interferometry \citep[e.g.][]{Blundell2003, HerreraRuiz2016, Jarvis2019} -- then the cumulative effect (via the output of kinetic and thermal energy) may have an impact on the evolution of the host galaxies.  

The radio-quiet population is also being well-studied at 144\,MHz with the LOw Frequency ARray (LOFAR), where the LOFAR Two-metre Sky Survey (LoTSS; \citealt{Shimwell2019}) allows for the ($0 < z < 5$) SDSS quasars studied by \citet{Gurkan2019} to be detected with median 144-MHz flux-density $>0.42$\,mJy. They find no evidence of bimodality in the radio-loudness distribution \citep[in agreement with e.g.][]{Lacy2001}, and suggest that the physical mechanisms giving rise to the radio emission from quasars (namely, radio jets and star formation) must be smoothly varying as a function of radio-loudness. When considering a subset of quasars at $0 < z < 3$, and seeing where they lie with respect to the far-infrared--radio correlation (FIRC) determined by \citet{Gurkan2018} for star-forming galaxies detected at 150\,MHz, they conclude that the majority of the radio emission for the radio-quiet(er) quasars is due to star formation. Furthermore, \citet{Gurkan2018} find that the FIRC is dependent on the stellar mass of the quasar host-galaxy, whilst there is also a dependence on redshift, as suggested by the evolving star-formation rates (SFRs) found for quasars by \citet{Macfarlane2021}. The latter study indicates a strong correlation between SFR and quasar bolometric-luminosity that is seen out to $z \sim 2$, and is consistent with far-infrared (FIR) studies \citep{Serjeant2009, Bonfield2011}.  

More recently, \citet{Rankine2021} studied the C\textsc{iv} $\lambda 1549$\,\AA\ emission-line for SDSS quasars (at $1.5 < z< 3.5$) that are detected in LoTSS ($5\sigma = 0.35$\,mJy; \citealt{Rosario2020}). Their sample has typical radio luminosities of $L_{\mathrm{144\,MHz}} \sim 10^{25}$\,W\,Hz$^{-1}$, and whilst they consider star formation, quasar winds, and compact radio-jets as the possible explanations for the radio emission from radio-quiet quasars, they are unable to rule out any particular mechanism (instead concluding that overlapping contributions are possible). 

There may also be a distinction in the physical mechanism that produces the radio emission as a function of quasar {\it colour}, which has been explored by, for example, \citet{Klindt2019} and \citet{Fawcett2020} [who combine SDSS with VLA surveys], and \citet{Rosario2020} [again based on the overlap of SDSS with LoTSS]. Each of these studies find an enhanced radio-detection fraction in the quasars with redder $g - i$ colours, which is suggested to be driven by sources with compact radio-morphology (where small-scale jets may indicate the young evolutionary-phase of the quasar). Whilst simulations by \citet{Rosario2020} show that star formation may be the main origin of the radio emission in the radio-quiet regime, this mechanism is ruled out as an explanation for the enhancement in `red' quasars. However, they do not have far-infrared data to help explore this further, on a source-by-source basis [unlike \citet{White2017} and the current study]. Finally, we note that the MIGHTEE (MeerKAT International GHz Tiered Extragalactic Exploration; \citealt{Jarvis2016, Heywood2022}) unobscured, Type-1 quasars presented in this work are more akin to the `blue' quasars studied by other groups (having `median' $g - i$ colours; \citealt{Klindt2019}), with a more-comparable analysis to be presented in an accompanying paper (White et al., in prep.).

\subsection{Paper outline}

Our quasar sample allows us to investigate intrinsically-fainter radio emission than previous work (with median radio-luminosity, $L_{\mathrm{1.4\,GHz}} = 2.8 \times 10^{23}$\,W\,Hz$^{-1}$, at redshifts up to $z = 3.4$). This is thanks to the unprecedented sensitivity of the MeerKAT telescope, with the 1.3-GHz radio data used in this work provided via the MIGHTEE survey. Whilst there are multiple avenues for exploring the properties of these sources, in this paper we focus on determining the level of radio emission that could be attributed to star formation, which requires careful consideration of how the SFR is estimated. For example, the SFR derived via the radio-luminosity relation of \citet{Yun2001} does not take redshift evolution into account, unlike the infrared--radio correlation (IRRC) found by \citet{Delhaize2017} for star-forming galaxies. The former relation was used by \citet{White2015} for a sample of radio-quiet quasars selected over the XMM-LSS field. Their analysis involved propagating observed-frame measurements (e.g., radio flux-densities) through probability distribution functions for the photometric redshifts, in order to calculate rest-frame properties (e.g., radio luminosities), whereas our sample is completely spectroscopically-confirmed.

In Section~\ref{sec:data}, we describe the datasets used for this work, and in Section~\ref{sec:sample_selection} we explain how we select our sample of quasars (which includes obtaining new spectroscopic redshifts). This is followed by in-depth analysis and discussion of the origin of their radio emission (Section~\ref{sec:radioanalysis}), with our conclusions presented in Section~\ref{sec:conclusions}. J2000 co-ordinates and AB magnitudes are used throughout this work, and we use a $\Lambda$CDM cosmology, with $H_{0} = 70$\,km\,s$^{-1}$\,Mpc$^{-1}$, $\Omega_{m}=0.3$, $\Omega_{\Lambda}=0.7$.

\section{Data}
\label{sec:data}

In this section we describe the deep, multiwavelength datasets that are used for the selection and analysis of sources over two fields:
\begin{enumerate}
    \item The Cosmic Evolution Survey region (COSMOS; \citealt{Scoville2007}), centred at 10:00:28.6, +02:12:21.0 and covering 2.0\,deg$^2$. 
    \item The XMM-{\it Newton} Large Scale Structure (XMM-LSS) field \citep{Pierre2004}. This is centred at 02:18:00.0, $-$7:00:00.0, and we consider the area over which deep near-infrared data is available, which covers 3.5\,deg$^2$. 
\end{enumerate}

\subsection{Near-infrared: (i) UltraVISTA and (ii) VIDEO}
\label{sec:NIRdata}

Our near-infrared (NIR) photometry, {\it YJHK$_{s}$}, is obtained using the Visible and Infrared Survey Telescope for Astronomy (VISTA; \citealt{Sutherland2015}), with coverage over the COSMOS and XMM-LSS fields being provided by the   UltraVISTA \citep{McCracken2012} and
VIDEO surveys \citep{Jarvis2013}, respectively. We use fluxes extracted from 2-arcsec apertures \citep{Adams2020} to calculate apparent magnitudes in the AB system, with limiting magnitudes shown in Table~\ref{tab:mag_limits}. In the {\it K$_{s}$} band, where we use the NIR morphology to aid our quasar selection (see Section~\ref{sec:sample_selection}), the typical spatial resolution (full-width half maximum; FWHM) is 0.8\,arcsec. 

\begin{table}
    \centering
    \begin{tabular}{c|c|c|c}
    \hline
    Filter & Survey  & 
    COSMOS  & XMM-LSS  \\ 
     &   & 
    mag. limit & mag. limit  \\ 
       \hline

        {\it g} & HSC &  27.2 &  26.4 \\
           {\it r} & HSC &  26.8 &  26.1 \\
             {\it i} & HSC &  26.6 &  25.4 \\
               {\it z} & HSC  &  25.9 &  24.6 \\
                 {\it y} & HSC &  25.5 &  24.1 \\
       {\it Y} & UltraVISTA   & 25.5 & -- \\
        {\it J} & UltraVISTA   & 25.3 & -- \\
          {\it H} &  UltraVISTA   & 25.0 & -- \\
            {\it K$_{s}$} &  UltraVISTA   & 24.8 & -- \\
   {\it Y} & VIDEO & --  & 25.1 \\
        {\it J} & VIDEO & --  & 24.7 \\
          {\it H} & VIDEO & --  & 24.2 \\
            {\it K$_{s}$} & VIDEO & --   & 23.8 \\
    \hline
    \end{tabular}
    \caption{AB magnitude limits ($5\sigma$), measured using 2-arcsec-diameter apertures, for the optical and near-infrared photometry provided via the Hyper Suprime-Cam (HSC) Subaru Strategic Program \citep{Aihara2018a, Aihara2018b}, UltraVISTA \citep{McCracken2012}, and the VIDEO survey \citep{Jarvis2013}. There is variation (up to 1.7\,mag) in the magnitude limits over the three XMM-LSS tiles (see table 1 of \citealt{Adams2020}) and so we provide the conservative limits here.} 
    \label{tab:mag_limits}
\end{table}

\subsection{Optical: Hyper Suprime-Cam Subaru Strategic Program}
\label{sec:opticaldata}

The Hyper Suprime-Cam \citep[HSC;][]{Miyazaki2012, Miyazaki2018} Subaru Strategic Program \citep{Aihara2018a, Aihara2018b} provides {\it grizy} photometry over the COSMOS and XMM-LSS fields through its `UltraDeep' and `Deep' tiers, respectively. The magnitude limits are provided in Table~\ref{tab:mag_limits} to help compare the depths of the optical data over the two fields, and the high image quality of the data is signified by achieving 0.65-arcsec seeing for the COSMOS field, and 0.85-arcsec seeing for the XMM-LSS field (as measured in the {\it i} band). A part of the image processing for this data release was the use of artificial sources to assess the impact of source blending on the photometry, with \citet{Aihara2018b} quoting the typical uncertainty to be 0.05\,mag in magnitude and colour estimates for galaxies.

\subsection{Radio: MIGHTEE}
\label{sec:radiodata}

This work is based on Data Release 1 \citep{Hale2025} of $\sim$1.3-GHz radio data from the MIGHTEE survey \citep{Jarvis2016}. This is one of the Large Survey Projects being conducted on MeerKAT \citep{Jonas2016}, and with 139.6 hours of on-source time for the COSMOS field (22 MeerKAT pointings, covering 4.2\,deg$^2$), and 297.9 hours of on-source time for the XMM-LSS field (45 MeerKAT pointings), we reach a typical {\it thermal noise} of $<2$\,\textmu Jy\,beam$^{-1}$. Like \citet{Heywood2022}, \citet{Hale2025} provide two sets of images, where different \citet{Briggs1995} robust parameters have been used to promote either spatial resolution ($-$1.2) or sensitivity (0.0). Given that the present work is concerned with the detection of `radio-quiet' quasars (RQQs), we use the deeper radio images (rms $\approx$ 3.0\,\textmu Jy\,beam$^{-1}$; see Section~\ref{sec:faintemission}) at the expense of spatial resolution (these images being at $\sim$9 arcsec), with the acknowledgement that these images are limited by confusion noise. With the synthesised beam being oversampled to satisfy Nyquist's theorem, the resulting pixel scale is 1.1\,arcsec\,pixel$^{-1}$. Accompanying the radio images are the effective-frequency maps provided by \citet{Hale2025}, following \citet{Heywood2022}. These take into account the deviation of the measured frequency, associated with a  single pixel, from the nominal band-centre frequency -- which is a result of multi-frequency synthesis (MFS) imaging techniques -- as a function of position. When extracting pixel-values from the radio images (Section~\ref{sec:faintemission}), the effective frequency is considered and a power-law function ($S_{\nu} \propto \nu ^{\alpha}$, with spectral index, $\alpha = -0.7$) is applied in order to scale all measurements to exactly 1.3\,GHz.

\subsection{Far-infrared: {\it Herschel} observations}
\label{sec:infrareddata}

For both fields, where available, we use FIR data from the {\it Herschel} Space Observatory \citep{Pilbratt2010}. These data are in two of the Photodetector Array Camera and Spectrometer \citep[PACS;][]{Poglitsch2010} bands, at 100 and 160\,\textmu m, and in the three Spectral and Photometric Imaging REceiver \citep[SPIRE;][]{Griffin2010} bands, at 250, 350 and 500 \textmu m, and have corresponding beam sizes of 6.7, 11.0\,arcsec and 18.1, 25.2, 36.6\,arcsec (at FWHM), respectively. 

Specifically, for the COSMOS field we use `super-deblended' FIR measurements \citep{Jin2018} extracted from the PACS Evolutionary Probe \citep[PEP;][]{Lutz2011}, CANDELS-{\it Herschel} (PI: M. Dickinson), and {\it Herschel} Multi-tiered Extragalactic Survey \citep[HerMES;][]{Oliver2012}, via the method of \citet{Liu2018}. This provides photometry with 1-$\sigma$ detection limits of 1.44, 3.55\,mJy at 100, 160\,\textmu m, and 1.77, 2.68, 2.91\,mJy at 250, 350 and 500 \textmu m.
Meanwhile, for the XMM-LSS field we use the `main' catalogue\footnote{\url{https://herschel-vos.phys.sussex.ac.uk/herschelhelp/q/cone/info}} from the {\it Herschel} Extragalactic Legacy Project \citep[HELP;][]{Shirley2021}, for which PACS and SPIRE flux-densities have been extracted using the Bayesian tool, {\sc XID}+ \citep{Hurley2017}. The uncertainties in these measurements cease to be Gaussian where the error is dominated by confusion, and the flux-density level at which this happens can serve as a proxy for a 1-$\sigma$ detection limit. These limits are 12.5, 17.5\,mJy at 100, 160\,\textmu m, and 4.0, 4.0, 4.0\,mJy at 250, 350 and 500 \textmu m.

\section{Sample selection}
\label{sec:sample_selection}

For this first MIGHTEE paper on the investigation of the radio emission from radio-quiet AGN, we wish to consider the {\it quasar} subset of the total AGN population, and so need to create a robust sample that ensures high reliability (at the expense of completeness). The wider radio-quiet AGN population is investigated in a companion paper (White et al., in prep.) and considers the impact of different selection criteria on the conclusions that are drawn, as to whether black-hole accretion or star formation is dominating the radio emission from these objects.  

For the current work, we select our quasar sample based on $g - J$ and $J - K_{s}$ colours (\citealt{Maddox2008}; their figure 1), using a catalogue \citep{Bowler2020, Adams2020}\footnote{Released internally to the MIGHTEE collaboration.} that combines data from VIDEO \citep{Jarvis2013} over the XMM-LSS field, UltraVISTA \citep{McCracken2012} over the COSMOS field, and HSC \citep{Aihara2018a, Aihara2018b} over both fields. This optical/NIR-based selection is chosen because the $J - K$ colour has been shown by \citet{Maddox2008} to be very effective in selecting quasars, with less than 10 per cent being missed on account of dust reddening. In addition, as demonstrated in figure 3 by \citet{White2015}, the box-shaped region defined by $-0.30 < (J - K_{s}) < 1.10$, $-1.00 < (g - J) < 0.65$ is where `Type 1' (i.e. unobscured) quasars tend to occupy $gJK_{s}$ space. This is also the region over which the optical/NIR colours for a `pure quasar', with no host-galaxy contribution towards the emission (P.C. Hewett, private communication\footnote{A refined version of the quasar template, and a link to Python code, is now available in \citet{Temple2021}.}), evolve over the redshift range $0 < z \lesssim 3.0$.

\begin{figure*}
    \centering
    \includegraphics[width=0.75\linewidth]{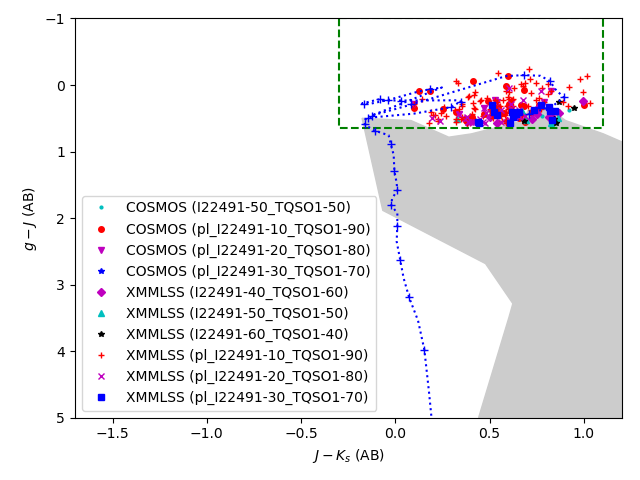}
    \caption{Quasar selection in $gJK_{s}$ colour-colour space (Section~\ref{sec:sample_selection}), following the work of \citet{Maddox2008} and \citet{White2015}. The green, dashed line demarcates the selection box defined by \citep{White2015}, who also use grey shading to indicate the parameter space that is occupied by evolving (inactive-)galaxy templates with varying dust extinction \citep{Adams2020}. The blue, dotted line shows the evolutionary track of a `pure quasar' with no host-galaxy contribution towards the optical/NIR colours (P.C. Hewett, private communication -- see main text.). The start of this track (i.e. where $z = 0$) is within the selection box, and blue `+' symbols are used to indicate redshift steps of $\Delta z = 0.2$ towards $z = 4.8$ (beyond the plot range). For a description of the different AGN templates that appear in the legend, please see Table~\ref{tab:AGN_templates}. [Note that axis ranges are chosen to ease comparison with figure 3 of \citet{White2015}.]}
    \label{fig:gJK_space}
\end{figure*}

In combination with the optical/NIR magnitudes -- where the $g, r, J$ and $K_{s}$ bands were used to assess the quality of the photometry --  we consider results from photometric template-fitting \citep{Adams2020} conducted using {\sc LePhare} \citep{Ilbert2006}. This employs a $\chi^2$-minimisation method for fitting spectral energy distributions (SEDs) to the multiband photometry, and for our sample selection we retain objects (within the $gJK_{s}$ selection region demarcated in Figure~\ref{fig:gJK_space}) that are best-modelled by one of 18 AGN templates \citep[][and outlined in Table~\ref{tab:AGN_templates}]{Salvato2009}, rather than a template belonging to the stellar library \citep{ Hamuy1992, Hamuy1994, Bohlin1995, Pickles1998, Chabrier2000} or the galaxy library \citep{Bruzual2003, Polletta2007, Ilbert2009}. The templates were evolved to different redshifts, $z$, from 0.00 to 6.00 in steps of 0.04, and then in steps of 0.10 up to a maximum of $z = 9.00$. The galaxy templates also have varying degrees of extinction applied to them, giving rise to galaxy tracks that span a large portion of $gJK_{s}$ colour-colour space, as indicated by the grey polygon in Figure~\ref{fig:gJK_space}. We refer the reader to section 3.1 of \citet{Adams2020} for further details of the template-fitting. 


\begin{table*}
    \centering
    \begin{tabular}{c|c|r|r|r}
    \hline
   Template name & Description & \multicolumn{2}{c|}{Number of selected quasars}  & Total \\
    & & in COSMOS & in XMM-LSS & \\
   \hline

        {\bf pl\_I22491-10\_TQSO1-90}  & Starburst/ULIRG (10\%) and Type-1 quasar (90\%) hybrid, plus power-law & 25 & 105 & 130 \\ 
         {\bf pl\_I22491-20\_TQSO1-80} & Starburst/ULIRG (20\%) and Type-1 quasar (80\%) hybrid, plus power-law & 9 & 23 & 32 \\ 
             {\bf pl\_I22491-30\_TQSO1-70} & Starburst/ULIRG (30\%) and Type-1 quasar (70\%) hybrid, plus power-law & 2 & 16 & 18 \\ 

    {\bf I22491-40\_TQSO1-60} & Starburst/ULIRG (40\%) and Type-1 quasar (60\%) hybrid & 0 & 14 & 14 \\ 
     {\bf I22491-50\_TQSO1-50} & Starburst/ULIRG (50\%) and Type-1 quasar (50\%) hybrid & 3 & 5 & 8 \\  
      {\bf I22491-60\_TQSO1-40} & Starburst/ULIRG (60\%) and Type-1 quasar (40\%) hybrid & 0 & 6 & 6 \\ 
       I22491-70\_TQSO1-30 & Starburst/ULIRG (70\%) and Type-1 quasar (30\%) hybrid & 0  & 0 & 0 \\  
I22491-80\_TQSO1-20 & Starburst/ULIRG (80\%) and Type-1 quasar (20\%) hybrid & 0  & 0 & 0 \\  
I22491-90\_TQSO1-10 & Starburst/ULIRG (90\%) and Type-1 quasar (10\%) hybrid & 0   & 0 & 0 \\  

S0-10\_QSO2-90 & Passive galaxy \citep[][10\%]{Polletta2007} and Type-2 quasar (90\%) hybrid & 0  & 0 & 0 \\  
S0-20\_QSO2-80 & Passive galaxy \citep[][20\%]{Polletta2007} and Type-2 quasar (80\%) hybrid & 0  & 0 & 0 \\  
S0-30\_QSO2-70 & Passive galaxy \citep[][30\%]{Polletta2007} and Type-2 quasar (70\%) hybrid & 0  & 0 & 0 \\  
S0-40\_QSO2-60 & Passive galaxy \citep[][40\%]{Polletta2007} and Type-2 quasar (60\%) hybrid & 0  & 0 & 0 \\  
S0-50\_QSO2-50 & Passive galaxy \citep[][50\%]{Polletta2007} and Type-2 quasar (50\%) hybrid & 0  & 0 & 0 \\  
S0-60\_QSO2-40 & Passive galaxy \citep[][60\%]{Polletta2007} and Type-2 quasar (40\%) hybrid & 0  & 0 & 0 \\  
S0-70\_QSO2-30 & Passive galaxy \citep[][70\%]{Polletta2007} and Type-2 quasar (30\%) hybrid & 0  & 0 & 0 \\  
S0-80\_QSO2-20 & Passive galaxy \citep[][80\%]{Polletta2007} and Type-2 quasar (20\%) hybrid & 0  & 0 & 0 \\  
S0-90\_QSO2-10 & Passive galaxy \citep[][90\%]{Polletta2007} and Type-2 quasar (10\%) hybrid & 0  & 0 & 0 \\  
\hline
 &  & 39 & 169 & 208 \\
\hline

    \end{tabular}
    \caption{The AGN templates, courtesy of \citet{Salvato2009},  that were used by \citet{Adams2020} for the photometric fitting (see Section~\ref{sec:sample_selection}). Where the description includes ``plus power-law", this is referring to a power-law extrapolation into the ultra-violet, based upon work by \citet{Scott2004}. The Type-1 quasar (QSO1) component specifically originates from the `top quartile' (`TQSO1') of quasar templates compiled by \citet{Hatziminaoglou2005}, where the ranking is according to their infrared/optical flux ratios. The templates that have their names in bold are those that satisfy all of the selection criteria and so appear in Figure~\ref{fig:gJK_space}.}
    \label{tab:AGN_templates}
\end{table*}

We also have information about the morphology of the quasars in the $K_{s}$ band, via the K\_CLASS\_STAR parameter [obtained through running {\sc SourceExtractor} software \citep{Bertin1996} over the VISTA images]. We apply the criterion of K\_CLASS\_STAR $> 0.9$, to select sources with point-like morphology, and also restrict the selection to sources with $K_{s} < 22.4$ (following \citealt{White2015}) to ensure high reliability for this morphology metric. This results in 208 objects in our parent quasar sample, with a summary of the selection steps (and the number of sources that remain selected after each step) in Table~\ref{tab:quasar_sample_selection}. 

\begin{table}
    \centering
    \begin{tabular}{l|c}
    \hline
    Selection criterion that is applied  & Number  \\ 

       \hline

        1. NIR sources with good-quality photometry  &  627,169 \\
        2. Best-fit by an AGN template (Table~\ref{tab:AGN_templates})  &  57,216 \\
        3. Within the selection-box of $gJK_{s}$ colour space (Figure~\ref{fig:gJK_space}) &  1,262 \\
        4. K\_CLASS\_STAR $> 0.9$ &   544 \\
        5. $K_{s} < 22.4$ &  208 \\
        6. Ensuring spectroscopic completeness (Section~\ref{sec:salt_spectra}) &  104 \\
      
    \hline
    \end{tabular}
    \caption{A summary of the steps taken in selecting our sample of quasars, with the number of quasars being those that remain in selection {\it after} the criterion has been applied (Section~\ref{sec:sample_selection}).} 
    \label{tab:quasar_sample_selection}
\end{table}

\subsection{Existing spectroscopy in the literature}
\label{sec:existing_spectra}

Thorough checks against existing spectroscopy over the COSMOS and XMM-LSS fields (\citealt{Lilly2009, Cool2013, Melnyk2013, Davies2015, White2015, Tie2017, Lyke2020, Flesch2021}, and via the COSMOS Archive) allow us to compile robust redshifts for 147 sources. As part of this assessment we cross-reference redshifts from the Million Quasars (Milliquas) Catalogue (update v7.2; \citealt{Flesch2019,Flesch2021}) with redshift-quality flags made available via:

\begin{itemize}
    \item{the G10/COSMOS region of Galaxy And Mass Assembly (GAMA) survey \citep{Davies2015}, zCOSMOS Data Release DR3 (January 2016; \citealt{Lilly2009}), and the PRIsm MUlti-object Survey (PRIMUS; \citealt{Cool2013}), for the COSMOS field,}
    \item{and via the Australian Dark Energy Survey (OzDES; \citealt{Lidman2020}), the deep XMM-Large Scale Structure catalogue (2XLSSd; \citealt{Melnyk2013}), and (again) PRIMUS \citep{Cool2013}, for the XMM-LSS field.}
\end{itemize}
See Appendix~\ref{app:assess_speczs} for an appreciation of the different redshifts that exist in the literature for a given quasar, and the specific values used for this work (these being highlighted in bold in Tables~\ref{tab:cosmos_speczs} and ~\ref{tab:xmmlss_speczs}).

\subsection{New redshifts via SALT longslit spectroscopy}
\label{sec:salt_spectra}

We aim to collate spectroscopic redshifts for as large a sample as possible, and maintain completeness in the $K_{s}$-band magnitude distribution. The latter allows us to probe lower stellar-mass/black-hole systems, and the old stellar population of quasar host-galaxies, than is possible with a purely optically-selected sample. For our parent sample of 208 objects, we first focus on those with $r < $ 22.3\,mag to ensure that the quasar can be observed with a good signal-to-noise ratio during a single track on the Southern African Large Telescope (SALT), noting the targets' equatorial declinations.
We find that existing spectroscopic coverage (Section~\ref{sec:existing_spectra}) peters out at $K_{s} \sim $ 21.2\,mag, leaving one quasar candidate with $K_{s} < $ 21.0\,mag in COSMOS and three in XMM-LSS lacking optical spectra. We therefore followed up these four quasars using SALT (proposal 2019-2-SCI-046; PI: White) -- see Table~\ref{tab:new_speczs} -- which results in a final sample of 104 quasars (Table~\ref{tab:quasar_sample_selection}).

We use the pg0900 grating for longslit spectroscopy on the Robert Stobie Spectrograph (RSS), and use a slitwidth of 2~arcsec to accommodate the typical seeing conditions\footnote{For example, see Figure~3.1 of the 2022-2 SALT Call for Proposals: \url{https://pysalt.salt.ac.za/proposal\_calls/current/ProposalCall.html}}. Grating angles are chosen to adjust the wavelength range and so optimise the detection of emission lines, based on photometric redshift estimates \citep{Adams2020}. We reduce the resulting spectra using the RSSMOSPipeline\footnote{\url{https://github.com/mattyowl/RSSMOSPipeline}} \citep{Hilton2018}, and present the spectra (from three exposures per quasar) in Appendix~\ref{app:new_speczs}.


For the quasar at R.A. = 02:24:44.33, Dec. = $-$04:45:36.0, it is unfortunate that the peak of the Lyman-$\alpha_{1216}$ line coincides with the first chip gap of the RSS CCD, but we can still see the redder shoulder of the N\textsc{v}$_{1240}$ line (Figure~\ref{fig:SALT_spectra}). The approximate observed wavelength for Lyman-$\alpha_{1216}$ (4275\,\AA), combined with the line at 5452\,\AA, leads to an emission-line ratio that indicates that the latter line is C\textsc{iv}$_{1549}$. From this we calculate a redshift of $z = 2.52$.

We take care to compare the target spectrum with the sky spectrum when identifying distinctive emission lines for the quasar at R.A. = 02:24:57.57, Dec. = $-$04:56:59.5. The photometric redshift ($z = 1.72$; \citealt{Adams2020}) was used to position the grating in order to detect the C\textsc{iv}$_{1549}$ emission line, and (indeed) the two lines labelled in Figure~\ref{fig:SALT_spectra} indicate that they belong to C\textsc{iii}]$_{1909}$ and  C\textsc{iv}$_{1549}$ at $z = 1.76$. 

Due to the quasar at R.A. = 02:25:08.57, Dec. = $-$04:25:12.8 being observed with a less-standard grating angle, the wavelength calibration does not perform optimally. However, the presence of a distinctive Lyman-$\alpha_{1216}$/N\textsc{v}$_{1240}$ profile allows us to unambiguously identify emission lines and apply a calibration correction (equating to $\delta z =0.05$). In doing so, we estimate the redshift for this source to be $z = 2.15$.

The signal-to-noise for the quasar at R.A. = 09:57:58.40, Dec. = +02:17:29.1 was lower than expected, but we can still distinguish emission lines at 5298\,\AA~and 6529\,\AA. A spectroscopic redshift of $z = 2.42$ is therefore determined from C\textsc{iv}$_{1549}$ and C\textsc{iii}]$_{1909}$. This is in close agreement with the photometric redshift, $z = 2.48$ \citep{Adams2020}. 

\begin{figure}
    \centering
    \includegraphics[width=1.0\linewidth]{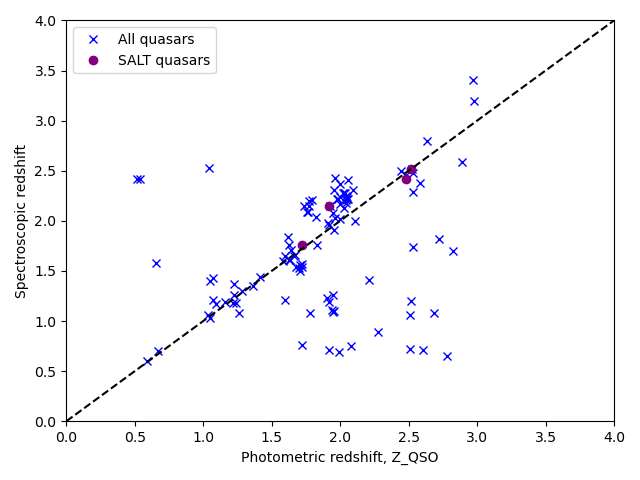}
    \caption{A comparison of the spectroscopic redshift with the photometric redshift \citep[Z\_QSO;][]{Adams2020} for the (reduced) parent sample of 104 quasars (Table~\ref{tab:quasar_sample_selection}). New redshifts from SALT are highlighted in purple.}
    \label{fig:comparing_redshifts}
\end{figure}

\begin{figure}
    \centering
    \includegraphics[width=1.0\linewidth]{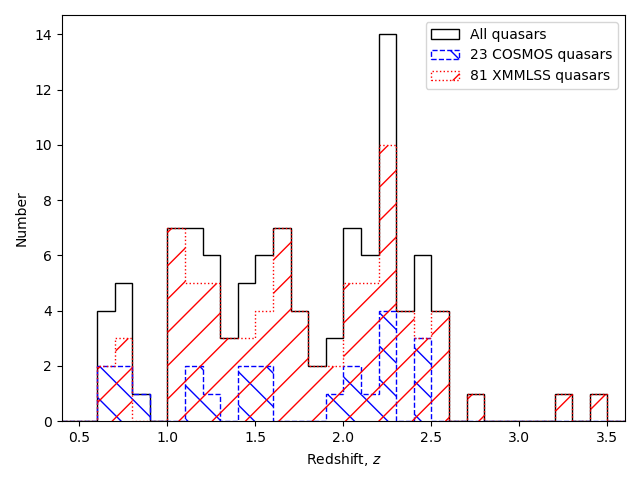}
    \caption{The redshift distribution for the full sample of 104 quasars (solid-black histogram), with the COSMOS and XMM-LSS subsets shown as blue-dashed and red-dotted histograms, respectively. }
    \label{fig:redshifts}
\end{figure}

 In Figure~\ref{fig:comparing_redshifts} we compare the spectroscopic redshift ($z_{s}$) to the photometric redshift ($z_{p}$) for the 104 quasars that compose the final sample. Like \citet{White2015}, we follow \citet{Ilbert2006} in defining a ``catastrophic outlier as having $| \Delta z|/(1+z) > 0.15$'', where $\Delta z = (z_{s} - z_{p})$. For our Type-1 quasars, the percentage of `catastrophic outliers' is 26\%, and so we have good justification to proceed with the {\it spectroscopic}-redshift values (rather than using photometric-redshift values to analyse a larger sample of 208 quasars -- see criterion 5 in Table~\ref{tab:quasar_sample_selection}). We present the distribution of the spectroscopic redshifts in Figure~\ref{fig:redshifts}.

\subsection{Estimates of IR luminosities and SFRs through SED-fitting}
\label{sec:sedfitting}

The combination of spectroscopic redshifts from the literature and SALT (this work) allows us to create a final sample of 23 quasars from the COSMOS field and 81 quasars from the XMM-LSS field. We obtain infrared luminosities and (`infrared-derived') SFR estimates for the entire quasar sample by running SED3FIT software \citep{Berta2013} over the optical, near-infrared, mid-infrared, and far-infrared data (Appendix~\ref{app:sed_fitting}). This software is a modified version of the MAGPHYS SED-fitting code \citep{daCunha2008}, where a dusty-torus/AGN component is included, in addition to the stellar-emission component and the dust component that is related to star formation. (The energy balance applied by MAGPHYS, regarding the optical and infrared emission, assumes that the only significant source of dust heating is starlight.) Hence, all parameters concerning the galaxy template are plainly taken from MAGPHYS\footnote{The public code and documentation for MAGPHYS can be found here: \url{https://www.iap.fr/magphys/}.}. Like \citet{daCunha2008}, \citet{Berta2013} adopt a \citet{Chabrier2003} initial mass function (IMF)\footnote{This is one of the most-popular IMFs, and easily re-scaled to other IMFs, such as those derived by \citet{Salpeter1955} and \citet{Kroupa2001}.} and employ stellar libraries from \citet{Bruzual2003}, with AGN templates taken from the libraries of \citet{Fritz2006} and \citet{Feltre2012}.

\section{Analysis of the radio emission}
\label{sec:radioanalysis}

The next step is to investigate the radio emission from the 104 quasars in our sample, by first extracting flux-density measurements from the MIGHTEE images, determining their radio luminosities, and calculating estimates of the (`radio-derived') SFRs within their host galaxies. We can then determine the fraction of sources that have their radio emission dominated by AGN-related processes.

\subsection{Detection of faint radio emission} \label{sec:faintemission}

\begin{figure*}
    \centering
    \includegraphics[width=0.7\linewidth]{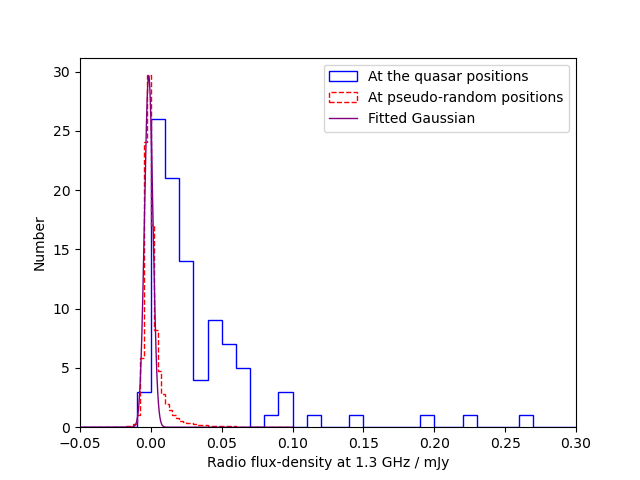}
    \caption{The distributions for radio flux-density measurements (Section~\ref{sec:faintemission}) extracted: (i) at the positions of the quasars and corrected for confusion (solid, blue histogram), and (ii) at pseudo-random positions in the radio map (dashed, red histogram; downscaled by a factor of 2000). The latter is to assess the noise level (3.0\,\textmu Jy\,beam$^{-1}$), via a Gaussian fit (purple curve) to this distribution. Note that 99 quasars are presented in this histogram, whilst 5 quasars have a radio flux-density above 0.3\,mJy.}
    \label{fig:radio_flux_densities}
\end{figure*}

We extract the radio pixel-values in the MIGHTEE-DR1 COSMOS and XMM-LSS images \citep{Hale2025} -- scaled via the effective frequency (see Section~\ref{sec:radiodata}) -- at the positions of the quasars, provided by their NIR photometry (Section~\ref{sec:NIRdata}). Given the point-like/unresolved radio-morphology for all but one of our sources (see below) -- i.e. smaller than, or approximately equal to, the 8.9-arcsec synthesised-beam -- we can interpret the extracted radio pixel-value (in units of Jy\,beam$^{-1}$) as the integrated flux-density (in units of Jy) at 1.3 GHz, $S_{\mathrm{1.3\,GHz}}$. The MIGHTEE images provide coverage for all 104 of the quasars in our sample, and we present their radio flux-densities in Figure~\ref{fig:radio_flux_densities}.

In order to estimate the level of the noise in the radio images, and therefore assess the confidence with which we detect the low-level radio emission from the quasars, we measure the radio flux-density at pseudo-random positions on the sky. We do this by extracting 1,000 1.3-GHz flux-density measurements from annuli (of radii 24 to 56\,arcsec) centred on each quasar position in the radio images, and the resulting flux-density distribution is shown by the red, dashed-line histogram in Figure~\ref{fig:radio_flux_densities}. The angular separation for the offset positions is chosen to be more than three times the size of the synthesised beam in order to avoid the possibility of correlated noise, or any artefacts that may be associated with nearby bright sources. Through fitting a Gaussian profile to this distribution (where we assume symmetric errors in the noise), we determine that the typical noise of the radio maps is 3.0\,\textmu Jy\,beam$^{-1}$. Noting that the thermal noise for these radio observations is 1.5--1.6\,\textmu Jy\,beam$^{-1}$ \citep{Hale2025}, we subtract this in quadrature from the total noise to estimate a confusion noise-level of 2.5--2.6\,\textmu Jy\,beam$^{-1}$.  

As there is a significant level of bias in the radio images (likely due to confusion), we calculate the median radio flux-density at the pseudo-random positions {\it per quasar}. This median value is then subtracted from the measured flux-density at the position of the quasar to give a {\it corrected} flux-density, which is presented as the blue, solid-line histogram in Figure~\ref{fig:radio_flux_densities}. The error in this corrected flux-density is calculated as the median absolute deviation of the pseudo-random flux-densities (again, per quasar). Considering these individual error-estimates, we have statistical detection of radio emission at the position of the quasars, with respect to random background noise, at: the $>1\sigma$ level for 97 quasars, the $>2 \sigma$ level for 93 quasars, and the $>3 \sigma$ level for 81 quasars. The 23 sources that are not detected at the 3-$\sigma$ level are retained for the subsequent analysis (wherever the ``full sample'' or ``whole sample'' is mentioned).

Note that the quasars at R.A. = (10:00:01.44; 02:18:27.29; 02:20:01.64; 02:25:03.12; 02:25:25.68), Dec. = (+02:48:44.7; $-$05:34:57.5; $-$05:22:17.1; $-$04:40:25.3; $-$04:35:09.6) have $S_{\mathrm{1.3\,GHz}} = $ 0.67, 3.40, 10.88, 0.47, and 0.52\,mJy, respectively, and so are beyond the plot range of Figure~\ref{fig:radio_flux_densities}. The second and fourth of these brighter quasars have possible extended emission\footnote{At the position of these quasars, any diffuse emission in the higher-sensitivity radio image (robust weighting of $0.0$), appearing to connect the radio core with potential radio lobes, is below the detection limit of the higher-resolution radio-image (robust weighting of $-1.2$) that is available for the XMM-LSS field \citep{Hale2025}.}, and so we note them as candidate `radio-loud' quasars. 

\subsection{The radio-loudness fraction}
\label{sec:radioloudness}

We calculate the radio-loudness parameter, $R$, that is defined by \citet{Kellermann1989} as $R = S_{\mathrm{5\,GHz}}/S_{\mathrm{4400\,\mathring{A}}}$, with both flux densities being in units of mJy. This definition is most-appropriate for our unobscured (Type-1) quasars, where the AGN is dominating the optical light from the source. 4400\,\AA\ is the effective observed wavelength of the optical $B$ band, and we calculate $S_{\mathrm{4400\,\mathring{A}}}$ in accordance with the AB magnitude system:
\begin{equation}
    m_{B} = -2.5 \log_{10}[S_{\mathrm{4400\,\mathring{A}}} \times 10^{-26}] - 48.60
\end{equation}

\begin{figure}
    \centering
\includegraphics[width=1.0\linewidth]{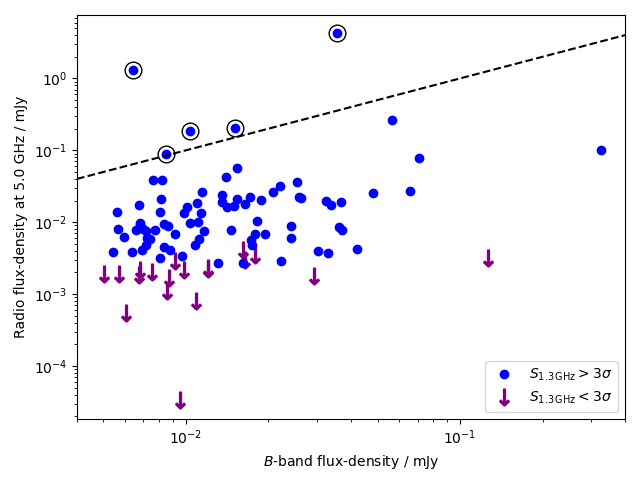}
    \caption{The distribution in 5-GHz flux-densities, and at 4400\,\AA\ in the optical, for the quasar sample. The dashed line demarcates the divide between radio-quiet sources ($R < 10$) and radio-loud sources ($R > 10$; circled), following the radio-loudness definition by \citet{Kellermann1989} (see Section~\ref{sec:radioloudness}). Arrows represent the quasars that are undetected at the 3-$\sigma$ level in the radio images. Note that, due to the log-log scales that are used, the four quasars that have negative radio flux-densities are not shown in this plot.}
    \label{fig:radioloudness}
\end{figure}

In order to apply the above definition for radio-loudness, we assume that the flux density ($S_{\nu}$) of the synchrotron radiation follows a power-law function ($S_{\nu} \propto \nu ^{\alpha}$, with spectral index, $\alpha = -0.7$), which allows us to extrapolate our measurements at 1.3\,GHz to 5\,GHz. We also estimate the apparent $B$-band magnitudes, $m_{B}$, from the apparent $g$-band magnitudes, $m_{g}$, that are provided via HSC photometry (Section~\ref{sec:opticaldata}). This is based upon assuming that the quasar continuum is well-described by a power-law function (with spectral-index, $\alpha_{c} = -0.5$):
\begin{equation}
m_{B}  = m_{g} + 2.5\ \alpha_{c}\ \log_{10}[\mathrm{4400\,\mathring{A}/4816\,\mathring{A}}],
\end{equation}
where 4816\,\AA\ is the effective observed wavelength of the optical $g$-band. We find that five of the quasars in our sample are classified as `radio loud', with $R > 10$ (lying above the dashed line in Figure~\ref{fig:radioloudness}), and so the radio-loudness fraction is $\sim$5 per cent. (These five quasars include the two quasars that are considered `radio loud' on account of their extended radio-morphology; end of Section~\ref{sec:faintemission}.)

\begin{figure}
    \centering
\includegraphics[width=1.0\linewidth]{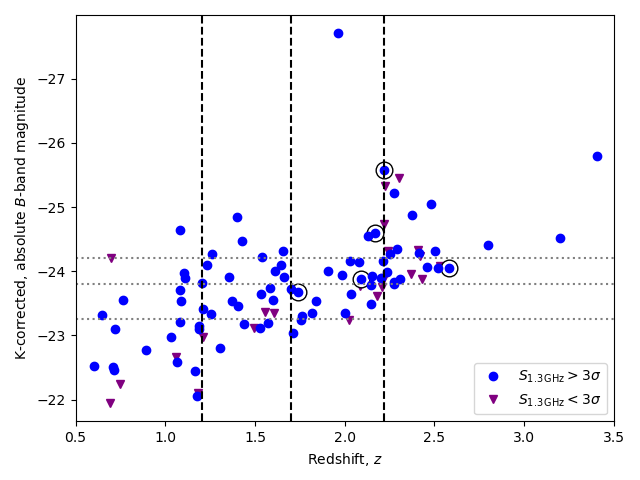}
    \caption{The K-corrected absolute $B$-band magnitude as a function of spectroscopic redshift, $z$. The vertical, black, dashed lines delineate the four redshift bins that aid later analysis (Section~\ref{sec:agn_fraction_discussion_Delhaize}), and the horizontal, grey, dotted lines indicate the four magnitude bins applied to the sample (Section~\ref{sec:binning}). With either type of binning, there are $\sim$26 quasars per bin, and triangles indicate the sources that are {\it not} detected at the 3-$\sigma$ level in the radio images. Radio-loud quasars (see Figure~\ref{fig:radioloudness}) are circled in black.}
    \label{fig:redshiftbinning}
\end{figure}

We also calculate the $K$-corrected absolute $B$-band magnitudes, $M_{B}$, as follows:
\begin{equation}
    M_{B} = m_{B} + 5.0 - 5.0 \log_{10}[ D_L ] + 2.5 ( 1 + \alpha_{c} ) \log_{10}[1 + z] 
\end{equation}
where the luminosity distance, $D_L$, is in units of pc. These magnitudes are presented as a function of redshift in Figure~\ref{fig:redshiftbinning}.

\subsection{Accretion versus star-formation contributions}
\label{sec:SFR_comparisons}

\begin{figure*}
    \centering
    
    \subfigure[Kennicutt-SFR versus the Yun-SFR]{
	\includegraphics[width=0.7\linewidth]{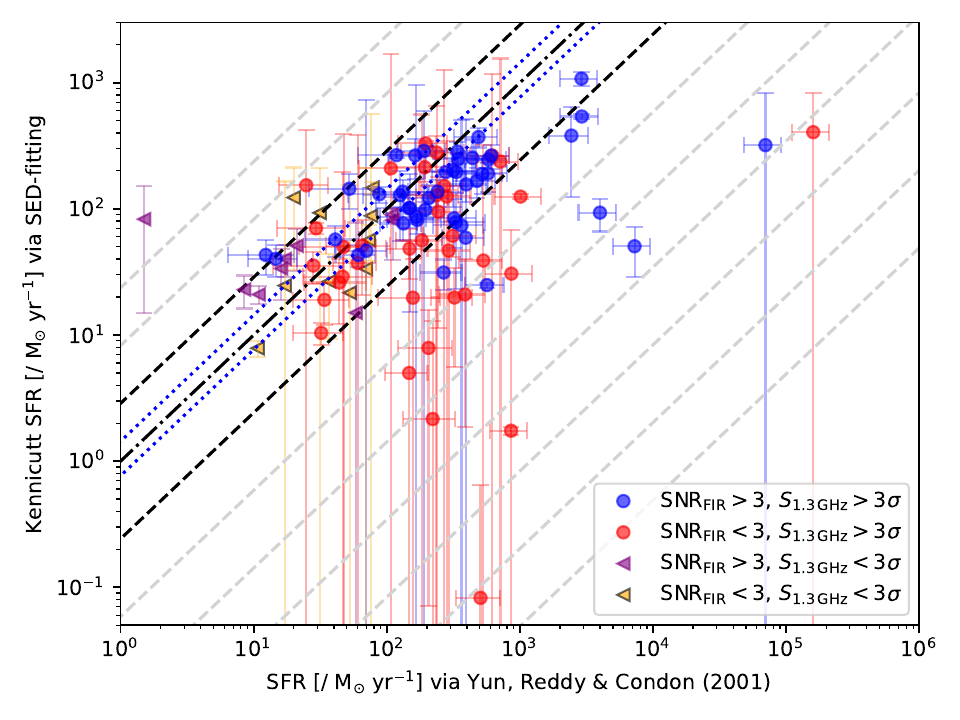} 
	}

    \subfigure[SED3FIT-SFR versus the Yun-SFR]{
	\includegraphics[width=0.7\linewidth]{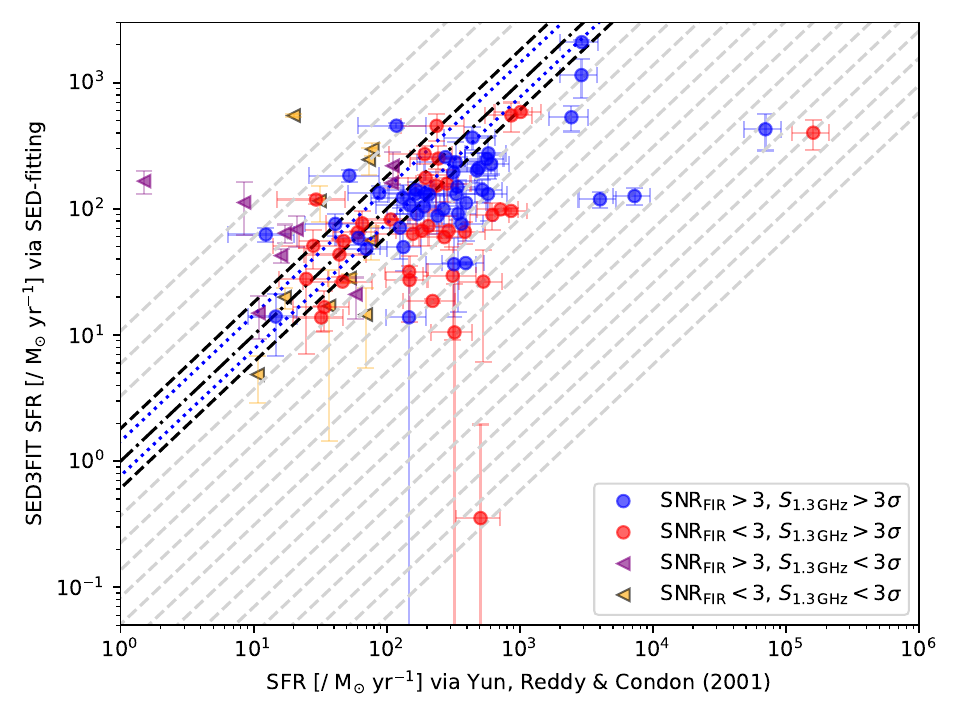}
	}
	
    \caption{A comparison of star-formation rates (SFRs) for quasars in the COSMOS and XMM-LSS fields. The SFR estimate on the ordinate axis is derived through SED-fitting of the infra-red photometry (Section~\ref{sec:infrareddata}), (a) using Equation~\ref{eqn:Kennicutt} \citep{Kennicutt1998b}, or (b) using SED3FIT \citep{Berta2013} [which is a modified version of MAGPHYS \citep{daCunha2008}]. The SFR estimate on the abscissa axis is derived via Equation~\ref{eqn:Yunrelation} \citep{Yun2001}, assuming that all of the radio emission can be attributed to star formation. The dash-dotted line represents where the two SFR estimates are equal to each other, with the blue-dotted lines indicating the error associated with the Yun-SFR, and the dashed lines indicating the total error in the one-to-one relation (i.e. the 1-$\sigma_{1:1}$ calibration error in both the Yun-SFR and the Kennicutt-SFR/SED3FIT-SFR, shown in black, and grey dashed lines for 2\,$\sigma_{1:1}$, 3\,$\sigma_{1:1}$, etc.). The legend is the same for both panels, with datapoints colour-coded by the signal-to-noise ratio (SNR$_{\mathrm{FIR}}$) for the FIR data (see Section~\ref{sec:SFR_comparisons}), and symbols indicating the detection level (above or below $3\sigma$) for the radio data at 1.3\,GHz. }
    \label{fig:SFRs_Yun_on_xaxis}
\end{figure*}

\begin{figure*}
    \centering
    
    \subfigure[Kennicutt-SFR versus the Delhaize-SFR]{
	\includegraphics[width=0.7\linewidth]{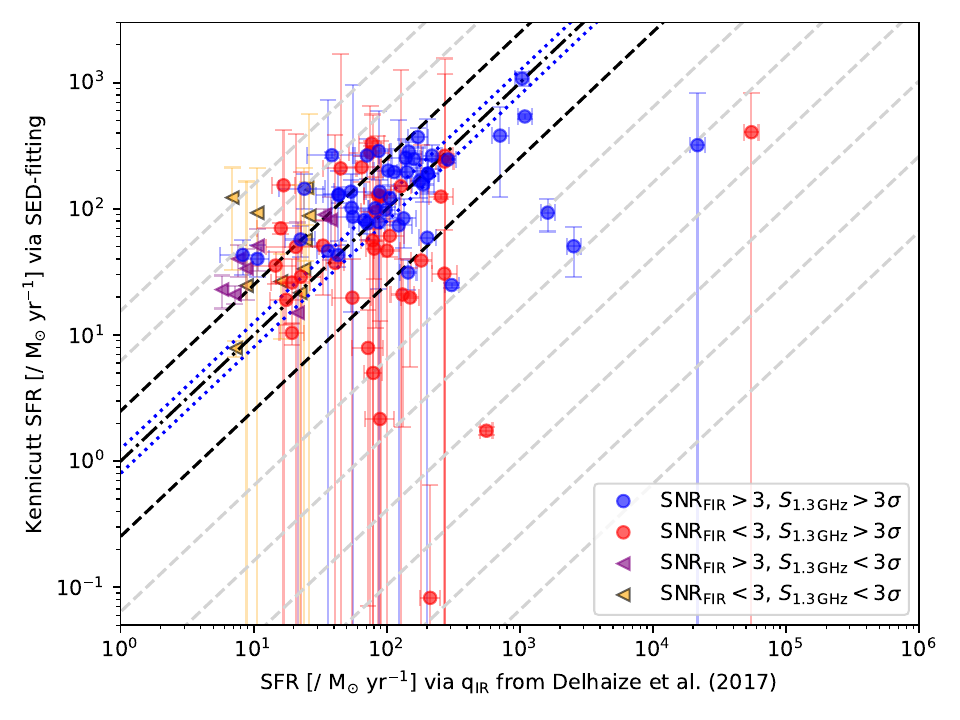} 
	} 
    
    \subfigure[SED3FIT-SFR versus the Delhaize-SFR]{
	\includegraphics[width=0.7\linewidth]{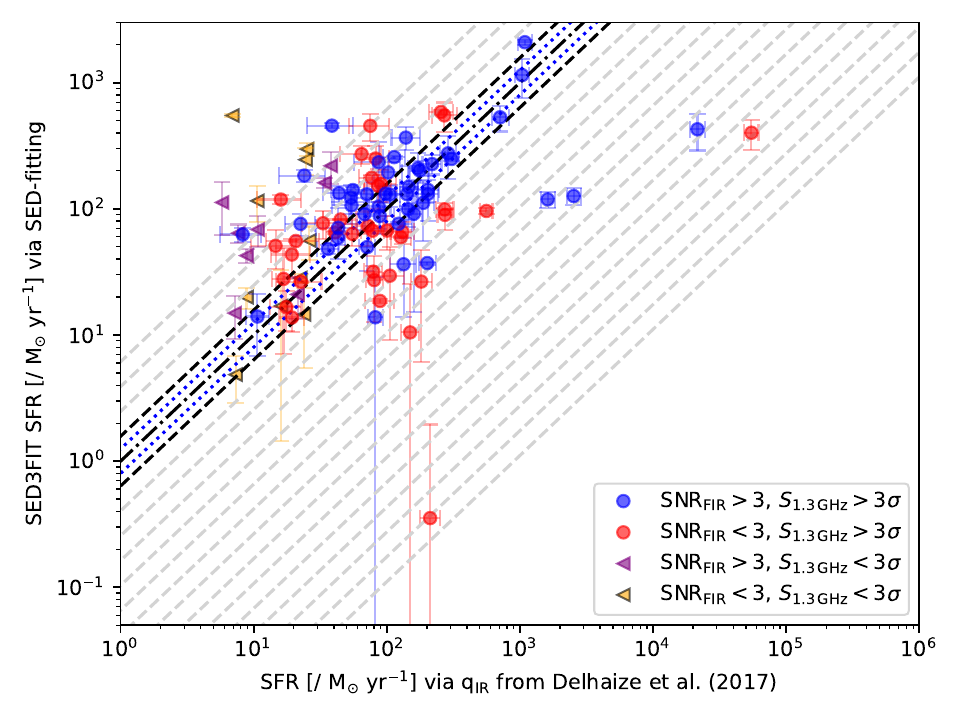}
	}
    \caption{A comparison of star-formation rates (SFRs) for quasars in the COSMOS and XMM-LSS fields. The SFR estimate on the ordinate axis is derived through SED-fitting of the infra-red photometry (Section~\ref{sec:infrareddata}), (a) using Equation~\ref{eqn:Kennicutt} \citep{Kennicutt1998b}, or (b) using SED3FIT \citep{Berta2013} [which is a modified version of MAGPHYS \citep{daCunha2008}]. The SFR estimate on the abscissa axis assumes that all of the radio emission can be attributed to star formation, and so lies on the IRRC (Equation~\ref{eqn:bestq_delhaize}) determined by \protect\cite{Delhaize2017} at the median redshift of the sample ($z = 1.728$). The dash-dotted line represents where the two SFR estimates are equal to each other, with the blue-dotted lines indicating the error associated with the Delhaize-SFR, and the dashed lines indicating the total error in the one-to-one relation (i.e. the 1-$\sigma_{1:1}$ calibration error in both the Delhaize-SFR and the Kennicutt-SFR/SED3FIT-SFR, shown in black, and grey dashed lines for 2\,$\sigma_{1:1}$, 3\,$\sigma_{1:1}$, etc.). The legend is the same for both panels, with datapoints colour-coded by the signal-to-noise ratio (SNR$_{\mathrm{FIR}}$) for the FIR data (see Section~\ref{sec:SFR_comparisons}), and symbols indicating the detection level (above or below $3\sigma$) for the radio data at 1.3\,GHz.}
\label{fig:SFRs_Delhaize_on_xaxis}
\end{figure*}

We calculate the radio luminosities (Table~\ref{tab:luminosities_sfrs_allquasars}) of the 104 quasars, including those that have negative radio pixel-values so that they can be included in our statistical analyses. These analyses involve a comparison of independent star-formation rate (SFR) estimates, where one of the estimates is derived from the radio luminosity {\it under the assumption} that all of the radio emission is the result of star formation. A mismatch between the two SFR estimates therefore indicates that this assumption does not hold, and the deviation from the one-to-one relation (lower diagonal-lines in Figures~\ref{fig:SFRs_Yun_on_xaxis}--\ref{fig:SFRs_Delhaize_on_xaxis}) can be used to calculate how much of the radio emission may in fact arise from processes connected to black-hole accretion.  

For comparison with previous work in the literature (e.g. \citealt{White2015, White2017}), we scale the K-corrected MeerKAT radio luminosities, $L_{\mathrm{1.3\,GHz}}$, to luminosities at frequency, $\nu = 1.4$\,GHz ($L_{\mathrm{1.4\,GHz}}$), again assuming a power-law function with $\alpha = -0.7$. This then allows us to apply the relation of \citet{Yun2001}, which was empirically derived for a sample of local star-forming galaxies:
\begin{equation}
\mathrm{SFR_{\mathrm{Yun}}/(\Msol\ yr}^{-1}) = (5.9 \pm 1.8) \times 10^{-22} L_{\mathrm{1.4\,GHz}}/(\mathrm{W\ Hz}^{-1})
\label{eqn:Yunrelation}
\end{equation}
We refer to this SFR estimate as the `Yun-SFR' (Table~\ref{tab:luminosities_sfrs_allquasars}). Whilst this does not take redshift dependence into account [like the SFR derived via the IRRC of \citet{Delhaize2017}; see Equations~\ref{eqn:bestq_delhaize} and \ref{eqn:qir}], it does allow us to compare our results more broadly, given the wider use of the \citet{Yun2001} relation in the literature. 


For the second SFR estimate, we complete spectral fitting over all of the available data using code developed by \citet{Jin2018}, where a two-component model describes the FIR/sub-mm emission. Noting that the fits are more reliable for the sources where we have FIR/sub-mm data available, we flag this through the signal-to-noise ratio, SNR$_{\mathrm{FIR}}$. For the COSMOS sources, we adopt SNR$_{\mathrm{FIR}}$ values from the catalogue of \citet{Jin2018}, as described by their equation 2. Since we do not have measurements at 850$\upmu$m, 1.1\,mm, nor 1.2\,mm for the XMM-LSS sources, we instead use the following approximation (for 5 bands rather than 8): 
\begin{align*}
    \mathrm{SNR}_{\mathrm{FIR}}^2 = \frac{8}{5} \lbrack \mathrm{SNR}_{100\mathrm{\upmu m}}^2 + \mathrm{SNR}_{160\mathrm{\upmu m}}^2 + \mathrm{SNR}_{250\mathrm{\upmu m}}^2 \\ + \mathrm{SNR}_{350\mathrm{\upmu m}}^2 +
    \mathrm{SNR}_{500\mathrm{\upmu m}}^2 \rbrack \\
    = \frac{8}{5} \lbrack (S_{100\mathrm{\upmu m}}/\sigma_{100\mathrm{\upmu m}})^2 + (S_{160\mathrm{\upmu m}}/\sigma_{160\mathrm{\upmu m}})^2 + (S_{250\mathrm{\upmu m}}/\sigma_{250\mathrm{\upmu m}})^2 \\ + (S_{350\mathrm{\upmu m}}/\sigma_{350\mathrm{\upmu m}})^2 +
    (S_{500\mathrm{\upmu m}}/\sigma_{500\mathrm{\upmu m}})^2 \rbrack
\end{align*} 
[These SNR$_{\mathrm{FIR}}$ aid the colour-coding of the datapoints in Figures~\ref{fig:SFRs_Yun_on_xaxis}--\ref{fig:SFRs_Delhaize_on_xaxis}, and also aid the assessment of the different `AGN-dominated' fractions (see Table~\ref{tab:AGNdominated_fractions}).] Across the full sample of 104 quasars, 53 per cent are detected with SNR$_{\mathrm{FIR}} > 3.0$.



As \citet{Jin2018} fitted the photometry with a {\it combination} of AGN and galaxy templates, it is possible to disentangle how much IR emission is coming from the AGN, and how much IR emission is due to star formation in the host galaxy. The IR luminosity, $L_{\mathrm{IR}}$, is then calculated by integrating only the fitted {\it galaxy}-spectrum over 8--1000$\upmu$m in the rest-frame (having already subtracted out the AGN contribution to the IR emission). This luminosity is in turn converted to an SFR estimate (which we refer to as the `Kennicutt-SFR'; Table~\ref{tab:luminosities_sfrs_allquasars}) via the following equation from \citet{Kennicutt1998b}:
\begin{equation}
\frac{\mathrm{SFR_{\mathrm{Kenn}}}}{1 \Msol \ \mathrm{yr}^{-1}}
= f_{\mathrm{IMF}} \frac{L_{\mathrm{IR}}}{2.2 \times 10^{43}\,\mathrm{erg\ s}^{-1}}
= f_{\mathrm{IMF}} \frac{L_{\mathrm{IR}}}{5.8 \times 10^9\,\Lsol}
\label{eqn:Kennicutt}
\end{equation}
where $f_{\mathrm{IMF}} = 1.00/1.72$. This factor allows us to scale the original conversion \citep{Kennicutt1998b} -- which is based upon the Salpeter initial mass function \citep[IMF;][]{Salpeter1955} -- to a Chabrier IMF \citep{Chabrier2003}, in line with \citet{Delhaize2017}. In addition, we note that the uncertainty in this $L_\mathrm{IR}$--SFR$_\mathrm{Kenn}$ calibration is $^{+0.3}_{-0.5}$\,dex, ``with the asymmetry reflecting the greater likelihood that the systematic errors tend to lead to overestimates'' \citep{Kennicutt1998b}.

In Figure~\ref{fig:SFRs_Yun_on_xaxis}a, we compare the Kennicutt-SFR estimate (derived from the IR emission) with the Yun-SFR (derived through the radio emission). A dash-dotted line (the `one-to-one relation') indicates where the two SFRs equal one another, with measurement errors in the fitted $L_{\mathrm{IR}}$ and radio flux-density, $S_{\mathrm{1.3\,GHz}}$, being propagated through. For datapoints below this line, the Yun-SFR exceeds the Kennicutt-SFR, illustrating that not all of the radio emission from the quasars can be explained by star formation alone. That is, there must be another process contributing to the radio emission, which we interpret as being connected to AGN activity. (Investigating the exact physical mechanism, connected to black-hole accretion, is beyond the scope of this work, but we direct the reader to a review by \citealt{Panessa2019}.) A possible explanation for the datapoints that lie above the line is that they are starbursts, with enhanced FIR emission over timescales that are not long enough for sufficient radio emission (associated with supernova remnants) to be produced \citep[e.g.][]{Cook2024}. Supporting this (see Appendix~\ref{app:lfir}), we note that 38 per cent of the sample have $>10^{12}$\,\Lsol\ IR luminosities that are typical of Ultraluminous Infrared Galaxies \citep[ULIRGs;][]{Sanders1996, Farrah2003}. This is unsurprising given that all of the best-fitted quasar models for our sample selection involve a combination of starbust/ULIRG and Type-1-quasar templates (Table~\ref{tab:AGN_templates}).

To explore the impact of how the IR-related SFR-estimate is calculated, on the SFR comparison (and therefore how much of the radio emission we attribute to the AGN), we also consider the `SED3FIT-SFR' (Table~\ref{tab:luminosities_sfrs_allquasars}) -- this being the SFR output by the SED-fitting code, SED3FIT \citep[][see Section~\ref{sec:sedfitting}]{Berta2013}. This code produces a probability distribution function for each of the fitted parameters, and for simplification (on account of possible asymmetry in this distribution) we estimate the measurement error by taking an average of the separation of the 16th and 84th percentiles from the median (the 50th percentile). The results are again plotted against the Yun-SFR (see Figure~\ref{fig:SFRs_Yun_on_xaxis}b), leading to some datapoints shifting in the vertical direction relative to their position in Figure~\ref{fig:SFRs_Yun_on_xaxis}a. The calibration error in the SED3FIT-SFR is found by \citet{Pacifici2023} to be 0.1\,dex, and we take this into account when plotting the uncertainty in the one-to-one relation (dashed lines in Figure~\ref{fig:SFRs_Yun_on_xaxis}b).


Next, we use the positions of the datapoints in the SFR-comparison plots to quantify how many (and what fraction of) sources have radio emission that is `AGN-dominated', i.e. where the radio-derived SFR exceeds the IR-derived SFR by more than 1\,$\sigma$. These numbers are presented in Table~\ref{tab:AGNdominated_fractions}, along with the fraction of sources that are possibly starbursts (on account of their datapoints lying more than 1\,$\sigma$ {\it above} the IRRC). We see that the total fraction of sources that are AGN-dominated (across the COSMOS and XMM-LSS fields, combined) increases from 23 per cent when using the Kennicutt-SFR, to 51 per cent when using the SED3FIT method. This is in large part due to the more-tightly-constrained calibration of the IR-related SFR calculated through SED3FIT \citep{Berta2013} SED-fitting compared with the calibration of the \citet{Kennicutt1998b} relation. Considering only the quasars that are detected at the 3-$\sigma$ level in both the radio and the FIR, the AGN-dominated fraction shows a similar increase, from 22 per cent to 58 per cent.

Meanwhile, for both combinations of IR-related SFR and (radio-related) Yun-SFR, the fraction of `possible starbursts' is relatively low, ranging from  5 per cent to 16 per cent (when the full sample is considered). However, again the more-tightly-constrained SED3FIT-SFRs lead to a narrower `range' within which quasars can have SFRs that are {\it consistent} with normal star-formation (Figure~\ref{fig:SFRs_Yun_on_xaxis}b), and so more datapoints lie above the one-to-one relation than in the Kennicutt-SFR case (Figure~\ref{fig:SFRs_Yun_on_xaxis}a).

\begin{table*}
    \centering
    \begin{tabular}{c|r|c|c}
    \hline
      & Median separation, $\sigma_{\mathrm{sep}}$,& Fraction of sources at & Fraction of sources that \\
      & with respect to the one-to-one &  that are AGN-dominated &  are possibly starbursts \\
      &  relation (Whole subset) & (i.e. $\sigma_{\mathrm{sep}} > 1.0$\,$\sigma_{1:1}$ &  (i.e. $\sigma_{\mathrm{sep}} > 1.0$\,$\sigma_{1:1}$ \\  
        &  & below the 1:1 relation) &  above the 1:1 relation) \\ 
       \hline
       
       {\bf Kennicutt SFR (via SED) \hspace{5mm} versus}  & {\bf SFR via Yun et al. (2001) } \\
       SNR$_{\mathrm{FIR}} > 3.0$, $S_{\mathrm{1.3\,GHz}} > 3\sigma$  & 0.39\,$\sigma_{1:1}$ & 10  /  45  =  22\% & 1 / 45 = 2\% \\
         SNR$_{\mathrm{FIR}} \leq 3.0$, $S_{\mathrm{1.3\,GHz}} > 3\sigma$      & 0.72\,$\sigma_{1:1}$ & 14  /  36  =  39\% & 1 / 36 = 3\% \\
         SNR$_{\mathrm{FIR}} > 3.0$, $S_{\mathrm{1.3\,GHz}} < 3\sigma$ & $-$0.56\,$\sigma_{1:1}$ & 0  /  10  =  0\% & 1 / 10 = 10\% \\
         SNR$_{\mathrm{FIR}} \leq 3.0$, $S_{\mathrm{1.3\,GHz}} < 3\sigma$      & $-$0.26\,$\sigma_{1:1}$ & 0  /  13  =  0\% & 2 / 13 = 15 \% \\
         Whole sample    & 0.37\,$\sigma_{1:1}$ & 24$_{14}^{31}$ /  104  =  23$_{13}^{30}$\% & 5$_{0}^{35}$ / 104 = 5$_{0}^{34}$\% \\

       {\bf SED3FIT SFR (via SED) \hspace{5mm} versus}  & {\bf SFR via Yun et al. (2001) } & & \\

          SNR$_{\mathrm{FIR}} > 3.0$, $S_{\mathrm{1.3\,GHz}} > 3\sigma$ & 1.50\,$\sigma_{1:1}$ & 26 / 45  =  58\% & 4 / 45 = 9\% \\
         SNR$_{\mathrm{FIR}} \leq 3.0$, $S_{\mathrm{1.3\,GHz}} > 3\sigma$      & 1.57\,$\sigma_{1:1}$ & 22 / 36  =  61\% & 3 / 36 = 8\% \\
         SNR$_{\mathrm{FIR}} > 3.0$, $S_{\mathrm{1.3\,GHz}} < 3\sigma$ & $-$2.17\,$\sigma_{1:1}$ & 1 / 10  =  10\% & 6 / 10 = 60\% \\
         SNR$_{\mathrm{FIR}} \leq 3.0$, $S_{\mathrm{1.3\,GHz}} < 3\sigma$      & $-$2.43\,$\sigma_{1:1}$ & 4 / 13  =  31\% & 4 / 13 = 31\% \\
         Whole sample    & 1.12\,$\sigma_{1:1}$ & 53$_{25}^{74}$  /  104  =  51$_{24}^{71}$\% & 17$_{8}^{30}$/ 104 = 16$_{8}^{29}$\% \\

{\bf Kennicutt SFR (via SED) \hspace{4mm} versus}  & {\bf SFR via Delhaize et al. (2017) } & & \\
                SNR$_{\mathrm{FIR}} > 3.0$, $S_{\mathrm{1.3\,GHz}} > 3\sigma$ & $-$0.14\,$\sigma_{1:1}$ & 5 / 45  =  11\% & 10 / 45 = 22\% \\
         SNR$_{\mathrm{FIR}} \leq 3.0$, $S_{\mathrm{1.3\,GHz}} > 3\sigma$      & 0.08\,$\sigma_{1:1}$ & 10 / 36  = 28\% & 6 / 36 = 17\% \\
         SNR$_{\mathrm{FIR}} > 3.0$, $S_{\mathrm{1.3\,GHz}} < 3\sigma$ & $-$0.99\,$\sigma_{1:1}$ & 0  /  10  =  0\% & 7 / 10 = 70\% \\
         SNR$_{\mathrm{FIR}} \leq 3.0$, $S_{\mathrm{1.3\,GHz}} < 3\sigma$      & $-$0.88\,$\sigma_{1:1}$ & 0  /  13  =  0\% & 5 / 13 =  38\% \\
         Whole sample    & $-$0.20\,$\sigma_{1:1}$ & 15$_{8}^{35}$  /  104  =  14$_{8}^{34}$\% & 28$_{7}^{55}$ / 104 = 27$_{7}^{53}$\% \\
          
       {\bf SED3FIT SFR (via SED) \hspace{3mm} versus}  & {\bf SFR via Delhaize et al. (2017) } & & \\
         SNR$_{\mathrm{FIR}} > 3.0$, $S_{\mathrm{1.3\,GHz}} > 3\sigma$ & $-$0.40\,$\sigma_{1:1}$ & 9 / 45  =  20\% & 15 / 45 = 33\% \\
         SNR$_{\mathrm{FIR}} \leq 3.0$, $S_{\mathrm{1.3\,GHz}} > 3\sigma$      & $-$0.17\,$\sigma_{1:1}$ & 13  /  36  = 36\% & 16 / 36 = 44\% \\
         SNR$_{\mathrm{FIR}} > 3.0$, $S_{\mathrm{1.3\,GHz}} < 3\sigma$ & $-$3.98\,$\sigma_{1:1}$ & 0  /  10  =  0\% & 8 / 10 = 80\% \\
         SNR$_{\mathrm{FIR}} \leq 3.0$, $S_{\mathrm{1.3\,GHz}} < 3\sigma$      & $-$5.08\,$\sigma_{1:1}$ & 1  /  13  = 8 \% & 6 / 13 = 46\%  \\
         Whole sample    & $-$0.63\,$\sigma_{1:1}$ & 23$_{18}^{40}$  /  104  =  22$_{17}^{38}$\% & 45$_{30}^{63}$ / 104 = 43$_{29}^{61}$\% \\
            
    \hline
    \end{tabular}
    \caption{The fraction of sources in the COSMOS and XMM-LSS fields that are described as `AGN-dominated' and `possible starbursts', based upon comparing different pairs of SFR estimates (Section~\ref{sec:SFR_comparisons}, Table~\ref{tab:luminosities_sfrs_allquasars}). The `median separation' indicates the median number of $\sigma_{1:1}$ {\it below} the one-to-one (1:1) relation in Figures~\ref{fig:SFRs_Yun_on_xaxis} and \ref{fig:SFRs_Delhaize_on_xaxis} that the quasars lie (see the lower, dashed lines), where $\sigma_{1:1}$ refers to the calibration error in the 1:1 relation. Therefore, negative values of `median separation' indicate that a larger fraction of the quasars lie {\it above} the 1:1 relation in Figures~\ref{fig:SFRs_Yun_on_xaxis} and \ref{fig:SFRs_Delhaize_on_xaxis}, and so are considered to be (possible) starbursts (Sections~\ref{sec:agn_fraction_discussion_Yun} and \ref{sec:agn_fraction_discussion_Delhaize}, and Appendix~\ref{app:lfir}). SNR$_{\mathrm{FIR}}$ is the combined signal-to-noise over the available IR bands (Section~\ref{sec:SFR_comparisons}), whilst upper and lower values are presented for the `Whole sample', for different SFR--SFR combinations. These upper and lower values (in fractions and percentages) are calculated by considering the intersection of the error-bars with respect to the lines indicating $\pm1\sigma_{1:1}$ away from the 1:1 relation.} 
    \label{tab:AGNdominated_fractions}
\end{table*}

To explore the impact (on the SFR comparison) of how the {\it radio}-related SFR-estimate is calculated, we consider a fourth and final SFR estimate. This time we apply the IRRC, which is a tight correlation that has been observed for star formation \citep[e.g.][]{Helou1985, deJong1985} and arises as a result of this process producing both infrared emission (from the associated dust) and radio emission (from supernova remnants). The dependence of the IRRC on redshift, $z$, has been determined empirically by \citet{Delhaize2017} as follows:
\begin{equation}
q_{\mathrm{IR}} (z) = (2.88 \pm 0.03) \times (1+z)^{(-0.19 \pm  0.01)}
\label{eqn:bestq_delhaize}
\end{equation}
for a typical spectral-index of $-0.7$, and where:
\begin{equation}
q_{\mathrm{IR}} = \log_{10} \left( \frac{L_{\mathrm{IR}} / \mathrm{W}}{3.75 \times 10^{12} / \mathrm{Hz} } \right) - \log_{10} [L_{\mathrm{1.4\,GHz}}/(\mathrm{W\ Hz}^{-1})],
   \label{eqn:qir}
\end{equation}
where $3.75 \times 10^{12}$\,Hz represents the central frequency over the FIR (rest-frame 42--122 \textmu m) domain, whilst $L_{\mathrm{IR}}$ refers to the full IR (rest-frame 8--1000 \textmu m) domain. We apply this relation by first assuming that each quasar lies exactly on the IRRC, and combining the result of Equation~\ref{eqn:bestq_delhaize} with the radio luminosity in order to obtain an estimate of the SFR via:
\begin{equation}
\mathrm{SFR} / (\Msol \ \mathrm{yr}^{-1}) =  10^{-24} 10^{q_{\mathrm{IR}}(z)} L_{\mathrm{1.4\,GHz}}/(\mathrm{W\ Hz}^{-1})
\label{eqn:sfr_delhaize}
\end{equation}
We refer to this estimate as the `Delhaize-SFR' (Table~\ref{tab:luminosities_sfrs_allquasars}) and plot this (against the Kennicutt-SFR and the SED3FIT-SFR) in Figure~\ref{fig:SFRs_Delhaize_on_xaxis}. For the measurement errors in the Delhaize-SFR, we propagate through the errors in the measured radio flux-density and redshift {\it per quasar} (Section~\ref{sec:faintemission}).

As before, we calculate the total fraction of sources that are AGN-dominated (Table~\ref{tab:AGNdominated_fractions}), this time with the Delhaize-SFR as the radio-derived SFR estimate. The fraction ranges from 14 per cent for the whole sample (when the IR-related SFR being compared against is the Kennicutt-SFR) to 22 per cent (when the IR-related SFR being compared against is the SED3FIT-SFR) -- although the large errors involved means that these quantities are consistent with one another. 

Considering the sources that lie more than 1\,$\sigma$ above the IRRC, the fractions of possible starbursts are (for the majority of the subsets in Table~\ref{tab:AGNdominated_fractions}) higher than the fractions that are described as being `AGN-dominated', ranging from 17 to 80 per cent. This is in large contrast with the fractions determined for the Yun-SFR comparisons, signifying that {\bf the relation that is used to derive the radio-related SFR has a greater impact on the results than the relation that is used to derive the IR-related SFR}.

\subsection{Discussion of the origin of the radio emission}
\label{sec:agn_fraction_discussion}

\subsubsection{When the (radio-derived) Yun-SFR is considered}
\label{sec:agn_fraction_discussion_Yun}

If we consider the combination of the \citet{Yun2001} and \citet{Kennicutt1998b} relations for the current sample (Table~\ref{tab:AGNdominated_fractions}), the fraction of AGN-dominated sources is 22 per cent (for quasars detected at $\mathrm{SNR}_{\mathrm{FIR}} > 3.0$ and $S_{\mathrm{1.3\,GHz}} > 3\sigma$). For comparison, when \citet{White2017} compare SFR estimates based upon the \citet{Yun2001} and \citet{Kennicutt1998,Kennicutt1998b} relations (the latter being specific for starbursts, rather than main-sequence star-formation), they find that the vast majority of the quasars have radio emission that exceeds that expected from star formation (see their figure 8). In addition, the exact fraction of radio-detected, FIR-detected quasars (that have radio emission that is dominated by the AGN) is shown to be 92 per cent \citep{White2017} when they apply a specific IRRC \citep{Smith2014} that is of most relevance for the sample of $z \sim 1$ quasars under investigation. (Note that their sample is of quasars detected to depths of $\sim$30.9\,$\upmu$Jy\,beam$^{-1}$, compared with the $\sim$3.0\,$\upmu$Jy\,beam$^{-1}$ rms for the current work.) Furthermore, \citet{White2017} find little difference in their results when employing the SFR-comparison method rather than using the IRRC to assess the level of star-formation-related radio emission.

However, 92 per cent may still be a lower limit on the `AGN-dominated' fraction because, as demonstrated by \citet{Wong2016} for a sample of 92 X-ray-selected radio-quiet AGN at $z < 0.05$ (see their figures 2 and 5), sources with significant accretion-related processes can have host galaxies that mimic the (far-)infrared properties of normal, star-forming galaxies (and so lie on the IRRC). As such, {\bf the `radio-excess' technique that is used to investigate the AGN-versus-SF fractions in radio-quiet quasars may underestimate the contribution of AGN activity towards the total radio-emission.}

\subsubsection{When the (radio-derived) Delhaize-SFR is considered}
\label{sec:agn_fraction_discussion_Delhaize}

The Delhaize-SFR values (Figure~\ref{fig:SFRs_Delhaize_on_xaxis}) tend to be lower than the Yun-SFR values (Figure~\ref{fig:SFRs_Yun_on_xaxis}), resulting in a greater number of quasars lying above the one-to-one relations, and therefore a smaller fraction of sources being classed as `AGN dominated' (see Table~\ref{tab:AGNdominated_fractions}). With the IRRC by \citet{Delhaize2017} (Equation~\ref{eqn:bestq_delhaize}) being used to estimate the SFR, instead of the relation by \citet{Yun2001} (Equation~\ref{eqn:Yunrelation}), we find that the AGN-dominated fraction ranges from 11 per cent to 20 per cent (for sources with $\mathrm{SNR}_{\mathrm{FIR}} > 3.0 $ and $S_{\mathrm{1.3\,GHz}} > 3\sigma$). {\bf This demonstrates the impact of the empirical relations that we use as part of our analysis of the radio emission from radio-quiet quasars, and is the purpose of this paper -- i.e., to prompt the underlying assumptions of the community researching the origin of faint radio-emission. } Again, we suggest that our AGN-dominated fractions may be underestimated, given the findings of \citet{Wong2016} [see Section~\ref{sec:agn_fraction_discussion_Yun}].

As mentioned in Section~\ref{sec:SFR_comparisons}, these low `AGN-dominated fractions' are accompanied by high `possibly-starburst fractions'. This could in part be due to the FIR luminosities being over-estimates themselves, as a result of the low-resolution of the FIR data. This means that unrelated FIR sources may be contributing to the FIR emission from the quasar in question, and so `pushing' the corresponding datapoint above the one-to-one relation. However, if that is the predominant explanation, then we would expect that the SFR comparisons involving the Yun-SFR to be similarly affected.  

We suggest that the reason that we find such a difference for the current work \citep[cf.,][]{White2017} is that \citet{Delhaize2017} find the IRRC to be dependent on redshift. However, {\bf the impact of stellar mass, as found by \citet{Delvecchio2021}, also needs to be disentangled}\footnote{The unavailability of reliable stellar-mass estimates for Type-1 quasars (due the strong contribution of the quasar continuum towards the optical emission) is the reason that we do not apply the IRRC by \citet{Delvecchio2021} for the current sample.}. They note that at larger masses the $q_{\mathrm{IR}}$ decreases (Equations~\ref{eqn:bestq_delhaize}--\ref{eqn:qir}), and so the difference in radio-derived SFRs with respect to \citet{Yun2001} increases. {\bf However, \citet{DeZotti2024} argue that a more-signifcant parameter than stellar mass is galaxy type, with starbursts tending to dominate the star-forming population at higher redshifts (rather than low-SFR ``normal, late-type galaxies'').} Furthermore, the less-pronounced contribution of the AGN towards the total radio emission may be a consequence of the sensitivity of MeerKAT towards diffuse emission, which enables a greater fraction of the star formation in the host galaxy to be detected. Hence, further work on larger samples selected in the MIGHTEE fields (White et al., in prep.) are needed to investigate how the fraction of AGN-dominated sources (with respect to the total radio-emission) varies as a function of stellar mass, as well as redshift and radio luminosity.



\subsubsection{Binning by redshift and absolute optical-magnitude}
\label{sec:binning}

\begin{figure*}
\centering
\subfigure[ Median redshift, $< z > = 1.064 $]{
\includegraphics[scale=0.53]{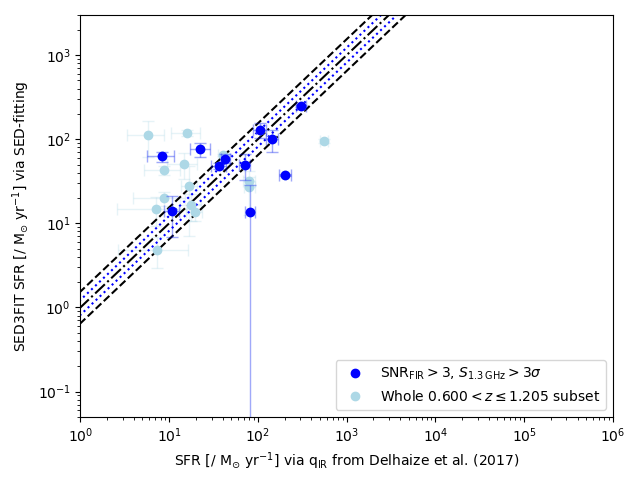} }
\subfigure[Median redshift, $< z > = 1.513 $]{
\includegraphics[scale=0.53]{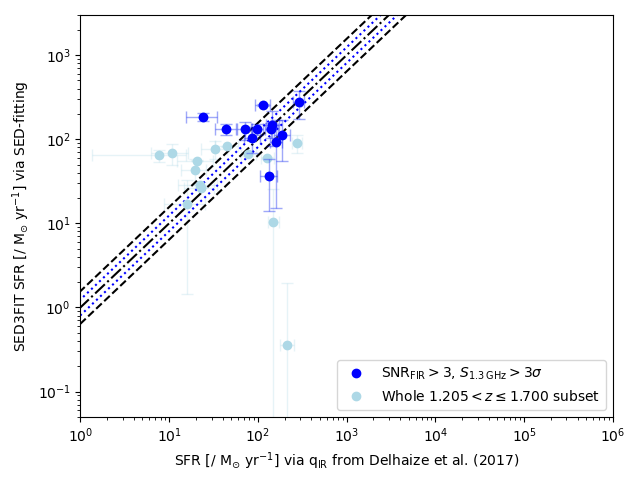} }
\subfigure[Median redshift, $< z > = 2.057 $]{
\includegraphics[scale=0.53]{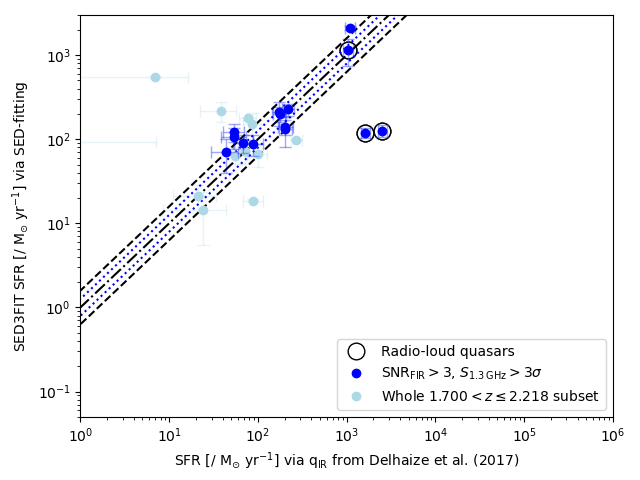} }
\subfigure[Median redshift, $< z > = 2.373 $]{
\includegraphics[scale=0.53]{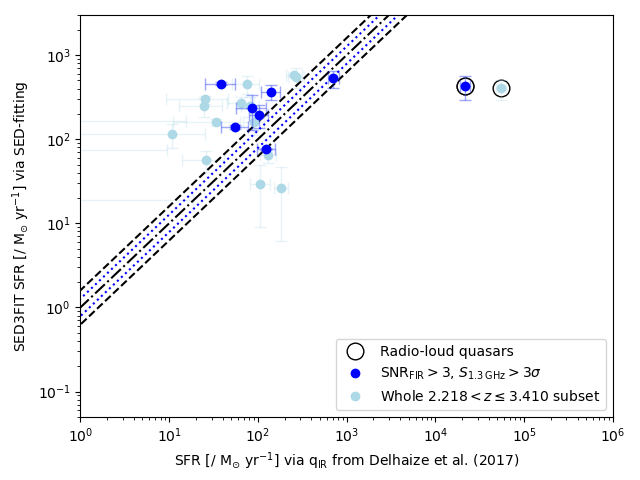} }
\caption{ A comparison of the same star-formation rates (SFRs) as in Figure~\ref{fig:SFRs_Delhaize_on_xaxis}b for quasars in the COSMOS and XMM-LSS fields. The SFR estimate on the abscissa axis assumes that all of the radio emission can be attributed to star formation, and so lies on the IRRC (Equation~\ref{eqn:bestq_delhaize}) determined by \citet{Delhaize2017}. Here the sample has been divided into four redshift bins, with each having its own calculation of the error in the Delhaize-SFR via the median redshift for that bin (blue, dotted lines). However, the redshift dependence is too small to be discernible in these log--log plots. The dash-dotted line represents where the two SFR estimates are equal to each other, and the black dashed lines indicating the total error in the one-to-one relation (i.e. the 1-$\sigma_{1:1}$ calibration error in both the Delhaize-SFR and the SED3FIT-SFR). In all panels, the darker-blue datapoints represent the quasars detected above 3\,$\sigma$ in both the radio and the FIR, and the light-blue datapoints represent all of the quasars within that redshift bin. }
\label{fig:SFRSFR_binned_by_redshift}
\end{figure*}

For this work, given the size of the quasar sample, we first investigate how the AGN-dominated fraction and possible-starburst fraction vary across four redshift bins. (See Figure~\ref{fig:redshiftbinning} for demarcation of these redshift bins.) This is done in order to take into account the impact of Malmquist bias on the sample properties, where {\it intrinsically} more-luminous sources tend to be detected at higher redshifts. For each of these four subsets, we simplify the analysis by calculating these fractions for all quasars within that redshift bin (without any detection thresholds being applied). The results are presented in Table~\ref{tab:AGNdominated_fractions_binned} and illustrated in Figure~\ref{fig:SFRSFR_binned_by_redshift}. 

For the AGN-dominated fraction, this is relatively constant across the redshift bins, with values ranging from 19 to 27 per cent. Surprisingly, for the third redshift bin, we see that one of the radio-loud quasars (identified in Figure~\ref{fig:radioloudness}) lies {\it on} the one-to-one relation rather than below it, as expected. This lends further support to the findings of \citet{Wong2016} that the properties of quasars may `conspire' in such a way as to mimic normal star-forming galaxies.

More significant than the (lack of) trend in the AGN-dominated fractions is the sudden increase in the possible-starburst fraction for the highest redshift bin (going from 31--38 per cent to 63 per cent). It is most likely that the FIR-derived SFR is being severely overestimated at these highest redshifts, as demonstrated by SED-fitting analyses \citep{Symeonidis2022a, Symeonidis2022b}. This is because, {\bf whilst it is {\it assumed} that all of the FIR emission is the result of star formation in the host galaxy\footnote{A reminder that our FIR luminosities have already had the AGN component subtracted as a part of SED-fitting.}, AGN heating of the dust becomes more significant and eventually becomes the dominant contribution towards the FIR emission}. As suggested by \citet{Symeonidis2022a}, the AGN is believed to be able to heat the dust at kpc scales, which is not taken into account for most AGN models that are used in SED-fitting. Overestimates of the FIR-derived SFRs therefore mean that the AGN contribution towards the total radio emission is {\it underestimated}, and so -- particularly for the fourth redshift bin, where the optical luminosities are the highest (Figure~\ref{fig:redshiftbinning}), and the disparity is more severe \citep{Symeonidis2022b}  -- our `AGN-dominated' fractions should be treated as lower limits \citep[see also][]{Cook2024}. Further support for this is provided by \citet{Symeonidis2021}, where the proportionality in the SFR--$L_{\mathrm{IR}}$ relation `breaks down' once the fraction of galaxies that are `AGN-powered' (albeit with respect to the {\it IR} emission) is $\gtrsim 0.35$ (see their figure 9).


{\bf However, the trend with redshift could be the result of a genuine increase in the typical SFR of galaxies, as would be expected from the evolution of the cosmic SFR density \citep[e.g.][]{Madau1996,Malefahlo2022}.} This is supported by the redshift-evolution (and quasar-luminosity-dependence) work by \citet{Bonfield2011} on the host galaxies of quasars, with $L_{\mathrm{IR}} \propto L_{\mathrm{QSO}}^{\theta} (1+z)^{\zeta}$ (where $\theta= 0.22 \pm 0.08$ and $\zeta = 1.6 \pm 0.4$). They also found that for the eight quasars ($1.1<z<2.3$) that had good FIR detections in all three SPIRE bands, the FIR luminosities spanned $6 \times 10^{12} < L_{\mathrm{IR}}/\Lsol< 2 \times 10^{13}$ (cf. Appendix~\ref{app:lfir}) for consistent grey-body dust temperatures of $T_{{\mathrm{d}}}\sim$30\,K. Given the redshifts of these quasars \citep[see, e.g.][]{DeZotti2024}, they are likely to be sources that exhibit both AGN and starburst activity, like our own sample. 

\begin{table*}
    \centering
    \begin{tabular}{c|c|r|c|c}
    \hline

          &  & Median separation, & Fraction of sources at & Fraction of sources that \\
      & &$\sigma_{\mathrm{sep}}$, with respect  &  that are AGN-dominated &  are possibly starbursts \\
     Range & Median &to the 1:1 relation & (i.e. $\sigma_{\mathrm{sep}} > 1.0$\,$\sigma_{1:1}$ &  (i.e. $\sigma_{\mathrm{sep}} > 1.0$\,$\sigma_{1:1}$ \\  
      of the bin & of the bin & (Whole subset)  & below the 1:1 relation) &  above the 1:1 relation) \\ 
       \hline

        $ 0.600 < z \leq 1.205 $ & 1.064 & $-$0.63\,$\sigma_{1:1}$ & 5  /  25  =  20\%,  & 9 / 25 =  36\%,  \\
            $ 1.205 < z \leq 1.700 $ & 1.513 & $-$0.38\,$\sigma_{1:1}$ & 7  /  26  =  27\%,  & 10  /  26  =  38\%,  \\ 
        $ 1.700 < z \leq 2.218 $ & 2.057 & $-$0.22\,$\sigma_{1:1}$ & 5 / 26 =  19\%, & 8 / 26 =  31\%,  \\
            $ 2.218 < z \leq 3.410 $ & 2.373 & $-$1.88\,$\sigma_{1:1}$ & 6 / 27 =  22\%,  & 17 / 27 =  63\%,  \\

\hline

        $ -23.25 < M_{B} \leq -21.90 $ & $-$22.98 & $-$0.63\,$\sigma_{1:1}$ & 5  /  25  =  20\%, & 12  /  25  =  48\%, \\
            $ -23.80 < M_{B} \leq -23.25 $ & $-$23.55 & $-$0.05\,$\sigma_{1:1}$ & 6  /  26  =  23\%,  & 7  /  26  =  27\%,  \\ 
        $ -24.20 < M_{B} \leq -23.80 $ & $-$23.97 & $-$0.26\,$\sigma_{1:1}$ & 7  /  26  =  27\%,  & 8 /  26  =  31\%,  \\
            $ -27.80 < M_{B} \leq -24.20 $ & $-$24.52 & $-$2.18\,$\sigma_{1:1}$ & 5  / 27 =  19\%,  & 18 / 27  =  67\%,  \\
            
    \hline
    \end{tabular}
    \caption{The fraction of sources in the COSMOS and XMM-LSS fields that are described as `AGN-dominated' and `possible starbursts', based upon comparing the SED3FIT-SFR with the Delhaize-SFR estimate, binned by redshift and absolute $B$-band magnitude (Section~\ref{sec:binning}). The `median separation' indicates the median number of $\sigma_{1:1}$ {\it below} the one-to-one (1:1) relation in Figures~\ref{fig:SFRSFR_binned_by_redshift} and \ref{fig:SFRSFR_binned_by_mag} that the quasars lie, where $\sigma_{1:1}$ refers to the calibration error in the 1:1 relation. Therefore, negative values of `median separation' indicate that a larger fraction of the quasars lie {\it above} the 1:1 relation in Figures~\ref{fig:SFRSFR_binned_by_redshift} and \ref{fig:SFRSFR_binned_by_mag}, and so are considered to be possible starbursts (see also Section \ref{sec:agn_fraction_discussion_Delhaize} and Figure~\ref{fig:fir_luminosities}) .} 
    \label{tab:AGNdominated_fractions_binned}
\end{table*}

Next we consider how the AGN-dominated and possible-starburst fractions vary with absolute optical-magnitude, $M_{\mathrm{B}}$. (Again, see Figure~\ref{fig:redshiftbinning} for demarcation of these bins.) For unobscured quasars (like the current sample), the optical luminosity can be used as a proxy for the accretion rate and, as shown by \citet{White2017}, this is more closely correlated with the accretion component of the total radio-luminosity than the star-formation component of the total radio-luminosity. However, that study was for a sample of quasars all at $z \sim 1$, whereas the optical-magnitude bins applied in the current work encompass a range of redshifts per bin. 

\begin{figure*}
\centering
\subfigure[Median magnitude, $<M_{B}> = -22.98$]{
\includegraphics[scale=0.53]{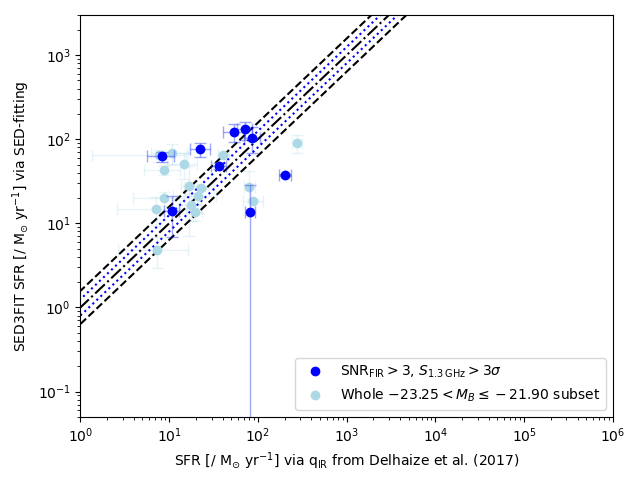} }
\subfigure[Median magnitude, $<M_{B}> = -23.55$]{
\includegraphics[scale=0.53]{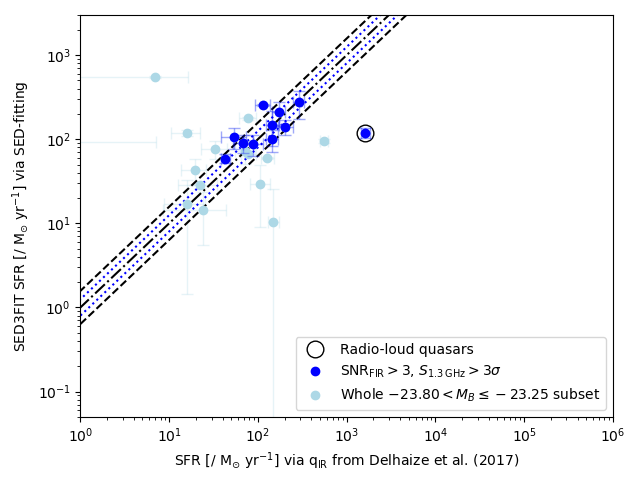} }
\subfigure[Median magnitude, $<M_{B}> = -23.97$]{
\includegraphics[scale=0.53]{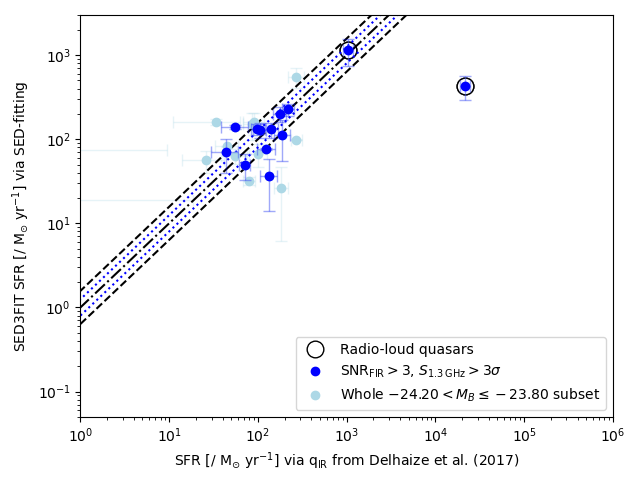} }
\subfigure[Median magnitude, $<M_{B}> = -24.52$]{
\includegraphics[scale=0.53]{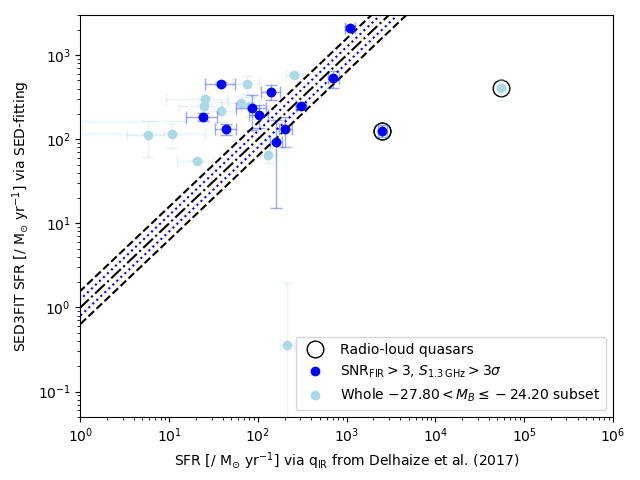} }
\caption{ A comparison of the same star-formation rates (SFRs) as in Figure~\ref{fig:SFRs_Delhaize_on_xaxis}b for quasars in the COSMOS and XMM-LSS fields. The SFR estimate on the abscissa axis assumes that all of the radio emission can be attributed to star formation, and so lies on the IRRC (Equation~\ref{eqn:bestq_delhaize}) determined by \citet{Delhaize2017}. Here the sample has been divided into four optical-magnitude bins, and the median redshift for the whole sample is used to calculate the error in the Delhaize-SFR (blue dotted lines). The dash-dotted line represents where the two SFR estimates are equal to each other, with the black dashed lines indicating the total error in the one-to-one relation (i.e. the 1-$\sigma_{1:}$ calibration error in both the Delhaize-SFR and the SED3FIT-SFR). In all panels, the darker-blue datapoints represent the quasars detected above 3\,$\sigma$ in both the radio and the FIR, and the light-blue datapoints represent all of the quasars within that magnitude bin.  }
\label{fig:SFRSFR_binned_by_mag}
\end{figure*}

Firstly, we note how all of the datapoints appear to move in unison towards higher SFRs with increasing $M_{B}$, generally in parallel with the one-to-one relation (see Figure~\ref{fig:SFRSFR_binned_by_mag}). This is as expected for two SFR estimates being derived for a flux-limited sample. However, the scatter in the FIR-related SFR appears to translate to the possible-starburst fraction showing no trend with optical magnitude (Table~\ref{tab:AGNdominated_fractions_binned}). What is notable, again, is that {\bf the possible-starburst fraction exceeeds 60 per cent for the fourth bin (in this case, $ -27.80 < M_{B} \leq -24.20 $). This is likely due to this bin containing the highest redshifts and therefore the highest FIR luminosities (via the Malmquist bias), and so is subject to the same breakdown in assumption described earlier for the highest-redshift bin} \citep{Symeonidis2021,Symeonidis2022a,Symeonidis2022b}. Furthermore, since the gas that fuels the accretion process is likely to also fuel star formation, we may expect the quasars with the largest stellar masses to be present within this brightest-magnitude bin. As such, the IRRC relation by \citet{Delhaize2017} -- derived for normal star-forming galaxies -- is most-discrepant for this bin \citep{Delvecchio2021}, and the different effects compound in such a way that we see the greatest separation of the datapoints from the one-to-relation, at 2.18\,$\sigma$ (Table~\ref{tab:AGNdominated_fractions_binned}) above the line. 

\subsubsection{Additional comparison with work in the literature}


Meanwhile, \citet{Jarvis2019} observe ten $z < 0.2$ obscured quasars, with the JVLA (at 0.25--1.0-arcsec resolution) and find extended radio structures (i.e. evidence of radio jets) for nine of them. They have radio luminosities of $L_{\mathrm{1.4\,GHz}} = 10^{23.3-24.4}$\,W\,Hz$^{-1}$, whilst our sample probes quasars at higher redshifts (Figure~\ref{fig:redshifts}) and with radio luminosities of $L_{\mathrm{1.4\,GHz}} = 10^{21.4-26.4}$\,W\,Hz$^{-1}$ (the median being $10^{23.5}$\,W\,Hz$^{-1}$ when considering the full sample, and $10^{23.8}$\,W\,Hz$^{-1}$ when considering the quasars that are detected at a $>$3-$\sigma$ level in the radio). \citet{Jarvis2019} find that their sources are significantly offset ($> 3 \sigma$) from the FIRC of \citet{Bell2003}, and the accompanying integral-field spectroscopy of the warm ionised gas shows co-spatiality of distinct kinematic components with these jet/lobe structures. \citet{Jarvis2021}  follow-up their work with a sample of 42 optically-luminous ($L_{\mathrm{[O\textsc{iii}]}} > 10^{42.11}$\,erg\,s$^{-1}$) quasars of median radio-luminosity, $L_{\mathrm{1.4\,GHz}} = 10^{23.8}$\,W\,Hz$^{-1}$ (also at $z < 0.2$). The 0.3--1.0-arcsec spatial resolution of the JVLA data again reveals extended radio structures for 67 per cent of the sample, noting that 10--21 per cent are classified as `radio loud' via traditional methods. 

\citet{Jarvis2021} also have FIR data for 13 of the 42 quasars, and within this subset (which includes only one `radio loud' quasar; see their table 2) they find that 77 per cent of the sources have radio emission that exceeds that expected from star formation alone. This is based on a threshold of $q_{\mathrm{IR}} < 1.64$ \citep{Helou1985}, and for quasars at $z = 0.1$--0.2 (to be consistent with \citealt{Jarvis2019,Jarvis2021}), we calculate $q_{\mathrm{IR}} < (2.806 \pm 0.010)$ via the IRRC of \citet{Delhaize2017} [Equations~\ref{eqn:bestq_delhaize} and \ref{eqn:qir}]. {\bf Our more-relaxed criterion means that our work may be exploring a more-representative sample of radio-quiet quasars (i.e., not restricted to the most optically-luminous sources), with a higher fraction of the radio emission being due to star formation.} This higher star-formation fraction is in agreement with the work of \citet{Gurkan2019}, who studied the 144\,MHz radio-emission of optically-selected quasars and applied a low-frequency-specific FIRC to their sample. However, we note that \citet{White2017} apply a similar $q$ threshold as ourselves ($ q_{125} = \log_{10} [ (L_{125 \mathrm{\upmu m}} / (\mathrm{ W\,Hz}^{-1})) / (L_{\mathrm{1.5\,GHz}} / (\mathrm{W\,Hz}^{-1})) ] $) of 2.65, and find that AGN-related processes account for at least 60 per cent of the radio emission from their sample of 70 quasars at $ 0.9 < z < 1.0 $ (see their table 4). As mentioned earlier within Section~\ref{sec:agn_fraction_discussion_Yun}, this monochromatic FIRC is based upon the work of \citet{Smith2014}, who find that the relation does not evolve with redshift (but has a strong dust-temperature dependence).


{\bf Another explanation for the smaller fraction of quasars with AGN-dominated radio emission found in this work (14--22 per cent; Section~\ref{sec:agn_fraction_discussion_Delhaize}) is that we are probing to a lower 1.4-GHz radio flux-density limit ($\approx$\,3.0\,\textmu Jy) than the work of both \citet{White2015} ($\approx$\,17.5\,\textmu Jy) and \citet{White2017} ($\approx$\,30.9\,\textmu Jy), where there may be a larger contribution from star-forming processes \citep{Wilman2008, Bonaldi2019, Gurkan2019, Hale2025}.} This is consistent with the interpretation by \citet{Fawcett2020} of the origin of the radio emission for radio-quiet quasars (see their section 4.1), who, in addition, find that 56 per cent of a 70-strong sample of {\it Herschel}-detected quasars have AGN-dominated radio emission (see their figure 12). This is based upon extrapolation of 3-GHz measurements to 1.4\,GHz, following a typical power-law function with spectral index, $\alpha = -0.7$. 

Meanwhile, \citet{Rosario2020} also suggest a more-significant contribution from star formation for lower-radio-luminosity sources, and we note that our sample includes luminosities that are $\sim$2 orders of magnitude lower (at $L_{\mathrm{144\,MHz}} \sim 4 \times 10^{22}$\,W\,Hz$^{-1}$, again scaling via $\alpha = -0.7$) than their work (e.g. see their figure 4). Furthermore, like \citet{Rosario2020}, \citet{Rankine2021} use the first data release of the LOFAR (LOw-Frequency ARray) Two-metre Sky Survey \citep[LoTSS;][]{Shimwell2019, Williams2019} to study the radio properties of quasars in the SDSS DR14Q catalogue of \citet{Paris2018}. Their alternative approach to simulating a SFR distribution for their quasar sample again suggests that star formation could be contributing a significant fraction of the radio emission, up to $L_{\mathrm{144\,MHz}} = 10^{26}$\,W\,Hz$^{-1}$, with their best-fitting quasar-sample model (see their figure 10) having a median SFR of $\approx 30$\,$\Msol$\,yr$^{-1}$ and a mean SFR of $\approx 420$\,$\Msol$\,yr$^{-1}$. \citet{Rankine2021} also consider jets and winds as possible mechanisms for the origin of the radio emission, but can only conclude that there are likely to be multiple overlapping contributions.   

In favour of the wind explanation,  \citet{Zakamska2014} study obscured, radio-quiet quasars at $z < 0.8$ and find that the velocity width of the [O\textsc{iii}] $\lambda 5007$\,\AA\ emission line (indicative of outflow in the narrow-line region) is positively correlated with both the mid-infrared luminosity (tracing the dust that is heated by the accretion disc) and the level of radio emission. The latter is thought to be the result of shock fronts generated via the outflow impacting upon the interstellar medium. These observations were extended by \citet{Hwang2018} to a sample of extremely red quasars (ERQs) at higher redshift ($ 2 < z < 4$), but the 1.3-arcsec resolution of their 6.2-GHz radio data is insufficient to rule out the possibility that the kinematics could be due to compact radio-jets rather than quasar-driven winds. High-resolution ($\sim$1.0\,arcsec) radio-images are also obtained by \citet{Smith2020} (at 22\,GHz) for 100 radio-quiet, X-ray-selected AGN ($0.003 < z < 0.049$), and they detect kiloparsec-scale radio structures for 11 of them.\footnote{For interest, these 11 sources have $S_{\mathrm{1.4\,GHz}}$ ranging from 6.6 to 454.6\,mJy.} Aided by the CO observations that they have for the entire sample, they are able to infer the molecular-gas mass of the system and estimate the energy required to eject this gas from the galaxy. Comparing this ejection-energy with output-energy estimates for (i) a radiative-AGN outflow scenario, and (ii) a small-scale radio-jet scenario, they reason that the former is the more-likely explanation for the radio structures that are seen, in order to achieve the level of suppressed star-formation that is observed in the host galaxies.

\subsubsection{Summary}
\label{sec:discussion_summary}

We have explored the impact of different SFR estimates on the analysis of the radio emission from quasars detected in the COSMOS and XMM-LSS fields of MIGHTEE. We find that the relation that is used to derive a SFR estimate through radio data has a greater impact than the FIR-based SFR relation (in terms of the results that we obtain via the `radio-excess' analysis method), but note that the latter may be more significant for high optical-luminosity AGN (towards which quasars at higher redshifts will be biased). Compared with previous work \citep[e.g.][]{White2015,White2017}, we find that a less-pronounced fraction of the radio emission can be attributed to accretion activity, and there is a greater prevalence of (possible) starbursts. As mentioned, this is likely a combination of the excellent sensitivity of MeerKAT towards star-formation activity, and that our sample is more representative of the wider quasar population. However, we caution that the low resolution of FIR data (which could blend together the FIR emission of unrelated sources) complicates the interpretation.

Given that \citet{Gurkan2018}, \citet{Delvecchio2021}, and \citet{McCheyne2022} find the IRRC to be strongly dependent on {\it stellar mass}, we explore the AGN versus star-formation fractions further with a larger sample of MIGHTEE quasars (White et al., in prep.) and encourage the consideration of dust temperature \citep[e.g.][]{Bonfield2011,Smith2014}, in addition.

\section{Conclusions}
\label{sec:conclusions}

In this work we study a sample of 104 unobscured (Type-1) quasars within the COSMOS and XMM-LSS fields of the MIGHTEE survey (Section~\ref{sec:radiodata}), selected via $gJK_{s}$ colour-space (Section~\ref{sec:sample_selection}) and reaching 1.3-GHz flux-densities of rms $\approx$ 3.0\,\textmu Jy\,beam$^{-1}$ (Section~\ref{sec:faintemission}). Thanks to the large, multiwavelength datasets that are available over these fields, the properties of radio-loud and radio-quiet quasars can be studied in a statistically-robust way (White et al., in prep.), with the emphasis of this work being on the AGN-related and star-formation-related contributions to the total radio emission. A standard method for such analysis is to compare a radio-derived SFR estimate with a (F)IR-derived SFR estimate, and we employ multiple combinations (Table~\ref{tab:AGNdominated_fractions}) so that our results can be compared more-easily with others in the literature.

\begin{enumerate}
\item{ Our quasar sample is spectroscopically-complete to $K_{s} = 21.0$\,mag, meaning that we can study systems that include old stellar populations (that may have lower levels of star formation).  }
\item{ These spectroscopic redshifts were carefully compiled from previous work (Appendix~\ref{app:assess_speczs}), and supplemented by four new redshifts via optical spectroscopy from the Southern African Large Telescope (Appendix~\ref{app:new_speczs}). The final redshift range of the sample is $0.60 < z < 3.41$ (Figure~\ref{fig:redshifts}), with median redshift, $z = 1.68$. }
\item{ The MIGHTEE images have a {\it thermal} noise of 1.5--1.6\,\textmu Jy\,beam$^{-1}$ \citep{Heywood2022,Hale2025}, and based on Gaussian fitting to pseudo-random positions in the radio maps, we determine the {\it confusion} noise to be at the level of 2.5--2.6\,\textmu Jy\,beam$^{-1}$. Considering the total-noise estimate {\it per quasar}, we detect 81 of the 104 quasars at the 3-$\sigma$ level (Section~\ref{sec:faintemission}). Those that are undetected remain in the subsequent statistical analysis. }
\item{ Based upon the radio-loudness parameter defined by \citet{Kellermann1989}, $R = S_{\mathrm{5\,GHz}}/S_{\mathrm{4400\,\mathring{A}}}$, we determine a radio-loudness fraction of 5 per cent (Section~\ref{sec:radioloudness}), which is consistent with other quasar samples in the literature \citep[e.g.][]{Rankine2021}. }
\item{ The SFR--$L_{\mathrm{1.4\,GHz}}$ relation of \citet{Yun2001} is often applied to the radio-luminosity of faint radio sources \citep[e.g.][]{White2015, White2017}, under the assumption that all of the radio emission is due to star-forming processes. Therefore, when this value exceeds the SFR estimated via FIR data (as an independent, best tracer of star formation), we can interpret this `radio excess' as the source having its radio emission dominated by AGN-related processes. For quasars that are detected at the 3-$\sigma$ level in both the radio data and the FIR data, we find that the fraction of sources that are AGN-dominated ranges from 22 per cent [when we apply the SFR relation of \citet{Kennicutt1998b} to the SED-fitted IR luminosities] to 58 per cent [when the SFR is instead estimated directly via SED-fitting with SED3FIT \citep{Berta2013}]. Alongside this, the fraction of sources (across the full sample) that are possible starbursts is at the level of 5--16 per cent (Section~\ref{sec:agn_fraction_discussion_Yun}, Table~\ref{tab:AGNdominated_fractions}).}
\item{ \citet{Delhaize2017} found the IRRC to be dependent on redshift, and we take this into account by applying Equations~\ref{eqn:bestq_delhaize} and \ref{eqn:sfr_delhaize} to the radio luminosity in order to estimate the expected `radio-derived' SFR. (Again, this is based on the assumption that all of the radio emission can be explained by star formation, and so the quasar lies directly on the IRRC.) For quasars that are detected at the 3-$\sigma$ level in both the radio data and the FIR data, we find that the fraction of sources that are AGN-dominated ranges from 11 per cent [when the \citet{Delhaize2017} SFR-estimate is compared with the \citet{Kennicutt1998b} SFR-estimate, the latter based on SED-fitting of only the FIR data] to 20 per cent [when the \citet{Delhaize2017} SFR-estimate is compared with the SED3FIT \citep{Berta2013} SFR-estimate, the latter being derived over optical-to-FIR data]. These lower fractions are evident from comparing Figure~\ref{fig:SFRs_Delhaize_on_xaxis} with Figure~\ref{fig:SFRs_Yun_on_xaxis}, where the latter has a greater fraction of datapoints below the one-to-one relation (dashed line). Meanwhile, the fraction of `possible starbursts' is higher than previously, ranging from 27 per cent to 43 for the full sample (Section~\ref{sec:agn_fraction_discussion_Delhaize}).  }
\item{ {\bf Compared with previous work \citep{White2015, White2017}, we find a lower fraction of the quasar sample has radio emission that is dominated by the AGN, which may be a combination of: (a) the lower radio flux-density limit of the radio images, enabling a greater sensitivity to the lower radio-luminosities associated with star formation; (b) the redshift-dependence of the IRRC \citep{Delhaize2017} leading to a more-accurate radio-derived SFR-estimate than the SFR--$L_{\mathrm{1.4\,GHz}}$ relation of \citet{Yun2001}, and (c) our selection resulting in a more-representative sample of the quasar population. } }
\item{ We also calculate the AGN-dominated fraction and the possible-starburst fraction across four redshift bins (Section~\ref{sec:binning}, Table~\ref{tab:AGNdominated_fractions_binned}). Whilst the AGN-dominated fraction is relatively consistent (19--27 per cent), the fraction of possible starbursts increases dramatically from 31--38 per cent to 63 per cent for the highest-redshift bin. This is likely due to {\bf the IRRC by \citet{Delhaize2017} being derived for normal star-forming galaxies, and so becoming less suitable for assessing the level of radio emission for starburst systems}, which are prevalent at higher redshift. }
\item{ We also investigate how the fractions vary with optical luminosity, and find that the fraction of possible starbursts is higher than the fraction of AGN-dominated sources for each optical-luminosity bin (Section~\ref{sec:binning}, Table~\ref{tab:AGNdominated_fractions_binned}). This is likely due to the varying redshifts sampled per bin, with {\bf the contribution of the AGN towards the {\it IR} emission becoming more significant at higher redshifts/FIR luminosities \citep{Symeonidis2021, Symeonidis2022a, Symeonidis2022b}, leading to an overestimation of the FIR-derived SFR and therefore an underestimation of the AGN-dominated fraction\footnote{Relatedly, we note that (for example) \citet{Ma2018} find, through SED-fitting, that the contribution of the AGN towards the IR emission of a gravitationally-lensed starbust, at $z \sim 1$, is ``negligible''.}.} Since our FIR luminosities have already had the AGN component subtracted, this means that the AGN contribution towards the IR emission is actually higher than accounted for in the SED-fitting. }
\item{ A consequence of the higher incidence of starbursts at higher redshifts is that it becomes more difficult to use the `radio-excess' method to determine whether or not the AGN is dominating the total radio emission in the system. As such, methods for selecting and analysing AGN will need to be considered more carefully. }
\end{enumerate}

In future work we will extend the analysis of AGN-versus-SF contributions to the total radio emission, as a function of stellar mass and redshift, with a larger sample of $\sim$1,600 sources that also includes obscured (Type-2) quasars. As to the origin of the AGN-related radio emission in radio-quiet quasars, higher-resolution radio data ($\lesssim 1$\,arcsec) could help to identify the presence of small-scale radio-jets \citep[e.g.][]{Morabito2025}. Where they are absent, the gas kinematics revealed by integral-field spectroscopy may still indicate an AGN outflow (generated via quasar winds, or relativistic jets that remain undetected in the radio images). How the radio properties vary with black-hole mass, accretion rate, and magnetic-field strength also warrants further investigation.

\section*{Acknowledgements}

We thank the anonymous referee for their careful reading of the manuscript and for their constructive, minor comments. We also thank Paul C. Hewett for providing a quasar-colour template, and Matt J. Jarvis for comments that helped to improve the manuscript. 

The financial assistance of the South African Radio Astronomy Observatory (SARAO) towards this research is hereby acknowledged (www.sarao.ac.za). Some of the observations reported in this paper were obtained with the Southern African Large Telescope (SALT), under program 2019-2-SCI-046 (PI: White). ID acknowledges support from INAF Mini-grant ``Harnessing the power of VLBA towards a census of AGN and star formation at high redshift''. RB acknowledges support from an STFC Ernest Rutherford Fellowship [grant number ST/T003596/1]. IHW and CLH acknowledge generous support from the Hintze Family Charitable Foundation through the Oxford Hintze Centre for Astrophysical Surveys. In addition, CLH acknowledges support from the Leverhulme Trust through an Early-Career Research Fellowship. NJA acknowledges support from the ERC Advanced Investigator grant EPOCHS (788113).

The MeerKAT telescope is operated by the South African Radio Astronomy Observatory, which is a facility of the National Research Foundation, an agency of the Department of Science and Innovation. We acknowledge the use of the ilifu cloud computing facility – www.ilifu.ac.za, a partnership between the University of Cape Town, the University of the Western Cape, Stellenbosch University, Sol Plaatje University and the Cape Peninsula University of Technology. The Ilifu facility is supported by contributions from the Inter-University Institute for Data Intensive Astronomy (IDIA – a partnership between the University of Cape Town, the University of Pretoria and the University of the Western Cape, the Computational Biology division at UCT and the Data Intensive Research Initiative of South Africa (DIRISA). The authors acknowledge the Centre for High Performance Computing (CHPC), South Africa, for providing computational resources to this research project.

This work is based on data products from observations made with ESO Telescopes at the La Silla Paranal Observatory under ESO programme ID 179.A-2005 (Ultra-VISTA) and ID 179.A-2006 (VIDEO) and on data products produced by CALET and the Cambridge Astronomy Survey Unit on behalf of the Ultra-VISTA and VIDEO consortia.

The Hyper Suprime-Cam (HSC) collaboration includes the astronomical communities of Japan and Taiwan, and Princeton University. The HSC instrumentation and software were developed by the National Astronomical Observatory of Japan (NAOJ), the Kavli Institute for the Physics and Mathematics of the Universe (Kavli IPMU), the University of Tokyo, the High Energy Accelerator Research Organization (KEK), the Academia Sinica Institute for Astronomy and Astrophysics in Taiwan (ASIAA), and Princeton University. Funding was contributed by the FIRST program from Japanese Cabinet Office, the Ministry of Education, Culture, Sports, Science and Technology (MEXT), the Japan Society for the Promotion of Science (JSPS), Japan Science and Technology Agency (JST), the Toray Science Foundation, NAOJ, Kavli IPMU, KEK, ASIAA, and Princeton University.

The {\it Herschel} Extragalactic Legacy Project (HELP) is a European Commission Research Executive Agency funded project under the SP1-Cooperation, Collaborative project, Small or medium-scale focused research project, FP7-SPACE-2013-1 scheme, Grant Agreement Number 607254. We thank Mattia Vaccari and Raphael Shirley for their help with HELP.

Funding for the Sloan Digital Sky Survey IV has been provided by the Alfred P. Sloan Foundation, the U.S. Department of Energy Office of Science, and the Participating Institutions. SDSS-IV acknowledges support and resources from the Center for High Performance Computing  at the University of Utah. The SDSS website is www.sdss.org. SDSS-IV is managed by the Astrophysical Research Consortium for the Participating Institutions of the SDSS Collaboration including the Brazilian Participation Group, the Carnegie Institution for Science, Carnegie Mellon University, Center for Astrophysics | Harvard \& Smithsonian, the Chilean Participation Group, the French Participation Group, Instituto de Astrof\'isica de Canarias, The Johns Hopkins University, Kavli Institute for the Physics and Mathematics of the Universe (IPMU) / University of Tokyo, the Korean Participation Group, Lawrence Berkeley National Laboratory, Leibniz Institut f\"ur Astrophysik Potsdam (AIP), Max-Planck-Institut f\"ur Astronomie (MPIA Heidelberg), Max-Planck-Institut f\"ur Astrophysik (MPA Garching), Max-Planck-Institut f\"ur Extraterrestrische Physik (MPE), National Astronomical Observatories of China, New Mexico State University, New York University, University of Notre Dame, Observat\'ario Nacional / MCTI, The Ohio State University, Pennsylvania State University, Shanghai Astronomical Observatory, United Kingdom Participation Group, Universidad Nacional Aut\'onoma de M\'exico, University of Arizona, University of Colorado Boulder, University of Oxford, University of Portsmouth, University of Utah, University of Virginia, University of Washington, University of Wisconsin, Vanderbilt University, and Yale University.

\section*{Data Availability}

A machine-readable version of Table~\ref{tab:luminosities_sfrs_allquasars} is available online as Supplementary Material.



\bibliographystyle{mnras}
\bibliography{MIGHTEEradioquiet}




\appendix

\section{Assessing existing spectroscopic redshifts}
\label{app:assess_speczs}

We collate spectroscopic redshifts for our sample of 104 quasars (at $K_{s} < 21.0$; Section~\ref{sec:existing_spectra}) using the following catalogues: Milliquas \citep{Flesch2021}, reanalysis of spectra for G10 \citep{Davies2015}, PRIMUS \citep{Cool2013}, zCOSMOS-bright \citep{Lilly2009}, OzDES \citep{Lidman2020}, and 2XLSSd \citep{Melnyk2013}. Due to the redshift values differing between catalogues by varying amounts (from source to source), we summarise our decisions through Tables~\ref{tab:cosmos_speczs} and \ref{tab:xmmlss_speczs}. Where there are multiple reliable redshifts (as deemed by the respective catalogues), we take the median redshift (except in the case of one COSMOS source, as explained below).

The G10 catalogue incorporates information from zCOSMOS-bright and PRIMUS, as well as SDSS -- which is redundant against Milliquas -- and the VIMOS-VLT Deep Survey (VVDS; \citealt{LeFevre2013}) -- which provides no redshifts for our COSMOS quasars. For the single COSMOS source where only a photometric redshift is available (R.A. = 09:57:58.40, Dec. = +02:17:29.1), we obtain SALT spectroscopy (Section~\ref{sec:salt_spectra}). For the quasar at R.A. = 10:01:16.95, Dec. = +01:38:04.1 we note significant discrepancy between the G10 `best' $z$ value (accompanied by a `use' flag of `1', meaning ``Reliable high resolution spectroscopic redshift -- spectrum has been visually inspected and redshift is good''; \citealt{Davies2015}) and the zCOSMOS-bright $z$ value (with a quality flag that confirms that it is a quasar, has ``a very secure redshift'', and that ``the spectroscopic and photometric redshifts are consistent to within  $0.08(1+z)$''; \citealt{Lilly2009}). Also provided in the G10 catalogue is the `class' (or `generation') flag, Z\_GEN, which (for this source) indicates that the newly-fitted {\textsc AUTOZ} redshift and the zCOSMOS-bright redshift ``agree at a $\delta z < 0.1 z$ level ($<10$\% error). \textsc{AUTOZ} redshift used''. This level of agreement is true for the {\it second} best-fit AUTOZ value (AUTOZ2 = 0.745) but not for the best-fit AUTOZ value (AUTOZ1 = 0.299) that has been adopted. Therefore, we proceed with the zCOSMOS-bright $z$ value for this source, and estimate its error to be $0.08(1+0.747) = 0.140$.

For the XMM-LSS source at R.A. = 02:17:05.30, Dec. = $-$04:23:13.6 we favour the DR16Q redshift, having checked the SDSS Science Archive Server \citep{Neilsen2008} and seen that multiple emission lines are identified. The alternative redshift, from OzDES, is based upon a single (albeit strong) spectral feature, as indicated by the quality flag `3'. The 2XLSSd quality flag for the quasar at R.A. = 02:23:32.14, Dec. = $-$04:57:39.4 indicates that the redshift (1.900) is ``good quality (two or more lines in the spectra)'' \citep{Melnyk2013}, but multiple surveys provide a redshift of 0.602. We err on the side of caution and proceed with the latter redshift value. A similar scenario occurs for the quasar at R.A. = 02:25:03.12, Dec. = $-$04:40:25.3, with $z=1.741$ and $z=0.270$ being possibilities. Checking the SDSS Science Archive Server again allows us to determine which redshift (1.741) to use for our work. In addition, there are three XMM-LSS quasars for which there are no spectroscopic redshifts available in the literature, and so we observe these using the Southern African Large Telescope (SALT) [Section~\ref{sec:salt_spectra}]. 

\begin{table*}
	\centering
	\caption{An overview of the redshifts considered for the {\bf 23 quasars within the COSMOS field}. All are {\it spectroscopic} redshifts except for two photometric ones, which are indicated via the G10 `use' flag being set to `4'. {\bf The redshift used in our analysis of the radio emission is shown in bold}, and its error (provided in the final column) encompasses the range of redshift values that are deemed to be {\it reliable} (via the various `use'/`quality' flags). Where there is only one reliable redshift value, the measurement error from that survey is quoted instead. Appendix~\ref{app:assess_speczs} lists the references for the catalogues involved (Milliquas, G10, zCOSMOS-bright and PRIMUS), where the Milliquas catalogue includes redshifts from: MQ = \citet{Flesch2021}, DR16Q = \citet{Lyke2020}, 0272 = \citet{Brusa2010}, 1483 = \citet{Prescott2006}, 1885 = \citet{Trump2007}, 1886 = \citet{Trump2009}, and C-COSM = \citet{Marchesi2016}.}
	\label{tab:cosmos_speczs}
	\begin{tabular}{lcccccccccr} 
		\hline
		R.A. & Dec. & Milliquas & Milliquas & G10 & G10 & zCOSMOS- & zCOSMOS-bright & PRIMUS & PRIMUS & Error\\
		(hms) & (dms) & $z$ & reference & `best' $z$ & `use' flag & bright $z$ & quality & $z$ & quality & in $z$  \\
		\hline
  09:57:58.40 & +02:17:29.1 &   --      &    MQ       &      2.220   &    4  &    -- &  --   &    -- &  --   & --      \\  
  09:58:22.18 & +01:45:24.1 &   1.964    &   DR16Q   &        {\bf 1.963} &  1  &    -- &  --    &   -- & -- & 0.001     \\     
  09:58:26.67 & +02:28:18.1 &   --  &        --     &         0.411 &  3  &   {\bf 0.692 } &   13.5   &       -- &  --     &  0.135    \\ 
  09:58:48.86 & +02:34:41.1 &  1.547 &      DR16Q    &       {\bf 1.541}   &   1  &    1.541 &  13.5    &      1.539 &  4    & 0.006  \\       
  09:59:34.82 & +02:02:49.9 &   {\bf  2.182} &       0272   &      2.181   &   1  &    2.181 &  18.5  &      2.203 & 4     &  0.021   \\       
  09:59:44.79 & +02:40:48.8 &   1.405   &    0272    &      {\bf 1.405}   &   1  &    1.405 &   18.1   &       1.407 & 4     &  0.002     \\      
  10:00:01.44 & +02:48:44.7 &   0.767 &      DR16Q   &       {\bf 0.765}  &  1  &    -- &  --  &     -- &  --      &  0.002         \\      
  10:00:11.23 & +01:52:00.2 & 2.411    &   0272      &      2.415   &   2  &    -- &  --   &   {\bf 2.415 } &   4      &  0.004          \\      
  10:00:12.44 & +01:40:57.9 &  2.277    &   1886     &    {\bf 2.276} &  1  &    2.272 &  18.5    &      -- &  --    &  0.004      \\      
  10:00:12.91 & +02:35:22.8 &  0.699     &  DR16Q    &  {\bf 0.698} &  1  &    0.696  &  13.5   &       -- &  --    &  0.002      \\      
  10:00:14.09 & +02:28:38.6 &  1.253 &       1483    &  {\bf  1.255}  &    1  &    1.255 &  13.5  &       1.269 &   3   &  0.014      \\     
  10:00:51.92 & +01:59:19.3 & {\bf  2.240 } &      1483     &  2.230   &   1 &     2.230 &   13.5  &        2.254 & 4    &  0.014      \\    
  10:00:55.39 & +02:34:41.4 &  1.404  &     DR16Q    &   {\bf  1.402} &   1  &    1.401 &  13.5   &      -- &  --  &  0.002      \\      
  10:00:56.70 & +02:17:20.9 &  2.078   &    0272     &       0.748  &  3  &   {\bf  2.077} &   12.1   &  -- &  --   &  0.001   \\   
  10:01:16.95 & +01:38:04.1 &  0.748    &   0272     &     0.299  &  1  &  {\bf  0.747} &  13.5   &      -- &  --  &  0.140    \\  
  10:01:23.98 & +02:14:46.0 &  0.894     &  1885     &  {\bf  0.895} &    1  &    0.895 &  3.5   &        0.887 & 4       &  0.002         \\      
  10:01:24.74 & +01:57:38.7 &  1.173 &       C-COSM   &       1.166 &  2  &    -- &  --  &    {\bf 1.166} &  4   &  0.007      \\       
  10:01:28.01 & +02:18:19.2 &    1.187 &      1885    &        1.177  &    2  &    -- &  --  &   {\bf  1.177} &   3     &  0.010         \\      
  10:01:32.81 & +01:57:59.9 &   1.530    &    1483    &   {\bf  1.536}   &   1 &    1.536 &  13.5    &      1.537  & 4     & 0.006      \\       
  10:01:41.41 & +02:00:50.9 & {\bf 2.277 }    &  1885   &  1.075  &   3 &     2.268 &   18.3   &       -- &  --   &  0.009  \\   
  10:01:43.04 & +01:49:31.9 &  2.076    &   1885   &         1.270  &   4  &  {\bf  2.084} &  14.5   &       -- &  --   &  0.008  \\                
  10:02:08.55 & +01:45:53.6 &  2.215     &  1885    &       0.223 &    3  &   {\bf 2.203} &  18.5   &       -- &  --   &  0.012  \\     
  10:02:51.85 & +01:46:09.7 &   2.459     &  DR16Q    &    {\bf  2.459 } &  1  &    -- & --   &    -- &  -- &  0.001 \\              
		\hline
	\end{tabular}
\end{table*}




\begin{table*}
	\centering

	\caption{An overview of the spectroscopic redshifts considered for the {\bf 81 quasars within the XMM-LSS field}.  {\bf The redshift used in our analysis of the radio emission is shown in bold}, and its error (provided in the final column) encompasses the range of redshift values that are deemed to be {\it reliable} (via the various `use'/`quality' flags). Where there is only one reliable redshift value, the measurement error from that survey is quoted instead. Appendix~\ref{app:assess_speczs} lists the references for the catalogues involved (Milliquas, OzDES, 2XLSSd and PRIMUS), where the Milliquas catalogue includes (further to those mentioned in Table~\ref{tab:cosmos_speczs}) redshifts from: 0646 = \citet{Garilli2014}, 0650 = \citet{Gavignaud2006}, 1042 = \citet{Lacy2013}, 1340 = \citet{Nakos2009}, 1566 = \citet{Rowan-Robinson2013}, 1689 = \citet{Sharp2002}, 1713 = \citet{Simpson2006}, 1718 = \citet{Smail2008}, 2032 = \citet{White2015}, DR16 = \citet{Ahumada2020}, VIPERS = \citet{Scodeggio2018}, XLSS = \citet{Stalin2010}, XMSS = \citet{Barcons2007}, and XWAS = \citet{Esquej2013}.}
	\label{tab:xmmlss_speczs}
	\begin{tabular}{lcccccccccr} 
		\hline
		R.A. & Dec. & Milliquas & Milliquas & OzDES & OzDES & 2XLSSd & 2XLSSd & PRIMUS & PRIMUS & Error\\
		(hms) & (dms) & $z$ & reference & $z$ & quality &  $z$ & quality & $z$ & quality & in $z$  \\
		\hline
 02:15:24.99 &          $-$04:53:54.0 &          0.713       &          DR16Q          &     {\bf 0.713} &          4         &          9.999     &          9            &          0.709 &          4               &           0.004 \\             
  02:15:43.28 &          $-$05:04:04.4 &          {\bf 1.602}    &          0646           &          --      &          --        &          --         &          --           &          --         &          --              &           0.001 \\ 
  02:16:14.40 &          $-$05:22:48.4 &          {\bf 2.253}       &          DR16Q          &          --      &          --        &          9.999     &          9            &          --         &          --              &           0.001 \\             
  02:16:41.63 &          $-$04:52:26.0 &         {\bf 1.061}       &          0646           &          1.064 &          4         &          --         &          --           &          0.160 &          4               &           0.003 \\             
  02:16:47.70 &          $-$04:16:51.7 &          2.026       &          DR16Q          &         {\bf 2.023} &          4         &          9.999     &          9            &          --         &          --              &           0.003 \\             
  02:16:55.32 &          $-$05:23:35.5 &          {\bf 2.221}       &          DR16Q          &          --      &          --        &          --         &          --           &          --         &          --              &           0.003 \\ 
  02:17:03.86 &          $-$05:31:40.7 &          2.281       &          1718           &          --      &          --        &     {\bf 2.290}       &          1            &          --         &          --              &           0.009 \\             
  02:17:05.30 &          $-$04:23:13.6 &          {\bf 2.222}       &          DR16Q          &          0.784 &          3         &          9.999     &          9            &          --         &          --              &           0.001 \\  
  02:17:11.99 &          $-$04:46:20.0 &          1.102       &          1689           &        {\bf  1.103} &          4         &          --         &          --           &          --         &          --              &           0.001 \\             
  02:17:30.58 &          $-$05:22:22.8 &          2.477       &          1718           &          --      &          --        &          --         &          --           &         {\bf 2.502}   &          4               &           0.025 \\             
  02:17:34.97 &          $-$04:29:51.7 &           {\bf 1.080}        &          DR16Q          &          $-$9.999  &          1         &   1.080       &          1            &          --         &          --              &           0.001 \\             
  02:17:46.63 &          $-$04:49:50.2 &          {\bf 2.145}       &          1566           &          --      &          --        &          2.140       &          1            &          2.149  &          4               &           0.005 \\             
  02:17:57.23 &          $-$05:02:16.4 &          1.087       &          1042           &          --      &          --        &         {\bf 1.088}      &          1            &          1.094  &          4               &           0.006 \\             
  02:18:01.24 &          $-$04:38:19.6 &          {\bf 1.064}       &          1566           &          --      &          --        &          --         &          --           &          1.062     &          3               &           0.002 \\             
  02:18:09.04 &          $-$04:47:50.0 &          2.215       &          1566           &          --      &          --        &          2.220       &          1            &         {\bf 2.217}  &          4               &           0.003 \\             
  02:18:17.44 &          $-$04:51:12.4 &          1.083       &          1689           &         {\bf 1.083} &          4         &          1.081      &          1            &          --         &          --              &           0.002 \\             
  02:18:20.48 &          $-$05:04:26.4 &          0.649       &          XMSS           &        {\bf  0.650} &          4         &          0.650       &          9            &          0.651  &          4               &           0.001 \\             
  02:18:24.43 &          $-$04:39:45.4 &          1.312       &          0646           &          --      &          --        &          --         &          --           &         {\bf 1.303}  &          4               &           0.009 \\             
  02:18:27.29 &          $-$05:34:57.5 &          {\bf 2.583}       &          1713           &          --      &          --        &          2.579      &          1            &          2.594  &          4               &           0.011 \\             
  02:18:30.45 &          $-$04:22:00.5 &          {\bf 2.371}       &          DR16Q          &          --      &          --        &          --         &          --           &          --         &          --              &           0.001 \\  
		\hline
	\end{tabular}
\end{table*}

\setcounter{table}{1} 

\begin{table*}
	\centering
	\caption{-- {\it continued} -- An overview of the spectroscopic redshifts considered for the {\bf 81 quasars within the XMM-LSS field}.}
	\begin{tabular}{lcccccccccr} 
		\hline
		R.A. & Dec. & Milliquas & Milliquas & OzDES & OzDES & 2XLSSd & 2XLSSd & PRIMUS & PRIMUS & Error\\
		(hms) & (dms) & $z$ & reference & $z$ & quality &  $z$ & quality & $z$ & quality & in $z$  \\
		\hline
		  02:18:45.94 &          $-$05:28:06.5 &          1.570        &          1718           &          --      &          --        &          --         &          --           & {\bf 1.570}  &          4               &           0.008 \\  
 		  02:19:24.30 &          $-$05:11:49.7 &          2.532       &          1718           &          --      &          --        &          2.540       &          1            &          {\bf 2.532}  &          4               &           0.008 \\  
  02:19:49.16 &          $-$05:29:25.9 &          {\bf 2.031}       &          VIPERS         &          --      &          --        &          9.999     &          9            &          --         &          --              &           0.002 \\ 
  02:19:58.77 &          $-$04:52:20.0 &          {\bf 1.204}       &          DR16Q          &          --      &          --        &          --         &          --           &          1.215  &          4               &           0.011 \\             
  02:20:01.64 &          $-$05:22:17.1 &          2.215       &          XLSS           &         {\bf 2.219} &          4         &          2.220       &          1            &          --         &          --              &           0.004 \\             
  02:20:15.60 &          $-$04:48:30.2 &          {\bf 2.800}         &          0646           &          --      &          --        &          --         &          --           &          --         &          --              &           0.002 \\   
  02:20:46.21 &          $-$04:20:38.4 &          2.135       &          XLSS           &         {\bf 2.129} &          3         &          2.120       &          1            &          --         &          --              &           0.009 \\             
  02:20:58.34 &          $-$04:11:49.9 &          3.200         &          XLSS           &          --      &          --        &          {\bf 3.200}        &          3            &          --         &          --              &           0.001 \\           
  02:21:37.73 &          $-$04:31:24.9 &           {\bf 1.714}       &          DR16Q          &          --      &          --        &          --         &          --           &         1.716  &          4               &           0.002 \\             
  02:21:37.90 &          $-$05:13:36.9 &          1.109       &          XLSS           &         {\bf 1.110} &          4         &          --         &          --           &          --         &          --              &           0.001 \\    
		
  02:21:39.48 &          $-$04:50:04.1 &          2.444       &          XLSS           &          --      &          --        &          --         &          --           &         {\bf 2.431}  &          4               &           0.013 \\             
  02:22:14.49 &          $-$04:31:33.7 &          {\bf 1.817}       &          DR16Q          &          --      &          --        &          9.999     &          9            &          1.807  &          4               &           0.010 \\             
  02:22:15.99 &          $-$05:16:01.3 &          1.230        &          XLSS           &          --      &          --        &    {\bf 1.230}       &          1            &          --         &          --              &           0.001 \\             
  02:22:36.10 &          $-$04:33:08.0 &          1.215       &          XLSS           &          --      &          --        &       {\bf 1.210}       &          1            &          --         &          --              &           0.005 \\             
  02:22:37.89 &          $-$04:11:40.4 &          {\bf 1.496}       &          DR16Q          &          --      &          --        &          9.999     &          9            &          1.490  &          4               &           0.006 \\             
  02:22:38.60 &          $-$04:43:22.8 &          {\bf 1.557}       &          XLSS           &          --      &          --        &          --         &          --           &          --         &          --              &           0.001 \\             
  02:22:39.18 &          $-$04:41:57.8 &          {\bf 1.189}       &          DR16Q          &          --      &          --        &          --         &          --           &          --         &          --              &           0.001 \\             
  02:22:39.99 &          $-$04:39:22.9 &          {\bf 1.698}       &          DR16Q          &          --      &          --        &          9.999     &          9            &          --         &          --              &           0.001 \\             
  02:22:54.40 &          $-$05:28:54.8 &          {\bf 1.213}       &          DR16Q          &          --      &          --        &          9.999     &          9            &          1.215  &          4               &           0.002 \\             
  02:22:56.45 &          $-$05:20:55.0 &          1.354       &          XLSS           &         {\bf 1.354} &          3         &          --         &          --           &          --         &          --              &           0.001 \\             
  02:23:06.05 &          $-$04:33:23.8 &        {\bf  1.084}       &          1340           &          1.086 &          3         &          1.080       &          1            &          --         &          --              &           0.004 \\             
  02:23:09.36 &          $-$04:21:50.3 &          --          &          --             &          --      &          --        &          --         &          --           &         {\bf 2.003}  &          4               &           0.009 \\  
  02:23:19.03 &          $-$04:46:14.0 &          {\bf 1.982}       &          XLSS           &          --      &          --        &          1.980       &          1            &          1.989  &          4               &           0.007 \\             
  02:23:32.14 &          $-$04:57:39.4 &          0.602       &          XLSS           &        {\bf  0.602} &          4         &          1.900        &          1            &          0.602     &          3               &           0.001 \\             
  02:23:35.05 &          $-$04:34:22.7 &          {\bf 2.208}       &          XLSS           &          --      &          --        &          --         &          --           &          --         &          --              &           0.001 \\             
  02:23:42.37 &          $-$05:22:43.8 &          {\bf 2.037}       &          DR16Q          &          --      &          --        &          9.999     &          9            &          2.035     &          3               &           0.002 \\             
  02:23:50.71 &          $-$04:27:03.7 &          1.190        &          XLSS           &          --      &          --        &          --         &          --           &         {\bf 1.180}  &          4               &           0.010 \\             
  02:23:55.07 &          $-$05:15:39.6 &          {\bf 2.408}       &          DR16Q          &          --      &          --        &          --         &          --           &          --         &          --              &           0.001 \\             
  02:23:55.13 &          $-$04:55:20.1 &          {\bf 1.529}       &          DR16Q          &          --      &          --        &          --         &          --           &          --         &          --              &           0.001 \\  
  02:24:01.60 &          $-$04:40:33.1 &          1.180        &          XLSS           &          {\bf 1.187} &          4         &          --         &          --           &          1.187  &          4               &           0.007 \\             
  02:24:13.46 &          $-$04:52:10.4 &          2.489       &          1340           &          {\bf 2.481} &          4         &          --         &          --           &          --         &          --              &           0.008 \\   		
  02:24:19.75 &          $-$04:03:15.7 &          {\bf 2.276}       &          XLSS           &          --      &          --        &          --         &          --           &          --         &          --              &           0.001 \\             
  02:24:44.29 &          $-$04:19:47.8 &          {\bf 1.607}       &          DR16Q          &          --      &          --        &          9.999     &          9            &          --         &          --              &           0.001 \\             
  02:24:44.33 &          $-$04:45:36.0 &          --          &          --             &          --      &          --        &           9.999    &          9            &          --         &          --              &           -- \\  
  02:24:48.69 &          $-$04:56:06.9 &          {\bf 1.654}       &          DR16Q          &          1.648 &          4         &          --         &          --           &          --         &          --              &           0.006 \\             
  02:24:57.57 &          $-$04:56:59.5 &          --          &          MQ             &          --      &          --        &          --         &          --           &          --         &          --              &          -- \\  
  02:25:03.12 &          $-$04:40:25.3 &          {\bf 1.741}       &          XLSS           &          --      &          --        &          0.270       &          1            &          --         &          --              &           0.002 \\             
  02:25:08.57 &          $-$04:25:12.8 &          --          &    MQ              &          --      &          --        &          --         &          --           &          --         &          --              &           -- \\   
  02:25:25.68 &          $-$04:35:09.6 &          2.166       &          0650           &          --      &          --        &          --         &          --           &         {\bf 2.168}   &          4               &           0.002 \\             
  02:25:43.53 &          $-$04:28:34.6 &          3.415       &          XLSS           &         {\bf 3.406} &          4         &          --         &          --           &          --         &          --              &           0.009 \\             
  02:25:52.16 &          $-$04:05:16.2 &          {\bf 1.441}       &          2032           &          --      &          --        &          1.438     &          2            &          --         &          --              &           0.003 \\             
  02:25:54.87 &          $-$05:13:54.4 &          1.256       &          XWAS           &         {\bf 1.258}  &          3         &          9.999     &          9            &          --         &          --              &           0.002 \\             
  02:25:55.43 &          $-$04:39:18.3 &          {\bf 1.030}        &          XLSS           &          --      &          --        &          1.030       &          3            &          --         &          --              &           0.001 \\   
  02:26:00.99 &          $-$05:06:58.2 &          {\bf 1.838}       &          DR16Q          &          --      &          --        &          --         &          --           &          --         &          --              &           0.001 \\             
  02:26:01.59 &          $-$04:59:19.0 &          {\bf 1.660}        &          0646           &          --      &          --        &          --         &          --           &          --         &          --              &           0.001 \\ 
  02:26:04.13 &          $-$04:50:20.4 &          2.387       &          OzDES          &          {\bf 2.373} &          4         &          --         &          --           &          --         &          --              &           0.014 \\             
  02:26:12.64 &          $-$04:34:01.4 &          2.301       &          XLSS           &          {\bf 2.305} &          4         &          --         &          --           &          --         &          --              &           0.004 \\             
  02:26:24.64 &          $-$04:20:02.4 &          2.236       &          0650           &          2.242 &          4         &          {\bf 2.236}     &          1            &          2.240  &          4               &           0.006 \\             
  02:26:33.31 &          $-$04:29:47.8 &          {\bf 2.147}       &          2032           &          --      &          --        &          --         &          --           &          2.148  &          4               &           0.001 \\             
  02:26:40.79 &          $-$04:16:36.3 &          {\bf 2.088}       &          DR16Q          &          --      &          --        &          --         &          --           &          --         &          --              &           0.001 \\             
  02:26:46.98 &          $-$04:18:38.1 &          {\bf 1.581}       &          0650           &          1.579  &          4         &          --         &          --           &          1.582   &          4               &           0.002 \\             
  02:26:52.14 &          $-$04:05:57.1 &          {\bf 1.430}        &          2032           &          1.426 &          4         &          --         &          --           &          --         &          --              &           0.004 \\             
  02:26:55.57 &          $-$04:42:17.9 &          {\bf 2.306}      &          0646           &          --      &          --        &          9.999     &          9            &          --         &          --              &           0.002 \\ 
  02:26:59.90 &          $-$04:44:30.6 &          {\bf 1.611}       &          0650           &          1.610 &          4         &          1.612      &          9            &          --         &          --              &           0.001 \\             
  02:27:16.12 &          $-$04:45:39.0 &          0.722       &          XLSS           &          0.721  &          4         &         {\bf 0.722}     &          1            &          --         &          --              &           0.001 \\             
  02:27:43.87 &          $-$05:09:52.6 &          2.211       &          OzDES2         &        {\bf  2.211} &          4         &          --         &          --           &          --         &          --              &           0.001 \\             
  02:27:46.51 &          $-$04:11:08.6 &          1.652       &          DR16Q          &       {\bf 1.647} &          4         &          9.999     &          9            &          --         &          --              &           0.005 \\             
  02:27:57.00 &          $-$04:01:30.8 &          {\bf 1.907}       &          DR16Q          &          --      &          --        &          9.999     &          9            &          --         &          --              &           0.001 \\             
  02:28:16.46 &          $-$04:59:38.5 &          0.706       &          OzDES          &   {\bf 0.706} &          3         &          9.999     &          9            &          --         &          --              &           0.001 \\             
  02:28:32.81 &          $-$04:22:45.8 &          {\bf 1.757}       &          DR16Q          &          --      &          --        &          --         &          --           &          --         &          --              &           0.001 \\             
  02:28:35.93 &          $-$04:15:56.1 &         {\bf 1.372}       &          DR16Q          &          --      &          --        &          --         &          --           &          --         &          --              &           0.001 \\
		\hline
	\end{tabular}
\end{table*}

\section{New spectroscopic redshifts}
\label{app:new_speczs}

\begin{table}
    \centering
    \begin{tabular}{c|c|c|c}
    \hline
       R.A.  & Dec. & photo- &SALT  \\
        (hms) & (dms) & $z$ & $z$  \\
        \hline
        02:24:44.33 & $-$04:45:36.0 & 2.52 $\pm$ 0.04 & 2.520 $\pm$ 0.008 \\ 
        02:24:57.57 & $-$04:56:59.5 & 1.72 $\pm$ 0.03 & 1.760 $\pm$ 0.001 \\ 
        02:25:08.57 & $-$04:25:12.8 & 1.92 $\pm$ 0.04 & 2.150 $\pm$ 0.011 \\ 
         09:57:58.40 & +02:17:29.1 & 2.48 $\pm$ 0.03 & 2.420 $\pm$ 0.001 \\ 

         
        \hline
    \end{tabular}
    \caption{Redshifts derived from SALT longslit spectroscopy (2019-2-SCI-046; PI: White) for four quasars (Appendix~\ref{app:new_speczs}).}
    \label{tab:new_speczs}
\end{table}

\begin{figure*}
    \centering
    \includegraphics[width=1.0\linewidth]{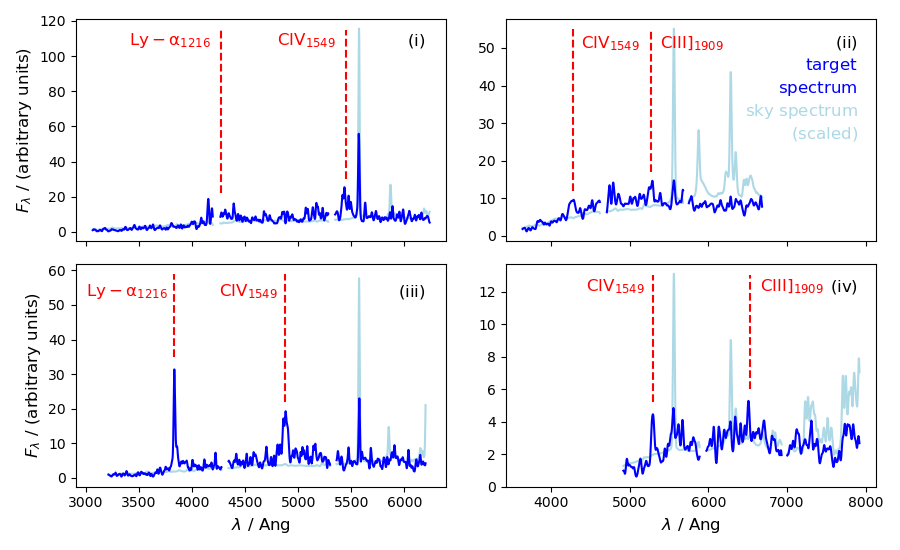}
    \caption{SALT spectra for four sources (blue lines): (i) A quasar at $z = 2.52$ (RA = 02:24:44.33, Dec. = $-$04:45:36.0). Template-fitting shows that a Ly-$\alpha$ line has unfortunately coincided with a chip gap on the CCD. (ii) A quasar at $z = 1.76$ (RA = 02:24:57.57, Dec. = $-$04:56:59.5). (iii) A quasar at $z = 2.15$ (RA = 02:25:08.57, Dec. = $-$04:25:12.8). (iv) A quasar at $z = 2.42$ (RA = 09:57:58.40, Dec. = +02:17:29.1). The light-blue lines provide an indication of the background-sky spectrum, which has been scaled using factors of: (i) 0.07, (ii) 0.07, (iii) 0.04, and (iv) 0.09, so that spectral features can be more-easily compared with those of the target spectrum (in order to avoid misidentifying sky emission-lines). }
    \label{fig:SALT_spectra}
\end{figure*}

We present new spectra from SALT for one quasar in the COSMOS field and three quasars in the XMM-LSS field (Figure~\ref{fig:SALT_spectra}). These are the sources for which the `Error in z' column in Tables~\ref{tab:cosmos_speczs} and \ref{tab:xmmlss_speczs} was left as `--'. The derived redshifts, and estimates of the spectroscopic-redshift errors, are presented in Table~\ref{tab:new_speczs}. These errors are based upon the median wavelength-offset for the observed sky emission-lines, with respect to their reference values. (The photometric-redshift errors are estimated via the 16th and 84th percentiles of the probability distribution function associated with the photometric redshift, as output by {\sc LePhare} -- see Section~\ref{sec:sample_selection}.)

\section{Fitting with SED3FIT}
\label{app:sed_fitting}

\begin{figure*}
    \centering
    \subfigure[R.A. = 02:17:03.86, Dec. = $-$05:31:40.7]{\includegraphics[width=0.4\linewidth]{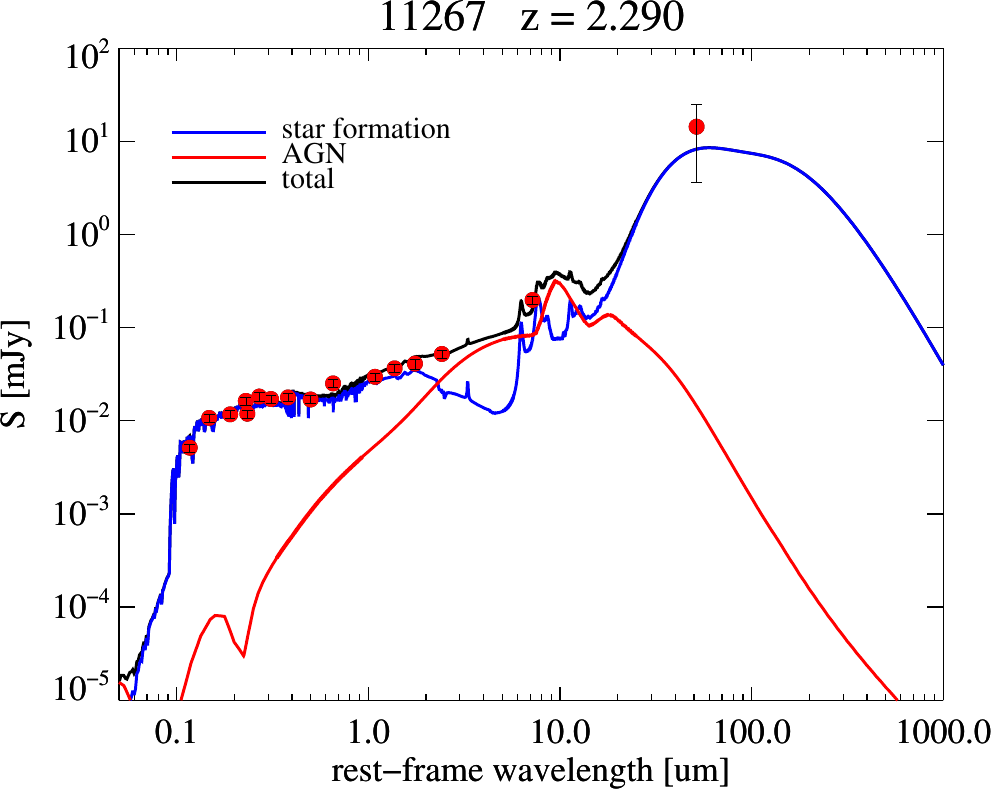}}
        \subfigure[R.A. = 02:18:17.44, Dec = $-$04:51:12.4]{\includegraphics[width=0.4\linewidth]{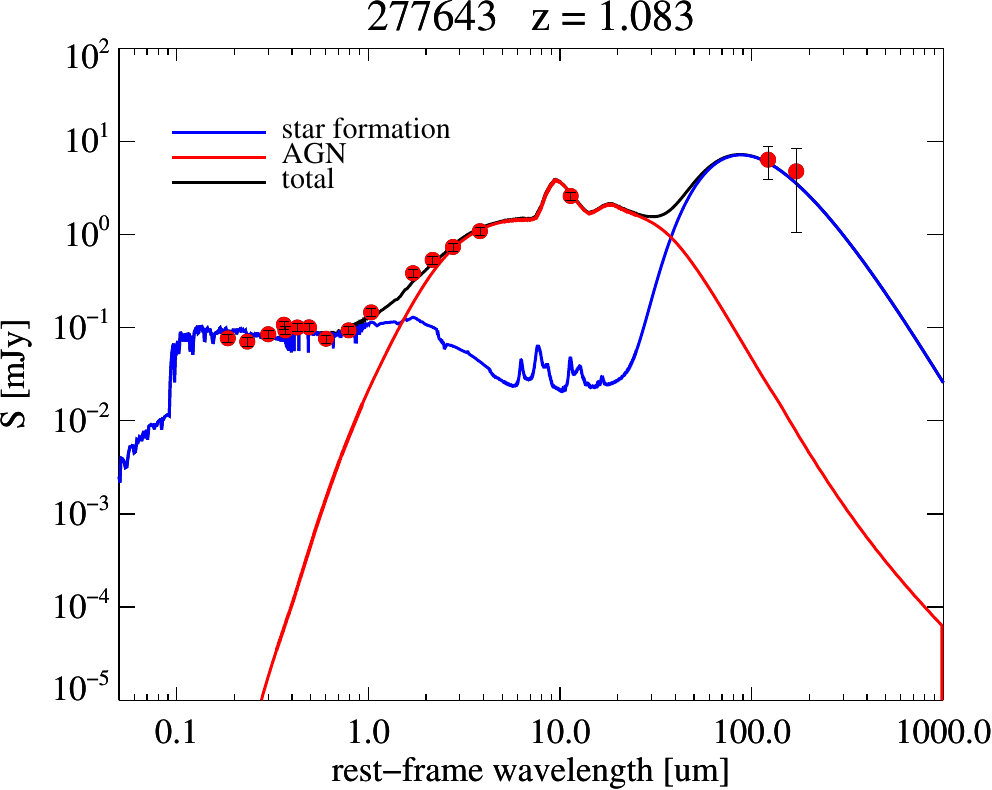}}\\ \vspace{5mm}
            \subfigure[R.A. = 02:17:34.97, Dec. = $-$04:29:51.7]{\includegraphics[width=0.4\linewidth]{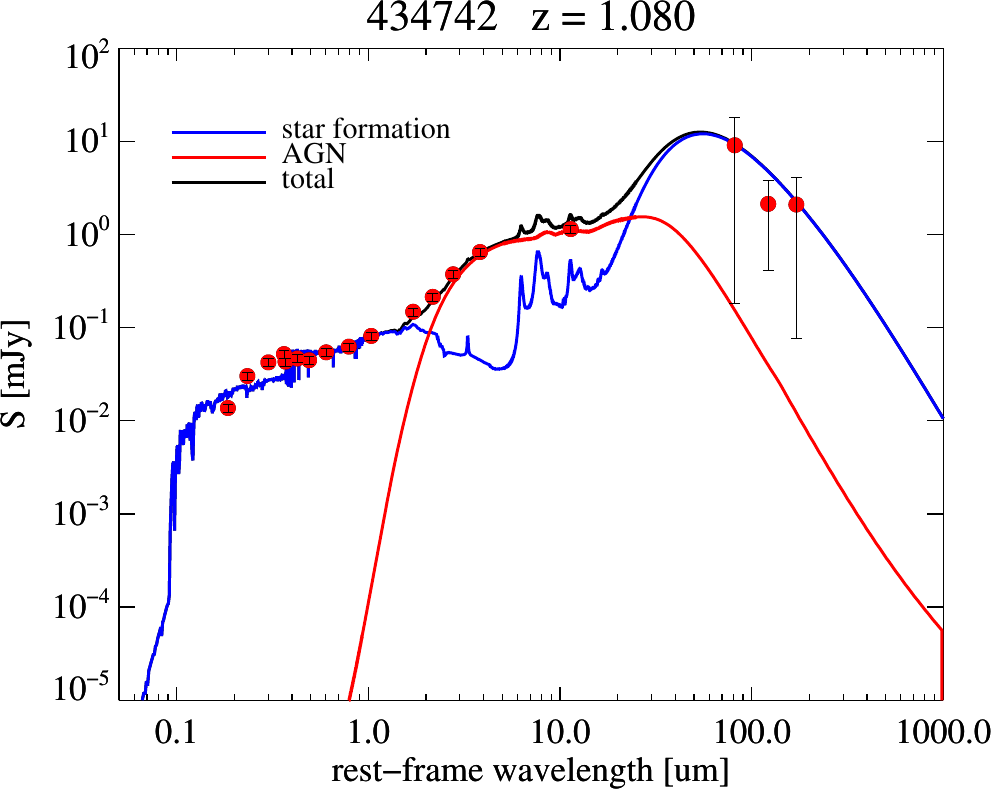}}
        \subfigure[R.A. = 02:26:33.31, Dec = $-$04:29:47.8]{\includegraphics[width=0.4\linewidth]{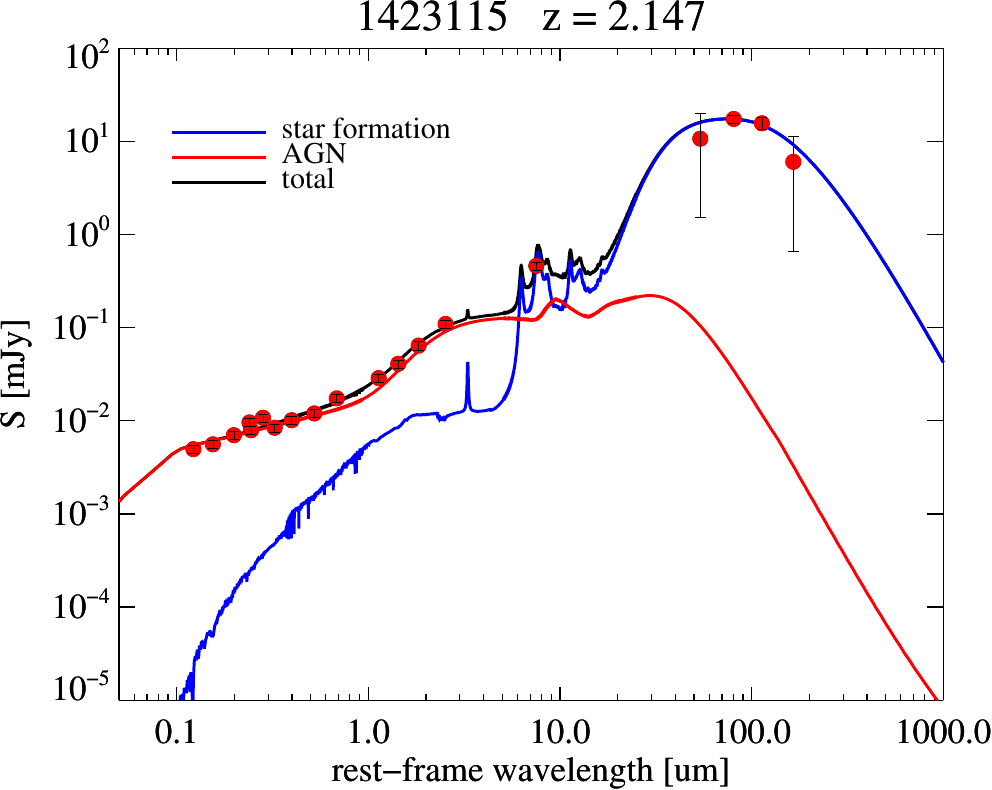}}\\ \vspace{5mm}
            \subfigure[R.A. = 02:25:55.43, Dec. = $-$04:39:18.3]{\includegraphics[width=0.4\linewidth]{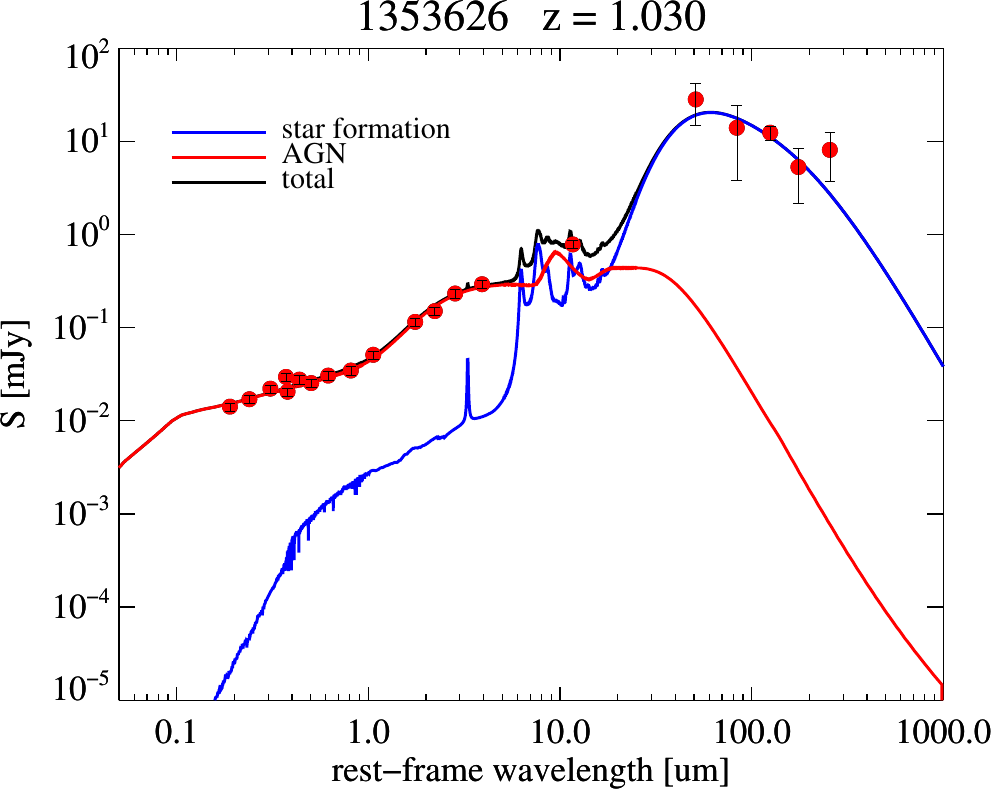}}
        \subfigure[R.A. = 02:26:40.79, Dec. = $-$04:16:36.3]{\includegraphics[width=0.4\linewidth]{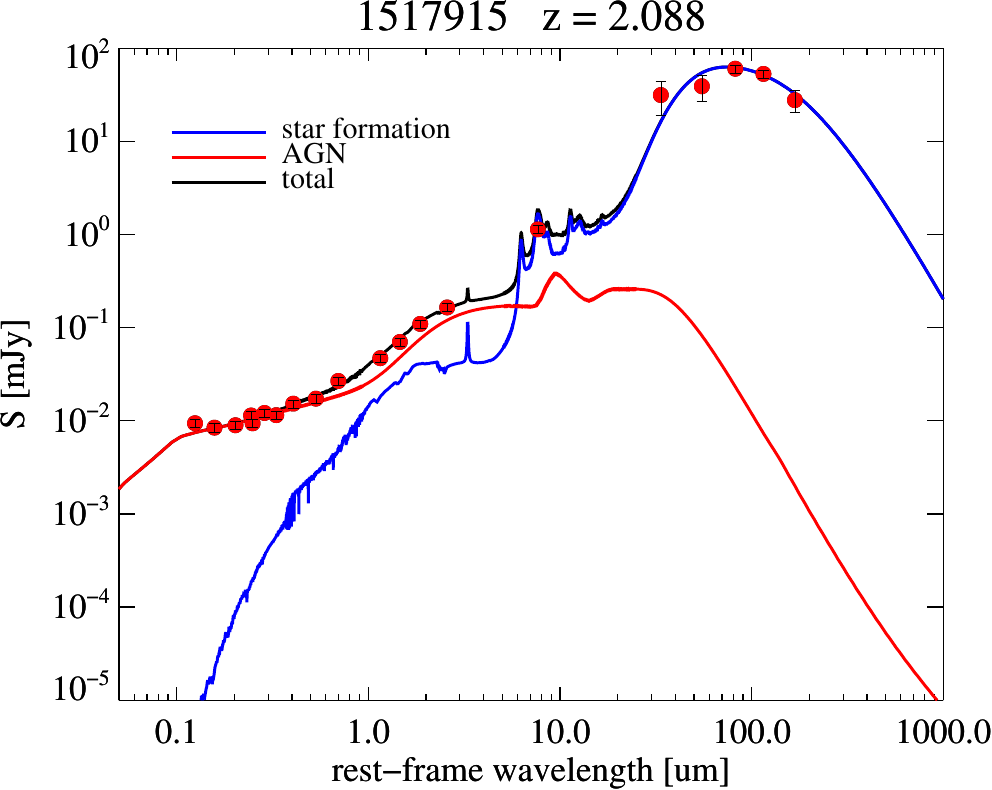}}
    \caption{Examples of SED3FIT fitting \citep{Berta2013} for six sources, with varying degrees of coverage in the FIR (Appendix~\ref{app:sed_fitting}). }
    \label{fig:sed_examples}
\end{figure*}

In Figure~\ref{fig:sed_examples}, we provide examples of the optical-through-to-FIR photometry that is fitted via SED3FIT \citep{Berta2013} to obtain IR luminosities and star-formation rates (Section~\ref{sec:sedfitting}). These examples are chosen to give an idea of the range of fitting that results from the different constraints provided by the available FIR-data.

\section{Radio luminosities and multiple star-formation rates}
\label{app:lradio_sfr_table}

In Table~\ref{tab:luminosities_sfrs_allquasars} (available online as Supplementary Material), we provide the K-corrected radio-luminosities at 1.4\,GHz (based upon measurements from the MIGHTEE images; Section~\ref{sec:faintemission}), and the SFRs that are used to assess the fraction of quasars that have their radio emission dominated by the AGN (Section~\ref{sec:SFR_comparisons}). These fractions are presented in Table~\ref{tab:AGNdominated_fractions} and discussed in Section~\ref{sec:agn_fraction_discussion}.

\begin{landscape}
\begin{table}
    \centering
    \begin{tabular}{c|c|c|c|c|c|c|c|c}
    \hline
    R.A. & Dec.  & spec-$z$  & $L_{\mathrm{1.4\,GHz}}$ / W\,Hz$^{-1}$ & Yun--SFR / $\Msol\,\mathrm{yr}^{-1}$ & Kennicutt--SFR  & SED3FIT--SFR  & Delhaize-SFR / $\Msol\,\mathrm{yr}^{-1}$ \\ 
        (hms) & (dms)  & $\pm$ error  & (+ error, $-$ error) &  (+ error, $-$ error) & $\pm$ error / $\Msol\,\mathrm{yr}^{-1}$ & $\pm$ error / $\Msol\,\mathrm{yr}^{-1}$ & (+ error, $-$ error) \\ 
       \hline

 09:57:58.40 &    +02:17:29.1 &    2.420  $\pm$ 0.050  &    1.32$\times 10^{23}$  (+8.41$\times 10^{22}$, $-$7.71$\times 10^{22}$) &     77.79     (+88.45,    $-$55.34)    &     147.17  $\pm$ 15.28   &     297.20  $\pm$ 33.72  &     15.85   (+41.27,    $-$21.27)   \\
  09:58:22.18 &    +01:45:24.1 &    1.963 $\pm$ 0.001 &    4.94$\times 10^{24}$  (+4.75$\times 10^{22}$, $-$4.74$\times 10^{22}$) &     2916.01   (+926.24,   $-$909.08)   &     539.59  $\pm$ 24.86   &     2094.00 $\pm$ 97.81  &     126.49  (+943.02,   $-$145.33)  \\
  09:58:26.67 &    +02:28:18.1 &    0.692 $\pm$ 0.135 &    1.81$\times 10^{22}$  (+1.88$\times 10^{22}$, $-$1.09$\times 10^{22}$) &     10.68     (+17.70,    $-$7.74)     &     7.85    $\pm$ 1.18    &     4.85   $\pm$ 1.95   &     4.67    (+5.41,     $-$9.05)    \\
  09:58:48.86 &    +02:34:41.1 &    1.541 $\pm$ 0.006 &    8.02$\times 10^{22}$  (+2.71$\times 10^{22}$, $-$2.66$\times 10^{22}$) &     47.30      (+35.28,    $-$25.34)    &     49.93   $\pm$ 338.84  &     55.57  $\pm$ 1.17   &     8.31    (+30.00,    $-$10.24)   \\
  09:59:34.82 &    +02:02:49.9 &    2.182 $\pm$ 0.021 &    3.38$\times 10^{22}$  (+3.63$\times 10^{22}$, $-$3.48$\times 10^{22}$) &     19.92     (+34.06,    $-$20.36)    &     122.71  $\pm$ 89.91   &     549.10  $\pm$ 16.92  &     7.11    (+9.82,     $-$9.24)    \\
  09:59:44.79 &    +02:40:48.8 &    1.405 $\pm$ 0.002 &    5.46$\times 10^{23}$  (+2.44$\times 10^{22}$, $-$2.42$\times 10^{22}$) &     322.24    (+117.08,   $-$108.24)   &     19.82   $\pm$ 14.24   &     10.48  $\pm$ 14.96  &     21.15   (+143.75,   $-$24.58)   \\
  10:00:01.44 &    +02:48:44.7 &    0.765 $\pm$ 0.002 &    1.45$\times 10^{24}$  (+1.28$\times 10^{22}$, $-$1.27$\times 10^{22}$) &     858.18    (+271.66,   $-$267.03)   &     1.74    $\pm$ 0.12    &     96.21  $\pm$ 5.72   &     55.37   (+533.74,   $-$61.99)   \\
  10:00:11.23 &    +01:52:00.2 &    2.415 $\pm$ 0.004 &    5.36$\times 10^{23}$  (+6.84$\times 10^{22}$, $-$6.80$\times 10^{22}$) &     316.22    (+149.18,   $-$124.34)   &     201.09  $\pm$ 98.48   &     194.30  $\pm$ 58.21  &     22.92   (+135.38,   $-$27.80)   \\
  10:00:12.44 &    +01:40:57.9 &    2.276 $\pm$ 0.004 &    4.14$\times 10^{23}$  (+5.01$\times 10^{22}$, $-$4.97$\times 10^{22}$) &     244.15    (+113.03,   $-$94.85)    &     94.79   $\pm$ 83.47   &     250.50  $\pm$ 65.54  &     17.87   (+124.22,   $-$21.62)   \\
  10:00:12.91 &    +02:35:22.8 &    0.698 $\pm$ 0.002 &    1.44$\times 10^{22}$  (+5.25$\times 10^{21}$, $-$5.18$\times 10^{21}$) &     8.48      (+6.63,     $-$4.71)     &     22.91   $\pm$ 6.69    &     112.30  $\pm$ 50.57  &     2.41    (+4.74,     $-$2.89)    \\
  10:00:14.09 &    +02:28:38.6 &    1.255 $\pm$ 0.014 &    9.82$\times 10^{23}$  (+3.55$\times 10^{22}$, $-$3.46$\times 10^{22}$) &     579.58    (+204.15,   $-$191.02)   &     246.02  $\pm$ 14.17   &     275.30  $\pm$ 101.54 &     37.83   (+321.12,   $-$43.85)   \\
  10:00:51.92 &    +01:59:19.3 &    2.240  $\pm$ 0.014 &    5.30$\times 10^{22}$   (+5.87$\times 10^{22}$, $-$5.71$\times 10^{22}$) &     31.26     (+54.70,    $-$32.93)    &     92.99   $\pm$ 116.26  &     115.60  $\pm$ 36.88  &     11.39   (+17.23,    $-$14.63)   \\
  10:00:55.39 &    +02:34:41.4 &    1.402 $\pm$ 0.002 &    5.83$\times 10^{23}$  (+2.29$\times 10^{22}$, $-$2.28$\times 10^{22}$) &     343.74    (+122.51,   $-$114.20)   &     247.65  $\pm$ 22.84   &     91.62  $\pm$ 76.48  &     21.82   (+176.36,   $-$25.31)   \\
  10:00:56.70 &    +02:17:20.9 &    2.077 $\pm$ 0.001 &    1.03$\times 10^{24}$  (+4.97$\times 10^{22}$, $-$4.96$\times 10^{22}$) &     608.04    (+223.75,   $-$205.82)   &     264.43  $\pm$ 58.31   &     226.30  $\pm$ 39.04  &     33.10    (+339.18,   $-$38.93)   \\
  10:01:16.95 &    +01:38:04.1 &    0.747 $\pm$ 0.140  &    1.85$\times 10^{22}$  (+1.88$\times 10^{22}$, $-$1.11$\times 10^{22}$) &     10.89     (+17.76,    $-$7.86)     &     20.93   $\pm$ 3.39    &     14.90   $\pm$ 5.58   &     4.57    (+5.55,     $-$8.76)    \\
  10:01:23.98 &    +02:14:46.0 &    0.895 $\pm$ 0.002 &    5.45$\times 10^{22}$  (+6.07$\times 10^{21}$, $-$6.01$\times 10^{21}$) &     32.13     (+14.48,    $-$12.27)    &     10.39   $\pm$ 2.05    &     13.75  $\pm$ 3.11   &     3.75    (+18.83,    $-$4.40)    \\
  10:01:24.74 &    +01:57:38.7 &    1.166 $\pm$ 0.007 &    4.76$\times 10^{22}$  (+1.37$\times 10^{22}$, $-$1.33$\times 10^{22}$) &     28.06     (+19.10,    $-$14.02)    &     35.50    $\pm$ 9.94    &     50.87  $\pm$ 17.27  &     5.12    (+20.08,    $-$6.27)    \\
  10:01:28.01 &    +02:18:19.2 &    1.177 $\pm$ 0.010  &    5.75$\times 10^{22}$  (+9.92$\times 10^{21}$, $-$9.57$\times 10^{21}$) &     33.94     (+17.99,    $-$14.28)    &     18.98   $\pm$ 6.92    &     16.62  $\pm$ 5.84   &     4.36    (+18.51,    $-$5.30)    \\
  10:01:32.81 &    +01:57:59.9 &    1.536 $\pm$ 0.006 &    7.46$\times 10^{22}$  (+1.65$\times 10^{22}$, $-$1.62$\times 10^{22}$) &     44.02     (+26.12,    $-$20.07)    &     26.06   $\pm$ 13.77   &     43.42  $\pm$ 14.58  &     5.75    (+27.32,    $-$7.02)    \\
  10:01:41.41 &    +02:00:50.9 &    2.277 $\pm$ 0.009 &    5.30$\times 10^{23}$   (+7.49$\times 10^{22}$, $-$7.37$\times 10^{22}$) &     312.54    (+153.06,   $-$125.57)   &     61.11   $\pm$ 19.30    &     29.50   $\pm$ 20.35  &     24.66   (+127.42,   $-$30.10)   \\
  10:01:43.04 &    +01:49:31.9 &    2.084 $\pm$ 0.008 &   $-$1.51$\times 10^{22}$ (+4.48$\times 10^{22}$, $-$4.40$\times 10^{22}$) &    $-$8.88     (+31.76,    $-$15.33)    &     75.81   $\pm$ 17.61   &     93.01  $\pm$ 18.69  &     7.95    (+5.47,    $-$10.24)   \\
  10:02:08.55 &    +01:45:53.6 &    2.203 $\pm$ 0.012 &    4.94$\times 10^{23}$  (+5.50$\times 10^{22}$, $-$5.38$\times 10^{22}$) &     291.17    (+131.21,   $-$110.90)   &     46.71   $\pm$ 95.38   &     67.21  $\pm$ 19.97  &     20.77   (+167.13,   $-$25.21)   \\
  10:02:51.85 &    +01:46:09.7 &    2.459 $\pm$ 0.001 &    4.79$\times 10^{23}$  (+7.35$\times 10^{22}$, $-$7.34$\times 10^{22}$) &     282.54    (+142.82,   $-$116.29)   &     125.19  $\pm$ 221.52  &     158.40  $\pm$ 45.01  &     22.37   (+164.92,   $-$27.20)   \\

  \hline 

  02:15:24.99 &   $-$04:53:54.0 &    0.713 $\pm$ 0.004 &    4.20$\times 10^{22}$   (+4.53$\times 10^{21}$, $-$4.42$\times 10^{21}$) &     24.81     (+11.06,    $-$9.38)     &     153.67  $\pm$ 264.78  &     27.83  $\pm$ 20.75  &     3.11    (+14.62,    $-$3.65)    \\
  02:15:43.28 &   $-$05:04:04.4 &    1.602 $\pm$ 0.001 &    3.10$\times 10^{23}$   (+2.60$\times 10^{22}$, $-$2.59$\times 10^{22}$) &     182.99    (+75.82,    $-$66.44)    &     56.12   $\pm$ 503.19  &     67.10   $\pm$ 3.23   &     13.97   (+122.18,   $-$16.52)   \\
  02:16:14.40 &   $-$05:22:48.4 &    2.253 $\pm$ 0.001 &    6.54$\times 10^{23}$  (+5.66$\times 10^{22}$, $-$5.65$\times 10^{22}$) &     385.95    (+161.35,   $-$140.92)   &     20.93   $\pm$ 19.08   &     65.74  $\pm$ 13.21  &     24.51   (+219.57,   $-$29.30)   \\
  02:16:41.63 &   $-$04:52:26.0 &    1.061 $\pm$ 0.003 &    2.73$\times 10^{22}$  (+9.79$\times 10^{21}$, $-$9.67$\times 10^{21}$) &     16.08     (+12.45,    $-$8.87)     &     33.66   $\pm$ 14.77   &     42.55  $\pm$ 5.42   &     3.68    (+11.71,    $-$4.47)    \\
  02:16:47.70 &   $-$04:16:51.7 &    2.023 $\pm$ 0.003 &    9.89$\times 10^{22}$  (+4.11$\times 10^{22}$, $-$4.08$\times 10^{22}$) &     58.33     (+49.44,    $-$34.54)    &     15.00    $\pm$ 62.93   &     20.98  $\pm$ 7.55   &     10.17   (+30.89,    $-$12.60)   \\
  02:16:55.32 &   $-$05:23:35.5 &    2.221 $\pm$ 0.003 &    1.29$\times 10^{23}$  (+5.30$\times 10^{22}$, $-$5.27$\times 10^{22}$) &     76.26     (+64.08,    $-$44.87)    &     88.11   $\pm$ 473.46  &     56.01  $\pm$ 16.46  &     12.36   (+33.70,    $-$15.36)   \\
  02:17:03.86 &   $-$05:31:40.7 &    2.290  $\pm$ 0.009 &    3.26$\times 10^{23}$  (+7.55$\times 10^{22}$, $-$7.42$\times 10^{22}$) &     192.47    (+116.82,   $-$89.13)    &     213.51  $\pm$ 49.60    &     271.50  $\pm$ 44.64  &     20.21   (+98.88,    $-$25.01)   \\
  02:17:05.30 &   $-$04:23:13.6 &    2.222 $\pm$ 0.001 &   $-$2.12$\times 10^{24}$ (+1.80$\times 10^{23}$, $-$1.80$\times 10^{23}$) &    $-$1251.48  (+-242.97,  $-$-308.03)  &     38.68   $\pm$ 327.34  &     87.45  $\pm$ 4.31   &    $-$14.76  (+807.36,  $-$-12.90)  \\
  02:17:11.99 &   $-$04:46:20.0 &    1.103 $\pm$ 0.001 &    2.49$\times 10^{23}$  (+1.30$\times 10^{22}$, $-$1.30$\times 10^{22}$) &     147.14    (+54.91,    $-$50.20)    &     5.00     $\pm$ 50.91   &     31.75  $\pm$ 3.55   &     11.38   (+112.43,   $-$13.17)   \\
  02:17:30.58 &   $-$05:22:22.8 &    2.502 $\pm$ 0.025 &    7.46$\times 10^{23}$  (+9.41$\times 10^{22}$, $-$9.06$\times 10^{22}$) &     440.30     (+206.80,   $-$171.47)   &     253.04  $\pm$ 101.60   &     367.00  $\pm$ 74.99  &     30.53   (+231.66,   $-$37.64)   \\
  02:17:34.97 &   $-$04:29:51.7 &    1.080  $\pm$ 0.001 &    4.98$\times 10^{22}$  (+1.36$\times 10^{22}$, $-$1.35$\times 10^{22}$) &     29.41     (+19.41,    $-$14.51)    &     70.12   $\pm$ 7.28    &     118.50  $\pm$ 9.47   &     5.45    (+15.28,    $-$6.55)    \\
  02:17:46.63 &   $-$04:49:50.2 &    2.145 $\pm$ 0.005 &    3.48$\times 10^{23}$  (+5.26$\times 10^{22}$, $-$5.21$\times 10^{22}$) &     205.59    (+103.22,   $-$84.07)    &     7.89    $\pm$ 7.82    &     73.13  $\pm$ 22.14  &     17.50    (+100.29,   $-$21.28)   \\
  02:17:57.23 &   $-$05:02:16.4 &    1.088 $\pm$ 0.006 &    4.52$\times 10^{23}$  (+1.99$\times 10^{22}$, $-$1.95$\times 10^{22}$) &     266.91    (+96.74,    $-$89.42)    &     31.28   $\pm$ 8.38    &     99.27  $\pm$ 27.91  &     19.61   (+168.10,   $-$22.72)   \\
  02:18:01.24 &   $-$04:38:19.6 &    1.064 $\pm$ 0.002 &    6.95$\times 10^{22}$  (+1.07$\times 10^{22}$, $-$1.06$\times 10^{22}$) &     41.01     (+20.71,    $-$16.84)    &     57.16   $\pm$ 16.00    &     76.01  $\pm$ 15.26  &     5.25    (+28.90,    $-$6.25)    \\
  02:18:09.04 &   $-$04:47:50.0 &    2.217 $\pm$ 0.003 &    1.87$\times 10^{23}$  (+6.39$\times 10^{22}$, $-$6.35$\times 10^{22}$) &     110.46    (+82.92,    $-$59.75)    &     81.71   $\pm$ 42.12   &     219.20  $\pm$ 60.79  &     15.63   (+55.49,    $-$19.35)   \\
  02:18:17.44 &   $-$04:51:12.4 &    1.083 $\pm$ 0.002 &    9.58$\times 10^{23}$  (+1.92$\times 10^{22}$, $-$1.91$\times 10^{22}$) &     565.48    (+187.32,   $-$180.34)   &     24.87   $\pm$ 1.15    &     248.60  $\pm$ 8.62   &     35.21   (+308.75,   $-$40.10)   \\
  02:18:20.48 &   $-$05:04:26.4 &    0.650  $\pm$ 0.001 &    1.03$\times 10^{23}$  (+3.13$\times 10^{21}$, $-$3.11$\times 10^{21}$) &     61.03     (+21.03,    $-$19.90)    &     43.01   $\pm$ 8.47    &     58.76  $\pm$ 7.06   &     4.98    (+33.66,    $-$5.64)    \\
  02:18:24.43 &   $-$04:39:45.4 &    1.303 $\pm$ 0.009 &    7.88$\times 10^{22}$  (+2.22$\times 10^{22}$, $-$2.16$\times 10^{22}$) &     46.49     (+31.31,    $-$23.03)    &     29.02   $\pm$ 165.31  &     26.48  $\pm$ 0.44   &     7.84    (+34.67,    $-$9.66)    \\
 {\bf  02:18:27.29 } &  {\bf  $-$05:34:57.5 } &   {\bf  2.583 $\pm$ 0.011 } &    1.19$\times 10^{26}$  (+1.19$\times 10^{24}$, $-$1.19$\times 10^{24}$) &     69949.14  (+22258.53, $-$21826.50) &     319.77  $\pm$ 508.27  &     427.30  $\pm$ 137.37 &     2620.78 (+31903.46, $-$3037.05) \\

    \hline
    \end{tabular}
    \caption{Positions, redshifts, MeerKAT (scaled to 1.4-GHz) radio luminosities, and SFR estimates for 23 COSMOS quasars and 81 XMM-LSS quasars in the final sample of 104 quasars. Those with positions and redshifts highlighted in bold are classed as `radio loud' (Figure~\ref{fig:radioloudness}) via the radio-loudness parameter, $R$, defined by \citet{Kellermann1989}. The `Yun-SFR' and the `Delhaize-SFR' are based upon the radio luminosity, and applying the relation by \citet{Yun2001} (Equation~\ref{eqn:Yunrelation}) and by \citet{Delhaize2017} (Equations~\ref{eqn:bestq_delhaize}--\ref{eqn:sfr_delhaize}), respectively. The `Kennicutt-SFR' is derived through two-component SED-fitting over 8--1000$\upmu$m \citep{Jin2018} and applying Equation~\ref{eqn:Kennicutt} \citep{Kennicutt1998b}, whilst the `SED3FIT-SFR' is that output by the SED-fitting code of \citet{Berta2013}. } 
    \label{tab:luminosities_sfrs_allquasars}
\end{table}
\end{landscape}

\setcounter{table}{0} 

\begin{landscape}
\begin{table}
    \centering
    \begin{tabular}{c|c|c|c|c|c|c|c|c}
    \hline
    R.A. & Dec.  & spec-$z$  & $L_{\mathrm{1.4\,GHz}}$ / W\,Hz$^{-1}$ & Yun--SFR / $\Msol\,\mathrm{yr}^{-1}$ & Kennicutt--SFR  & SED3FIT--SFR  & Delhaize-SFR / $\Msol\,\mathrm{yr}^{-1}$ \\ 
        (hms) & (dms)  & $\pm$ error  & (+ error, $-$ error) &  (+ error, $-$ error) & $\pm$ error / $\Msol\,\mathrm{yr}^{-1}$ & $\pm$ error / $\Msol\,\mathrm{yr}^{-1}$ & (+ error, $-$ error) \\ 
       \hline

        02:18:30.45 &   $-$04:22:00.5 &    2.371 $\pm$ 0.001 &   $-$1.21$\times 10^{22}$ (+5.58$\times 10^{22}$, $-$5.57$\times 10^{22}$) &    $-$7.13     (+40.78,    $-$20.65)    &     38.50    $\pm$ 36.91   &     74.56  $\pm$ 28.93  &     9.31    (+-4.67,    $-$11.86)   \\
  02:18:45.94 &   $-$05:28:06.5 &    1.570  $\pm$ 0.008 &    2.77$\times 10^{23}$  (+2.69$\times 10^{22}$, $-$2.63$\times 10^{22}$) &     163.53    (+70.59,    $-$60.68)    &     264.69  $\pm$ 93.25   &     130.50  $\pm$ 31.58  &     13.35   (+76.34,    $-$15.97)   \\
  02:19:24.30 &   $-$05:11:49.7 &    2.532 $\pm$ 0.008 &    1.84$\times 10^{23}$  (+1.19$\times 10^{23}$, $-$1.17$\times 10^{23}$) &     108.81    (+124.72,   $-$81.26)    &     89.88   $\pm$ 9.33    &     160.90  $\pm$ 14.35  &     23.02   (+53.50,    $-$29.15)   \\
  02:19:49.16 &   $-$05:29:25.9 &    2.031 $\pm$ 0.002 &    3.99$\times 10^{23}$  (+4.63$\times 10^{22}$, $-$4.61$\times 10^{22}$) &     235.70     (+107.53,   $-$90.79)    &     127.94  $\pm$ 248.10   &     152.40  $\pm$ 53.68  &     18.15   (+133.79,   $-$21.81)   \\
  02:19:58.77 &   $-$04:52:20.0 &    1.204 $\pm$ 0.011 &    3.51$\times 10^{23}$  (+2.30$\times 10^{22}$, $-$2.22$\times 10^{22}$) &     206.87    (+80.80,    $-$72.23)    &     121.84  $\pm$ 31.19   &     129.80  $\pm$ 25.64  &     16.45   (+125.63,   $-$19.38)   \\
  {\bf 02:20:01.64 } &   {\bf $-$05:22:17.1 } &    {\bf 2.219 $\pm$ 0.004} &    2.71$\times 10^{26}$  (+1.21$\times 10^{24}$, $-$1.21$\times 10^{24}$) &     159867.61 (+49704.91, $-$49267.97) &     405.06  $\pm$ 428.11  &     401.30  $\pm$ 107.06 &     6245.54 (+73609.86, $-$7172.54) \\
  02:20:15.60 &   $-$04:48:30.2 &    2.800   $\pm$ 0.002 &    4.13$\times 10^{24}$  (+1.03$\times 10^{23}$, $-$1.03$\times 10^{23}$) &     2436.51   (+822.73,   $-$785.49)   &     380.81  $\pm$ 257.57  &     532.00  $\pm$ 121.46 &     96.53   (+1196.41,  $-$113.22)  \\
  02:20:46.21 &   $-$04:20:38.4 &    2.129 $\pm$ 0.009 &    9.76$\times 10^{23}$  (+5.62$\times 10^{22}$, $-$5.52$\times 10^{22}$) &     576.05    (+219.00,   $-$198.39)   &     190.94  $\pm$ 55.72   &     131.20  $\pm$ 50.68  &     32.45   (+330.04,   $-$38.53)   \\
  02:20:58.34 &   $-$04:11:49.9 &    3.200   $\pm$ 0.001 &    5.57$\times 10^{23}$  (+1.47$\times 10^{23}$, $-$1.47$\times 10^{23}$) &     328.82    (+213.74,   $-$160.63)   &     78.32   $\pm$ 55.17   &     234.00  $\pm$ 105.59 &     30.37   (+155.53,   $-$37.74)   \\
  02:21:37.73 &   $-$04:31:24.9 &    1.714 $\pm$ 0.002 &    2.23$\times 10^{23}$  (+3.68$\times 10^{22}$, $-$3.66$\times 10^{22}$) &     131.62    (+68.50,    $-$55.17)    &     136.89  $\pm$ 33.42   &     122.50  $\pm$ 30.83  &     13.56   (+57.26,    $-$16.36)   \\
  02:21:37.90 &   $-$05:13:36.9 &    1.110  $\pm$ 0.001 &    2.26$\times 10^{23}$  (+8.41$\times 10^{21}$, $-$8.38$\times 10^{21}$) &     133.19    (+47.11,    $-$44.07)    &     76.54   $\pm$ 19.59   &     49.74  $\pm$ 17.43  &     9.31    (+58.57,    $-$10.71)   \\
  02:21:39.48 &   $-$04:50:04.1 &    2.431 $\pm$ 0.013 &   $-$1.17$\times 10^{22}$ (+9.81$\times 10^{22}$, $-$9.58$\times 10^{22}$) &    $-$6.88     (+73.40,    $-$37.16)    &     17.20    $\pm$ 138.94  &     18.97  $\pm$ 10.94  &     15.91   (+3.51,    $-$20.72)   \\
  02:22:14.49 &   $-$04:31:33.7 &    1.817 $\pm$ 0.010  &    3.31$\times 10^{23}$  (+3.58$\times 10^{22}$, $-$3.50$\times 10^{22}$) &     195.57    (+87.21,    $-$74.00)    &     330.23  $\pm$ 15.21   &     176.50  $\pm$ 8.42   &     15.42   (+113.86,   $-$18.59)   \\
  02:22:15.99 &   $-$05:16:01.3 &    1.230  $\pm$ 0.001 &    3.30$\times 10^{23}$   (+1.69$\times 10^{22}$, $-$1.69$\times 10^{22}$) &     194.61    (+72.41,    $-$66.29)    &     98.20    $\pm$ 14.75   &     132.00  $\pm$ 20.55  &     14.23   (+124.39,   $-$16.51)   \\
  02:22:36.10 &   $-$04:33:08.0 &    1.210  $\pm$ 0.005 &    3.56$\times 10^{22}$  (+1.28$\times 10^{22}$, $-$1.25$\times 10^{22}$) &     21.02     (+16.23,    $-$11.54)    &     50.97   $\pm$ 18.58   &     68.57  $\pm$ 18.98  &     4.44    (+13.07,    $-$5.44)    \\
  02:22:37.89 &   $-$04:11:40.4 &    1.496 $\pm$ 0.006 &    2.91$\times 10^{22}$  (+2.38$\times 10^{22}$, $-$2.34$\times 10^{22}$) &     17.15     (+23.56,    $-$14.82)    &     40.10    $\pm$ 11.22   &     64.17  $\pm$ 10.92  &     6.31    (+11.72,    $-$7.88)    \\
  02:22:38.60 &   $-$04:43:22.8 &    1.557 $\pm$ 0.001 &    6.25$\times 10^{22}$  (+2.50$\times 10^{22}$, $-$2.49$\times 10^{22}$) &     36.88     (+30.49,    $-$21.46)    &     26.59   $\pm$ 17.25   &     16.97  $\pm$ 15.53  &     7.39    (+25.94,    $-$9.04)    \\
  02:22:39.18 &   $-$04:41:57.8 &    1.189 $\pm$ 0.001 &    1.19$\times 10^{23}$  (+1.05$\times 10^{22}$, $-$1.05$\times 10^{22}$) &     70.27     (+29.55,    $-$25.74)    &     46.47   $\pm$ 684.93  &     48.11  $\pm$ 3.94   &     6.42    (+51.70,    $-$7.54)    \\
  02:22:39.99 &   $-$04:39:22.9 &    1.698 $\pm$ 0.001 &    4.68$\times 10^{23}$  (+3.58$\times 10^{22}$, $-$3.57$\times 10^{22}$) &     276.09    (+111.82,   $-$98.88)    &     195.46  $\pm$ 6.75    &     257.00  $\pm$ 7.14   &     19.68   (+136.82,   $-$23.26)   \\
  02:22:54.40 &   $-$05:28:54.8 &    1.213 $\pm$ 0.002 &    1.11$\times 10^{23}$  (+2.55$\times 10^{22}$, $-$2.53$\times 10^{22}$) &     65.27     (+39.56,    $-$30.30)    &     51.09   $\pm$ 30.38   &     76.85  $\pm$ 18.96  &     10.11   (+44.71,    $-$12.18)   \\
  02:22:56.45 &   $-$05:20:55.0 &    1.354 $\pm$ 0.001 &    6.69$\times 10^{23}$  (+5.98$\times 10^{22}$, $-$5.96$\times 10^{22}$) &     394.59    (+166.45,   $-$144.83)   &     156.00   $\pm$ 43.66   &     111.50  $\pm$ 55.96  &     33.93   (+163.51,   $-$39.99)   \\
  02:23:06.05 &   $-$04:33:23.8 &    1.084 $\pm$ 0.004 &    2.51$\times 10^{23}$  (+1.13$\times 10^{22}$, $-$1.12$\times 10^{22}$) &     148.09    (+53.90,    $-$49.75)    &     48.35   $\pm$ 20.62   &     27.26  $\pm$ 14.69  &     11.00    (+65.23,    $-$12.73)   \\
  02:23:09.36 &   $-$04:21:50.3 &    2.003 $\pm$ 0.009 &    4.07$\times 10^{23}$  (+5.92$\times 10^{22}$, $-$5.80$\times 10^{22}$) &     240.33    (+118.87,   $-$97.11)    &     136.76  $\pm$ 33.39   &     87.76  $\pm$ 25.04  &     20.80    (+104.00,   $-$25.34)   \\
  02:23:19.03 &   $-$04:46:14.0 &    1.982 $\pm$ 0.007 &    8.05$\times 10^{23}$  (+4.29$\times 10^{22}$, $-$4.23$\times 10^{22}$) &     475.24    (+178.06,   $-$162.35)   &     166.52  $\pm$ 44.61   &     201.20  $\pm$ 39.64  &     27.27   (+272.69,   $-$32.22)   \\
  02:23:32.14 &   $-$04:57:39.4 &    0.602 $\pm$ 0.001 &    2.48$\times 10^{22}$  (+2.63$\times 10^{21}$, $-$2.61$\times 10^{21}$) &     14.63     (+6.49,     $-$5.54)     &     40.15   $\pm$ 11.24   &     14.00   $\pm$ 7.17   &     1.96    (+8.91,     $-$2.27)    \\
  02:23:35.05 &   $-$04:34:22.7 &    2.208 $\pm$ 0.001 &    1.18$\times 10^{23}$  (+7.10$\times 10^{22}$, $-$7.09$\times 10^{22}$) &     69.52     (+75.88,    $-$50.26)    &     33.42   $\pm$ 177.48  &     14.52  $\pm$ 9.06   &     15.44   (+29.36,    $-$19.24)   \\
  02:23:42.37 &   $-$05:22:43.8 &    2.037 $\pm$ 0.002 &    2.53$\times 10^{23}$  (+5.20$\times 10^{22}$, $-$5.18$\times 10^{22}$) &     149.05    (+85.54,    $-$66.71)    &     101.83  $\pm$ 86.57   &     106.60  $\pm$ 30.42  &     15.80    (+74.06,    $-$19.26)   \\
  02:23:50.71 &   $-$04:27:03.7 &    1.180  $\pm$ 0.010  &    2.89$\times 10^{22}$  (+1.53$\times 10^{22}$, $-$1.47$\times 10^{22}$) &     17.08     (+16.96,    $-$11.23)    &     24.61   $\pm$ 140.20   &     19.94  $\pm$ 3.78   &     4.90     (+13.44,    $-$6.13)    \\
  02:23:55.07 &   $-$05:15:39.6 &    2.408 $\pm$ 0.001 &    2.54$\times 10^{21}$  (+6.94$\times 10^{22}$, $-$6.93$\times 10^{22}$) &     1.50       (+53.91,    $-$28.87)    &     82.79   $\pm$ 67.90    &     165.80  $\pm$ 33.98  &     11.83   (+0.83,     $-$15.03)   \\
  02:23:55.13 &   $-$04:55:20.1 &    1.529 $\pm$ 0.001 &    1.05$\times 10^{24}$  (+2.09$\times 10^{22}$, $-$2.09$\times 10^{22}$) &     621.09    (+205.61,   $-$198.05)   &     263.63  $\pm$ 903.65  &     89.53  $\pm$ 21.68  &     33.03   (+339.89,   $-$37.95)   \\
  02:24:01.60 &   $-$04:40:33.1 &    1.187 $\pm$ 0.007 &    6.63$\times 10^{23}$  (+2.88$\times 10^{22}$, $-$2.82$\times 10^{22}$) &     391.26    (+141.54,   $-$130.93)   &     59.05   $\pm$ 455.35  &     37.17  $\pm$ 1.02   &     27.52   (+192.56,   $-$31.96)   \\
  02:24:13.46 &   $-$04:52:10.4 &    2.481 $\pm$ 0.008 &    4.01$\times 10^{23}$  (+9.10$\times 10^{22}$, $-$8.98$\times 10^{22}$) &     236.55    (+142.25,   $-$108.97)   &     279.27  $\pm$ 375.46  &     454.80  $\pm$ 111.88 &     23.32   (+109.80,   $-$28.87)   \\
  02:24:19.75 &   $-$04:03:15.7 &    2.276 $\pm$ 0.001 &    6.16$\times 10^{23}$  (+7.07$\times 10^{22}$, $-$7.06$\times 10^{22}$) &     363.66    (+165.42,   $-$139.90)   &     73.91   $\pm$ 431.04  &     76.16  $\pm$ 1.71   &     26.02   (+244.68,   $-$31.34)   \\
  02:24:44.29 &   $-$04:19:47.8 &    1.607 $\pm$ 0.001 &    8.91$\times 10^{22}$  (+3.36$\times 10^{22}$, $-$3.35$\times 10^{22}$) &     52.60      (+41.89,    $-$29.77)    &     21.59   $\pm$ 131.95  &     28.31  $\pm$ 2.34   &     9.87    (+35.53,    $-$12.08)   \\
  02:24:44.33 &   $-$04:45:36.0 &    2.520  $\pm$ 0.050  &    1.46$\times 10^{24}$  (+1.38$\times 10^{23}$, $-$1.30$\times 10^{23}$) &     862.46    (+369.43,   $-$316.61)   &     30.64   $\pm$ 37.74   &     549.90  $\pm$ 146.17 &     51.81   (+417.84,   $-$63.73)   \\
  02:24:48.69 &   $-$04:56:06.9 &    1.654 $\pm$ 0.006 &    8.57$\times 10^{23}$  (+5.69$\times 10^{22}$, $-$5.61$\times 10^{22}$) &     505.54    (+198.05,   $-$177.22)   &     0.08    $\pm$ 0.56    &     0.35   $\pm$ 1.60    &     34.49   (+738.38,   $-$40.78)   \\
  02:24:57.57 &   $-$04:56:59.5 &    1.760  $\pm$ 0.050  &    2.86$\times 10^{23}$  (+4.55$\times 10^{22}$, $-$4.17$\times 10^{22}$) &     169.02    (+86.62,    $-$68.65)    &     81.57   $\pm$ 17.99   &     90.61  $\pm$ 22.31  &     16.00    (+108.93,   $-$20.23)   \\
   {\bf 02:25:03.12 } &  {\bf  $-$04:40:25.3} &   {\bf  1.741 $\pm$ 0.002 } &    6.77$\times 10^{24}$  (+5.07$\times 10^{22}$, $-$5.05$\times 10^{22}$) &     3994.13   (+1257.62,  $-$1239.27)  &     93.31   $\pm$ 27.23   &     118.90  $\pm$ 16.72  &     181.14  (+2494.54,  $-$207.07)  \\
  02:25:08.57 &   $-$04:25:12.8 &    2.150  $\pm$ 0.050  &    2.66$\times 10^{23}$  (+6.10$\times 10^{22}$, $-$5.60$\times 10^{22}$) &     156.99    (+94.90,    $-$70.86)    &     19.76   $\pm$ 20.23   &     63.37  $\pm$ 12.43  &     16.34   (+106.56,   $-$21.09)   \\
   {\bf 02:25:25.68 } &   {\bf $-$04:35:09.6 } &   {\bf  2.168 $\pm$ 0.002 } &    1.23$\times 10^{25}$  (+7.37$\times 10^{22}$, $-$7.35$\times 10^{22}$) &     7263.43   (+2272.75,  $-$2246.11)  &     50.50    $\pm$ 21.53   &     126.30  $\pm$ 19.67  &     290.73  (+3815.15,  $-$334.07)  \\
  02:25:43.53 &   $-$04:28:34.6 &    3.406 $\pm$ 0.009 &    1.71$\times 10^{24}$  (+1.45$\times 10^{23}$, $-$1.44$\times 10^{23}$) &     1010.87   (+420.17,   $-$367.27)   &     125.04  $\pm$ 5.76    &     585.90  $\pm$ 20.20   &     48.84   (+415.92,   $-$59.28)   \\
  02:25:52.16 &   $-$04:05:16.2 &    1.441 $\pm$ 0.003 &    3.22$\times 10^{23}$  (+3.20$\times 10^{22}$, $-$3.18$\times 10^{22}$) &     189.96    (+82.63,    $-$70.97)    &     287.57  $\pm$ 308.76  &     104.30  $\pm$ 35.00   &     16.54   (+116.90,   $-$19.64)   \\

    \hline
    \end{tabular}
    \caption{{\it Continued} -- Positions, redshifts, MeerKAT (scaled to 1.4-GHz) radio luminosities, and SFR estimates for 23 COSMOS quasars and 81 XMM-LSS quasars in the final sample of 104 quasars. Those with positions and redshifts highlighted in bold are classed as `radio loud' (Figure~\ref{fig:radioloudness}) via the radio-loudness parameter, $R$, defined by \citet{Kellermann1989}. The `Yun-SFR' and the `Delhaize-SFR' are based upon the radio luminosity, and applying the relation by \citet{Yun2001} (Equation~\ref{eqn:Yunrelation}) and by \citet{Delhaize2017} (Equations~\ref{eqn:bestq_delhaize}--\ref{eqn:sfr_delhaize}), respectively. The `Kennicutt-SFR' is derived through two-component SED-fitting over 8--1000$\upmu$m \citep{Jin2018} and applying Equation~\ref{eqn:Kennicutt} \citep{Kennicutt1998b}, whilst the `SED3FIT-SFR' is that output by the SED-fitting code of \citet{Berta2013}.} 
    
\end{table}
\end{landscape}

\setcounter{table}{0} 

\begin{landscape}
\begin{table}
    \centering
    \begin{tabular}{c|c|c|c|c|c|c|c|c}
    \hline
    R.A. & Dec.  & spec-$z$  & $L_{\mathrm{1.4\,GHz}}$ / W\,Hz$^{-1}$ & Yun--SFR / $\Msol\,\mathrm{yr}^{-1}$ & Kennicutt--SFR  & SED3FIT--SFR  & Delhaize-SFR / $\Msol\,\mathrm{yr}^{-1}$ \\ 
        (hms) & (dms)  & $\pm$ error  & (+ error, $-$ error) &  (+ error, $-$ error) & $\pm$ error / $\Msol\,\mathrm{yr}^{-1}$ & $\pm$ error / $\Msol\,\mathrm{yr}^{-1}$ & (+ error, $-$ error) \\ 
       \hline

  02:25:54.87 &   $-$05:13:54.4 &    1.258 $\pm$ 0.002 &    1.49$\times 10^{23}$  (+2.38$\times 10^{22}$, $-$2.36$\times 10^{22}$) &     87.90      (+45.11,    $-$36.49)    &     130.39  $\pm$ 19.59   &     133.10  $\pm$ 19.34  &     10.57   (+50.46,    $-$12.63)   \\
  02:25:55.43 &   $-$04:39:18.3 &    1.030  $\pm$ 0.001 &    2.49$\times 10^{23}$  (+6.97$\times 10^{21}$, $-$6.94$\times 10^{21}$) &     146.73    (+50.13,    $-$47.61)    &     100.93  $\pm$ 30.66   &     13.81  $\pm$ 14.98  &     9.92    (+99.99,    $-$11.33)   \\
  02:26:00.99 &   $-$05:06:58.2 &    1.838 $\pm$ 0.001 &    8.83$\times 10^{23}$  (+6.91$\times 10^{22}$, $-$6.89$\times 10^{22}$) &     521.23    (+212.22,   $-$187.28)   &     188.84  $\pm$ 57.38   &     141.40  $\pm$ 28.53  &     35.83   (+282.70,   $-$42.46)   \\
  02:26:01.59 &   $-$04:59:19.0 &    1.660  $\pm$ 0.001 &    5.42$\times 10^{23}$  (+6.09$\times 10^{22}$, $-$6.08$\times 10^{22}$) &     319.80     (+144.48,   $-$122.48)   &     84.04   $\pm$ 34.79   &     36.37  $\pm$ 22.39  &     27.36   (+185.11,   $-$32.62)   \\
  02:26:04.13 &   $-$04:50:20.4 &    2.373 $\pm$ 0.014 &    2.00$\times 10^{23}$   (+5.41$\times 10^{22}$, $-$5.27$\times 10^{22}$) &     118.09    (+77.66,    $-$57.64)    &     266.12  $\pm$ 15.33   &     456.40  $\pm$ 20.91  &     13.35   (+53.39,    $-$16.70)   \\
  02:26:12.64 &   $-$04:34:01.4 &    2.305 $\pm$ 0.004 &    1.24$\times 10^{23}$  (+5.18$\times 10^{22}$, $-$5.14$\times 10^{22}$) &     73.32     (+62.23,    $-$43.43)    &     56.71   $\pm$ 17.91   &     245.20  $\pm$ 60.8   &     11.72   (+40.16,    $-$14.60)   \\
  02:26:24.64 &   $-$04:20:02.4 &    2.236 $\pm$ 0.006 &    9.00$\times 10^{23}$   (+5.65$\times 10^{22}$, $-$5.59$\times 10^{22}$) &     531.27    (+205.58,   $-$184.99)   &     38.93   $\pm$ 283.27  &     26.50   $\pm$ 20.35  &     29.96   (+262.13,   $-$35.64)   \\
  02:26:33.31 &   $-$04:29:47.8 &    2.147 $\pm$ 0.001 &    8.27$\times 10^{23}$  (+4.91$\times 10^{22}$, $-$4.90$\times 10^{22}$) &     488.16    (+186.77,   $-$169.04)   &     370.65  $\pm$ 64.33   &     209.90  $\pm$ 67.18  &     27.78   (+265.09,   $-$32.86)   \\
   {\bf 02:26:40.79 } &   {\bf $-$04:16:36.3} &    {\bf 2.088 $\pm$ 0.001} &    4.92$\times 10^{24}$  (+4.91$\times 10^{22}$, $-$4.90$\times 10^{22}$) &     2899.96   (+922.57,   $-$904.84)   &     1077.01 $\pm$ 136.76  &     1148.00 $\pm$ 395.96 &     122.12  (+1263.03,  $-$140.59)  \\
  02:26:46.98 &   $-$04:18:38.1 &    1.581 $\pm$ 0.002 &    5.74$\times 10^{23}$  (+2.74$\times 10^{22}$, $-$2.73$\times 10^{22}$) &     338.84    (+124.51,   $-$114.57)   &     284.61  $\pm$ 52.72   &     149.40  $\pm$ 67.39  &     21.35   (+159.71,   $-$24.94)   \\
  02:26:52.14 &   $-$04:05:57.1 &    1.430  $\pm$ 0.004 &    8.86$\times 10^{22}$  (+2.55$\times 10^{22}$, $-$2.52$\times 10^{22}$) &     52.27     (+35.62,    $-$26.29)    &     144.25  $\pm$ 43.83   &     183.60  $\pm$ 20.45  &     8.58    (+29.66,    $-$10.48)   \\
  02:26:55.57 &   $-$04:42:17.9 &    2.306 $\pm$ 0.002 &    2.82$\times 10^{23}$  (+6.49$\times 10^{22}$, $-$6.46$\times 10^{22}$) &     166.21    (+100.65,   $-$77.20)    &     86.71   $\pm$ 872.90   &     140.80  $\pm$ 6.99   &     17.47   (+98.96,    $-$21.44)   \\
  02:26:59.90 &   $-$04:44:30.6 &    1.611 $\pm$ 0.001 &    5.63$\times 10^{23}$  (+3.00$\times 10^{22}$, $-$2.99$\times 10^{22}$) &     332.37    (+124.47,   $-$113.65)   &     195.78  $\pm$ 43.17   &     130.90  $\pm$ 26.21  &     21.43   (+125.98,   $-$25.08)   \\
  02:27:16.12 &   $-$04:45:39.0 &    0.722 $\pm$ 0.001 &    2.09$\times 10^{22}$  (+5.36$\times 10^{21}$, $-$5.32$\times 10^{21}$) &     12.32     (+7.88,     $-$5.94)     &     43.11   $\pm$ 13.1    &     62.60   $\pm$ 8.35   &     2.66    (+7.81,     $-$3.16)    \\
  02:27:43.87 &   $-$05:09:52.6 &    2.211 $\pm$ 0.001 &    2.13$\times 10^{23}$  (+5.06$\times 10^{22}$, $-$5.05$\times 10^{22}$) &     125.43    (+77.21,    $-$58.96)    &     128.22  $\pm$ 72.82   &     70.78  $\pm$ 30.97  &     13.86   (+50.50,    $-$16.97)   \\
  02:27:46.51 &   $-$04:11:08.6 &    1.647 $\pm$ 0.005 &    1.82$\times 10^{23}$  (+3.20$\times 10^{22}$, $-$3.16$\times 10^{22}$) &     107.17    (+57.35,    $-$45.65)    &     209.33  $\pm$ 1470.99 &     82.28  $\pm$ 1.39   &     11.70    (+67.71,    $-$14.19)   \\
  02:27:57.00 &   $-$04:01:30.8 &    1.907 $\pm$ 0.001 &    1.21$\times 10^{24}$  (+4.74$\times 10^{22}$, $-$4.73$\times 10^{22}$) &     714.19    (+254.36,   $-$237.26)   &     236.81  $\pm$ 1317.96 &     98.93  $\pm$ 1.03   &     38.60    (+433.52,   $-$45.10)   \\
  02:28:16.46 &   $-$04:59:38.5 &    0.706 $\pm$ 0.001 &    1.02$\times 10^{23}$  (+5.94$\times 10^{21}$, $-$5.90$\times 10^{21}$) &     60.30      (+22.97,    $-$20.81)    &     37.25   $\pm$ 349.84  &     64.99  $\pm$ 1.68   &     5.82    (+41.31,    $-$6.68)    \\
  02:28:32.81 &   $-$04:22:45.8 &    1.757 $\pm$ 0.001 &    3.74$\times 10^{23}$  (+5.06$\times 10^{22}$, $-$5.05$\times 10^{22}$) &     220.83    (+106.33,   $-$88.06)    &     2.16    $\pm$ 10.68   &     18.61  $\pm$ 0.71   &     20.09   (+177.64,   $-$24.10)   \\
  02:28:35.93 &   $-$04:15:56.1 &    1.372 $\pm$ 0.001 &    4.58$\times 10^{23}$  (+2.93$\times 10^{22}$, $-$2.92$\times 10^{22}$) &     270.48    (+105.06,   $-$94.48)    &     151.30   $\pm$ 1113.86 &     59.61  $\pm$ 1.23   &     20.17   (+181.96,   $-$23.60)   \\

    \hline
    \end{tabular}
    \caption{{\it Continued} -- Positions, redshifts, MeerKAT (scaled to 1.4-GHz) radio luminosities, and SFR estimates for 23 COSMOS quasars and 81 XMM-LSS quasars in the final sample of 104 quasars. Those with positions and redshifts highlighted in bold are classed as `radio loud' (Figure~\ref{fig:radioloudness}) via the radio-loudness parameter, $R$, defined by \citet{Kellermann1989}. The `Yun-SFR' and the `Delhaize-SFR' are based upon the radio luminosity, and applying the relation by \citet{Yun2001} (Equation~\ref{eqn:Yunrelation}) and by \citet{Delhaize2017} (Equations~\ref{eqn:bestq_delhaize}--\ref{eqn:sfr_delhaize}), respectively. The `Kennicutt-SFR' is derived through two-component SED-fitting over 8--1000$\upmu$m \citep{Jin2018} and applying Equation~\ref{eqn:Kennicutt} \citep{Kennicutt1998b}, whilst the `SED3FIT-SFR' is that output by the SED-fitting code of \citet{Berta2013}.} 
    
\end{table}
\end{landscape}

\section{IR luminosities of quasars}
\label{app:lfir}

\begin{figure}
    \centering

	\includegraphics[width=1.0\linewidth]{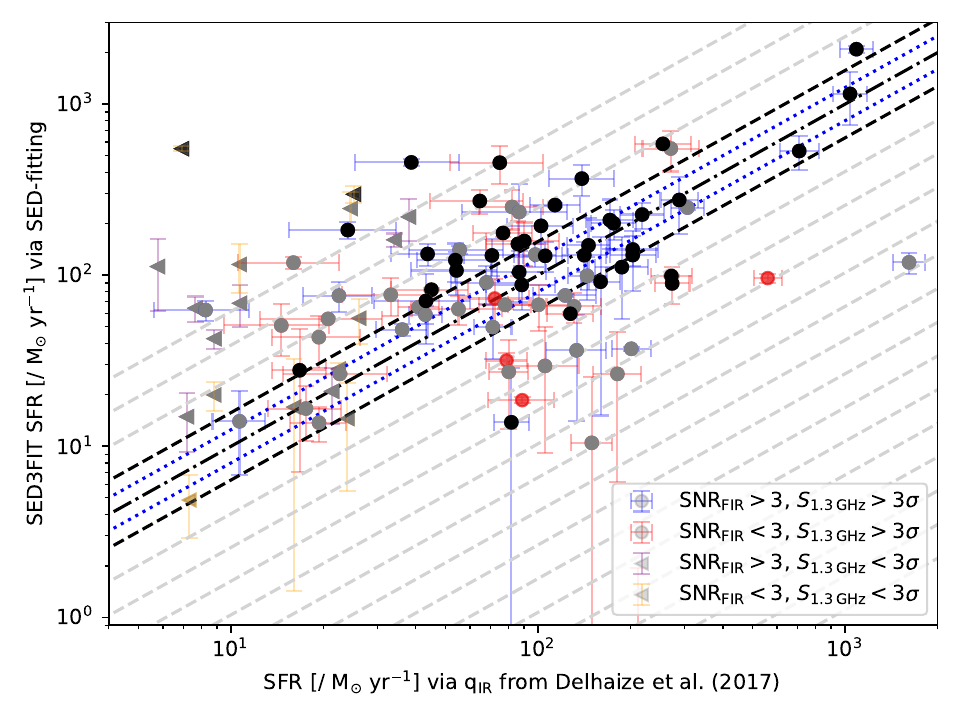}
	
    \caption{A comparison of star-formation rates (SFRs) for quasars in the COSMOS and XMM-LSS fields. The SFR estimate on the ordinate axis is derived through SED-fitting of the infra-red photometry (Section~\ref{sec:infrareddata}) using equation~3 of \citet{Kennicutt1998b}, which is specifically for starbursts. The SFR estimate on the abscissa axis assumes that all of the radio emission can be attributed to star formation, and so lies on the IRRC (Equation~\ref{eqn:bestq_delhaize}) determined by \citet{Delhaize2017} at the median redshift of the sample ($z = 1.698$). The dash-dotted line represents where the two SFR estimates are equal to each other, with the blue-dotted lines indicating the error associated with the Delhaize-SFR, and the dashed lines indicating the total error in the one-to-one relation (i.e. the 1-$\sigma_{1:1}$ calibration error in both the Delhaize-SFR and the SED3FIT-SFR, shown in black, and grey dashed lines for 2\,$\sigma_{1:1}$, 3\,$\sigma_{1:1}$, etc.). The error-bars are colour-coded by the signal-to-noise ratio (SNR$_{\mathrm{FIR}}$) for the FIR data (see Section~\ref{sec:SFR_comparisons}), and symbols indicate the detection level (above or below $3\sigma$) for the radio data at 1.3\,GHz. {\bf The grey markers represent quasars with IR luminosities that are typical of LIRGs ($>10^{11}$\,\Lsol), whilst the black markers represent quasars with IR luminosities that are typical of ULIRGs ($>10^{12}$\,\Lsol).  This `starbursty' nature  means that the radio emission is a poorer tracer of star formation than in the case of normal star-forming galaxies, due to radio emission being produced on longer timescales than the IR emission}.}
\label{fig:fir_luminosities}
\end{figure}

See Figure~\ref{fig:fir_luminosities} for validity of the suggestion (Section~\ref{sec:SFR_comparisons}) that datapoints lying above the one-to-one relation are possibly starburst systems. Such an interpretation also aligns with the evolutionary model of quasars, where they exhibit rapid star-formation as a result of undergoing gas-rich mergers \citep[e.g.][]{Hopkins2008}.


\bsp	
\label{lastpage}
\end{document}